\documentclass[useAMS,usenatbib]{mn2e}

\usepackage{graphicx}
\usepackage{rotating}
\usepackage{caption}

\title[Abundance drops in galaxy groups and clusters]{A volume-limited sample of X-ray galaxy groups and clusters - III. Central abundance drops}
\author[E. K. Panagoulia, J. S. Sanders \& A. C. Fabian]{E. K. Panagoulia$^{1}$\thanks{E-mail:
epanagoulia@ast.cam.ac.uk}, J. S. Sanders$^{2}$ and A. C. Fabian$^{1}$\\
$^{1}$Institute of Astronomy, Madingley Road, Cambridge CB3 0HA\\
$^{2}$Max-Planck-Institut f\"{u}r extraterrestrische Physik, 85748, Garching, Germany.}
\begin{document}

\renewcommand{\arraystretch}{1.7}

\date{Accepted . Received ; in original form }
\pagerange{\pageref{firstpage}--\pageref{lastpage}} \pubyear{2014}

\maketitle

\label{firstpage}

\begin{abstract}
We present the results of a search and study of central abundance drops in a volume-limited sample ($z$ $\leq$ 0.071) of 101 X-ray galaxy groups and clusters. These are best observed in nearby, and so best resolved, groups and clusters, making our sample ideal for their detection. Out of the 65 groups and clusters in our sample for which we have abundance profiles, 8 of them have certain central abundance drops, with possible central abundance drops in another 6. All sources with central abundance drops have X-ray cavities, and all bar one exception have a central cooling time $\leq$1 Gyr. These central abundance drops can be generated if the iron injected by stellar mass loss processes in the core of these sources is in grains, which then become incorporated in the central dusty filaments. These, in turn, are dragged outwards by the bubbling feedback process in these sources. 
%if the iron injected by stellar mass loss processes in the central regions is in the form of grains, which remain as such while the surrounding injected cold dusty gas blends with and, eventually, becomes incorporated in the dusty filaments, also seen in the cores of these sources. The bubbling feedback process, observed in the cores of all the sources, then drags these dusty filaments outwards, where, at some distance, they are destroyed and enrich the local ICM with metals. 
We find that data quality significantly affects the detection of central abundance drops, inasmuch as a higher number of counts in the central 20 kpc of a source makes it easier to detect a central abundance drop, as long as these counts are more than $\sim$13000. On the other hand, the magnitude of the central abundance drop does not depend on the number of these counts, though the statistical significance of the measured drop does. Finally, in line with the scenario briefly outlined above, we find that, for most sources, the location of X-ray cavities acts as an upper limit to the location of the peak in the radial metallicity distribution.   
\end{abstract}


\section{Introduction}
Many X-ray galaxy groups and clusters exhibit a strong central surface brightness peak and temperature drop, which are the fingerprint of a cooling flow \citep[for a review, see][]{Fabian94}. However, predictions of the simple cooling flow model often disagree with the observed properties of X-ray groups and clusters. First of all, despite the temperature declining towards the centre of groups and clusters, the amount of gas seen to be cooling at temperatures 2--3 times lower than the ambient temperature is much lower than that predicted \citep{McNamara07, McNamara12, Peterson06}. In addition, the expected star formation rates in group and cluster cores fall short of the predicted values, as revealed by UV and optical observations \citep{Nulsen87}. A similar discrepancy is present between the predicted amount of cold gas in cluster cores, and observations. 

It is therefore clear that, since the intracluster medium (ICM) is not cooling at the expected rate, one or more mechanisms must be heating it up. The main, and most favoured, heating mechanism is feedback from the central active galactive nuclei (AGNs) in group and cluster cores. AGN outbursts deposit large amounts of energy into the surrounding ICM, which are often sufficient to reduce, and sometimes quench, cooling flows \citep[for reviews, see][]{McNamara07, McNamara12, Fabian12}. A number of studies now support the idea that AGNs affect their environment out to large distances \citep{Birzan04, Birzan12, Dunn08}. Generally, it is believed that radio-loud AGNs produce strong outflows in the form of jets, which in turn inflate bubbles filled with radio-emitting plasma. These bubbles displace the hot ICM, creating depressions in the X-ray surface brightness, which are called X-ray cavities. If the jets are very powerful, the radio bubbles will expand supersonically, resulting in a shock that will heat and compress the surrounding gas. In most cases, the jets are relatively weak, meaning that the jets are growing subsonically without shock-heating the ICM. 

Many studies, both observational and theoretical, have examined the chemodynamical impact of AGN-driven X-ray cavities on their surrounding ICM. AGN-induced mixing has been studied using both 2-D and 3-D hydrodynamical simulations \citep[e.g.][]{Bruggen02, Heath07, Roediger07, Churazov13, Planelles14}. Generally, these studies concluded that low power, short-lived ($\sim$100 Myr) jets affect pre-existing metallicity gradients by only a few tens of percent, at best. On the other hand, quasar-like AGN outbursts can decrease metallicity gradients down to 10 percent of their initial value, by moving metals outwards, mainly through convection and entrainment of gas behind buoyantly rising bubbles in the ICM \citep[e.g.][]{Heath07}. Generally, simulations generate anisotropic metallicity distributions, which are aligned along the jet axis. This is chiefly due to the fact that simulations use jets that are launched at the same angle, into a static ICM. These anisotropic metallicity distributions can be made smoother through the inclusion of turbulent mixing and gas circulation \citep[e.g.][]{Mathews03}, and/or bubbles being launched in random directions through e.g. jet precession \citep{Gitti06}. 

Observationally, cool, metal-enriched X-ray gas is found around the cavities and radio sources in several clusters and groups \citep[e.g.][]{Simionescu09, Kirkpatrick09, OSullivan11}. In addition, there are quite a few spectacular examples of plumes and shells of metal-enriched gas, that appear to have been dragged outwards by jets and radio bubbles propagating through the ICM. Some such sources are the Perseus cluster \citep{Sanders04} and Hydra~A \citep{Nulsen02}. Finally, H$\alpha$ emission surrounding the X-ray cavities in several sources, such as NGC~4696 in the Centaurus cluster \citep{Crawford05a} and Abell~1795 \citep{Crawford05b}, has also been observed. 

Perhaps one of the most interesting metallicity effects of AGN feedback on cool core (CC) groups and clusters, is the presence of central abundance drops. These drops manifest themselves as sudden, large and statistically significant changes in the innermost regions of a radial metallicity profile. Central abundance drops have been observed in a few groups and clusters of galaxies, such as the Centaurus cluster \citep{Sanders02, Panagoulia13}, the Perseus cluster \citep[][though the Perseus cluster shows more of an abundance peak than a central abundance drop]{Sanders07}, the Ophiuchus cluster \citep{Million10}, Abell~2199 \citep{Johnstone02} and HCG~62 \citep{Rafferty13}.

It is therefore clear that AGN-driven outflows can have a profound and lasting impact on the chemodynamical make up of the cores of X-ray groups and clusters. Understanding the processes of chemical enrichment that take place in galaxy groups and clusters will help undestand more about the AGN feedback loop, and possibly even the accretion mechanisms onto the supermassive black holes (SMBHs) that lie at the core of central cluster galaxies. 

Motivated by the findings outlined above, we set out to examine whether central abundance drops are present in any other groups and clusters from our sample of 101 sources. 
%We define a steep central abundance drop as a difference of a factor of about 2 or more between the innermost spectral bins, which show the drop, and the spectral bin which is considered as the peak of the radial abundance distribution. 
Our sample is ideal for this kind of work, as central abundance drops are best observed in nearby, better-resolved groups and clusters. In \cite{Panagoulia13}, we interpreted the central iron abundance drop in NGC~4696 as the attenuation of iron in dust grains in the dusty filaments in the core of the cluster, with the sulphur and silicon abundance drops possibly sharing the same origin. The fact that these central drops persisted even when 2- or 3-temperature components were used to model the thermal emission in the core of NGC~4696 \citep{Panagoulia13}, is important in establishing the robustness of these results. This is because it is well known that, if a single-temperature model is used to account for more complex temperature structure, the measured metallicities can be biased low \citep[e.g.][]{Buote00, Buote03}. This particularly affects the cores of CC clusters, where cooling and multiphase gas are present. In this paper, we take a closer look at the metallicity profiles of 65 sources (including NGC~4696) out of a parent sample of 101 galaxy groups and clusters, which is discussed in detail in \cite{Panagoulia14a} (hereafter referred to as Paper I), for which we have abundance profiles. We find certain central abundance drops in 8 sources (including NGC~4696), and possible central abundance drops in 6 sources. Here, we define a central abundance drop as ``certain'' when the drop in abundance is sharp and statistically significant, and there are no similar features (i.e. a steep abundance drop, followed by an increase in abundance) further out in the radial abundance profile. Sources with ``possible'' central abundance drops have a central drop in metallicity, but the drop is not as statistically significant, and there are features of a similar statistical significance and shape at larger distances in the radial abundance profile. All 14 sources have a central cooling time of $\leq$3 Gyr and X-ray cavities. In \cite{Panagoulia14b} (hereafter referred to as Paper II), we find 30 sources with X-ray cavities and a central cooling time $\leq$3 Gyr. This means that up to $\sim$47 percent of the sources in our sample that have cavities, may also have a central abundance drop. We also check whether the abundance drops in the 13 additional sources are consistent with the interpretation of the abundance drop in NGC~4696. All 14 sources show certain or possible central drops in the radial abundance profile of iron, and some show a similar trend in their silicon and sulphur profiles. 
%These 17 sources are NGC~4636, NGC~4696, 2A0335+096, Abell~2052, Abell~1991, Abell~262, NGC~5813, NGC~5044, NGC~5846, Abell~3581, Abell~496, NGC~6338, IC~1262, Abell~4059, Abell~S1101, HCG~62 and NGC~4325. The first 9 sources have certain central abundance drops, while the last 8 have possible central abundance drops.

The paper is structured as follows: The sample and subsample selection are outlined in Section 2. Section 3 contains a brief discussion of the data preparation, and the spectral analysis. The resulting abundance profiles are presented in Section 4. In Section 5, we calculate the estimated mass of ``missing iron'' for each of the seventeen sources, and propose a theory that can explain the shape of the abundance profiles. In the same section, we investigate whether the detection of a central abundance drop, and its magnitude, depend on the quality of the available data, or the power of the cavities in a source. A brief summary of the paper and the main conclusions is given in Section 6. 

 In this paper, we adopt a flat ${\rm \Lambda}$CDM cosmology, with H$_{0}$ = 71 km s$^{-1}$ Mpc$^{-1}$, $\Omega_{\rm m}$ = 0.27 and $\Omega_{\Lambda}$ = 0.73. All abundances in this paper are relative to solar, as defined in \cite{Anders89}. In all the images shown in this paper, north is to the top and east is to the left. Unless otherwise stated, the errors are at the 90 percent confidence level.

\section{A volume limited sample of X-ray groups and clusters of galaxies}
The motivation behind the work in this paper, but also in Paper I, is the examination of the radial profiles of the properties of galaxy groups and clusters, such as metallicity and temperature. The ultimate goal of this process is to determine the relative importance and impact of non-gravitational processes, such as AGN feedback, on the ICM. The cores of groups and clusters are of a particular interest, as it is there that the effect of AGN feedback is most evident. In order to study the cores of groups and clusters, data of a high spatial resolution are needed. We therefore constructed a sample consisting of nearby X-ray groups and clusters, the vast majority of which have {\it Chandra} and/or {\it XMM-Newton} data available. 

We refer the reader to Section 2 of Paper I for details on the sample selection. We give a brief outline of the selection process below:
\begin{itemize}
  \item{We created a volume-limited sample of sources, using the Northern {\it ROSAT} All-Sky catalogue \citep[NORAS;][]{Bohringer00}, and the {\it ROSAT-ESO} Flux-Limited X-tay galaxy cluster survey \citep[REFLEX;][]{Bohringer04}. These sources were selected to lie at a distance $\leq$300 Mpc. In total, 289 sources from the two catalogues met this limit.} 
      \item{As we require a statistically complete sample, cuts in the X-ray luminosity, $L_{\rm {X}}$, and distance were made, to avoid the inclusion of groups of sources that have no available data.} 
          \item{After the cuts in both $L_{\rm {X}}$ and distance, our final sample contains 101 X-ray groups and clusters of galaxies. All but four of these sources have {\it Chandra} and/or {\it XMM-Newton} data. We use {\it Chandra} data whenever possible, in order to take advantage of the higher spatial resolution and lower background levels of the ACIS detectors on board {\it Chandra}. The details of all the sources in this parent sample, as well as the observations used in the data analysis, are listed in tables 1--4 in Paper I.}
\end{itemize}

We point out that the Perseus cluster, and the two Virgo BCGs, M86 and M87, are not contained in the NORAS and REFLEX catalogues. M86 and M87 are not included due to the difficulty in making individual flux measurements for them, as they are embedded in the Virgo cluster's diffuse emission. The Perseus cluster is excluded due to its low galactic latitude, which means it falls outside the region covered by the REFLEX and NORAS catalogues. 

The samples of \cite{Reiprich02} and \cite{Edge90} show some overlap with our own parent sample. After cross-checking the sources in the aforementioned two samples with those in our sample, we find that the sources listed in the \cite{Reiprich02} and \cite{Edge90} samples, but not ours, are too distant to be included in our sample, or are not present in the REFLEX and NORAS catalogues, or did not meet our selection criteria. 

\subsection{Sources with central abundance drops}
As previously mentioned, it is thought that AGN feedback can, at least in part, shape the radial abundance profiles of X-ray groups and clusters out to large distances from their cores, mainly through the outwards buoyant motion of X-ray cavities \citep{Bruggen03, Planelles14}. As our parent sample consists of nearby groups and clusters of galaxies, it is an ideal ``hunting ground'' for such abundance drops, which are detectable in the innermost regions of the nearest groups and clusters. In addition, our parent sample contains 30 groups and clusters which have X-ray cavities, all of which are listed in table 2 of Paper II. Motivated by these findings, we examine the radial abundance profiles of the 65 groups and clusters of our parent sample for which we have reliable abundance profiles, in search for sources with central abundance drops. We find that eight groups and clusters in our parent sample have certain central abundance drops, while an additional six have possible abundance drops. As previously mentioned, a ``certain`` central abundance drop has to be statistically significant and not replicated at larger distances in the radial abundance profile, while a ``possible'' central abundance drop is less statistically significant, and there are features of a similar significance and shape at larger distances from the source centre. All fourteen sources also have X-ray cavities and a central cooling time $\leq$3 Gyr. These 14 sources are NGC~4636, NGC~4696, 2A0335+096, Abell~1991, Abell~262, NGC~5813, NGC~5044, NGC~5846, Abell~3581, NGC~6338, IC~1262, Abell~S1101, HCG~62 and NGC~4325. The first 8 sources have certain central abundance drops, while the last 6 have possible central abundance drops. The properties of all these sources are listed in tables 1 and 3 in Paper I. All sources without X-ray cavities do not have central abundance drops. 
%The distance between an X-ray cavity and its host source is measured from the centre of the cavity to the centre of the source. The sizes of the X-ray cavities were calculated assuming that they are spherical in shape. In the case of Abell~1991, the X-ray cavities were not readily visible after generating a background-subtracted, exposure-corrected image for the cluster. In this case, we obtained the size of the X-ray cavities and their distance from the centre of the cluster from \cite{Pandge13}. 

\section{Observations and data analysis}
\subsection{Data preparation}

\begin{figure*}
\begin{center}
\includegraphics[trim = 1.2cm 11.5cm 1.2cm 10cm, clip, height=4.0cm, width=8.8cm]{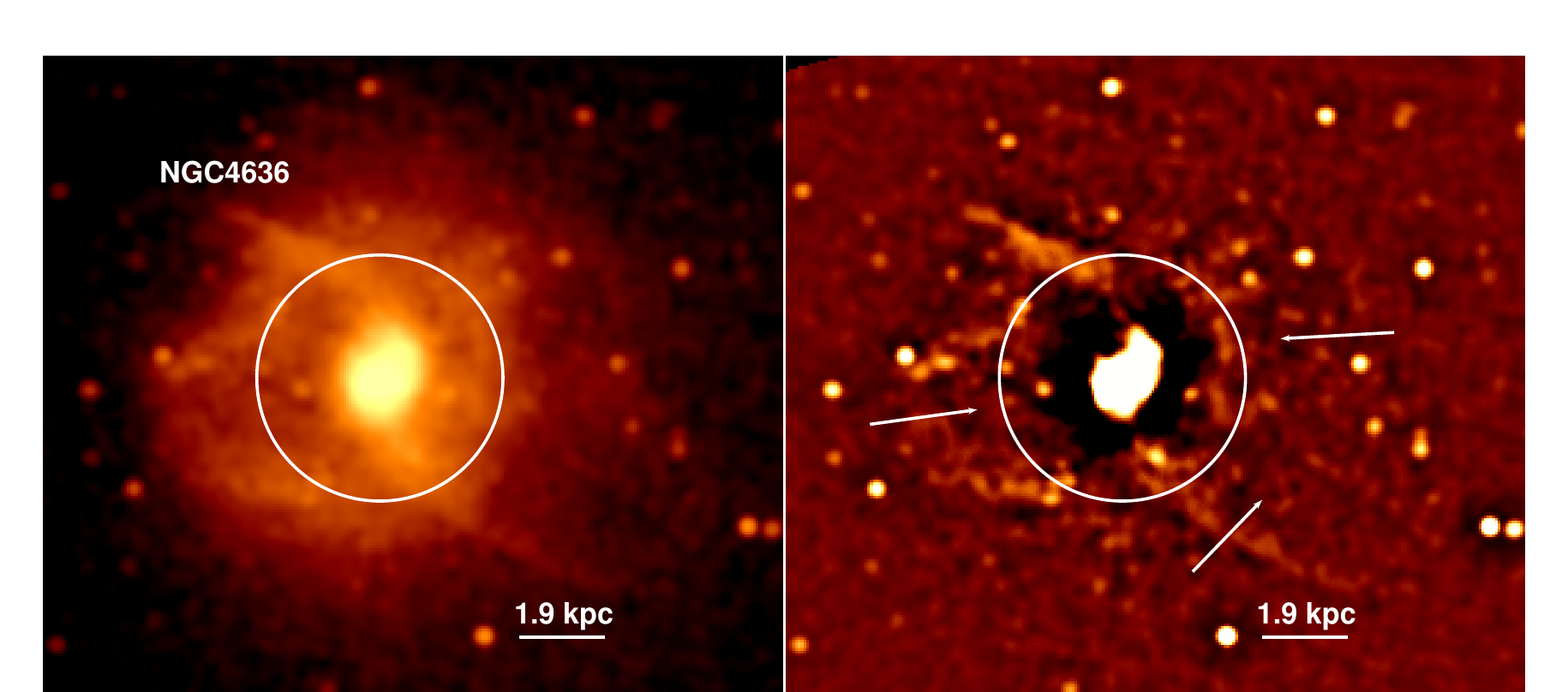}
\includegraphics[trim = 1.2cm 11.5cm 1.2cm 10cm, clip, height=4.0cm, width=8.8cm]{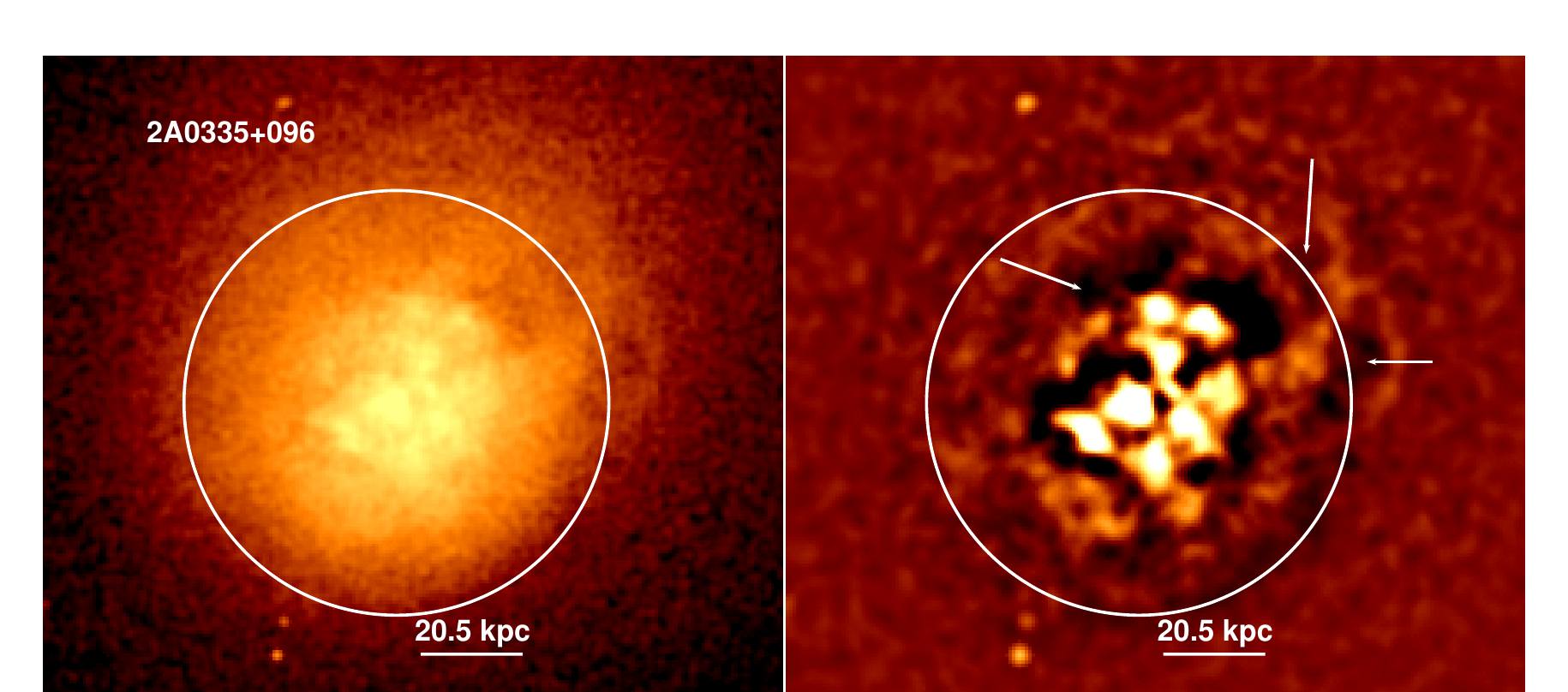}
\includegraphics[trim = 1.2cm 11.5cm 1.2cm 10cm, clip, height=4.0cm, width=8.8cm]{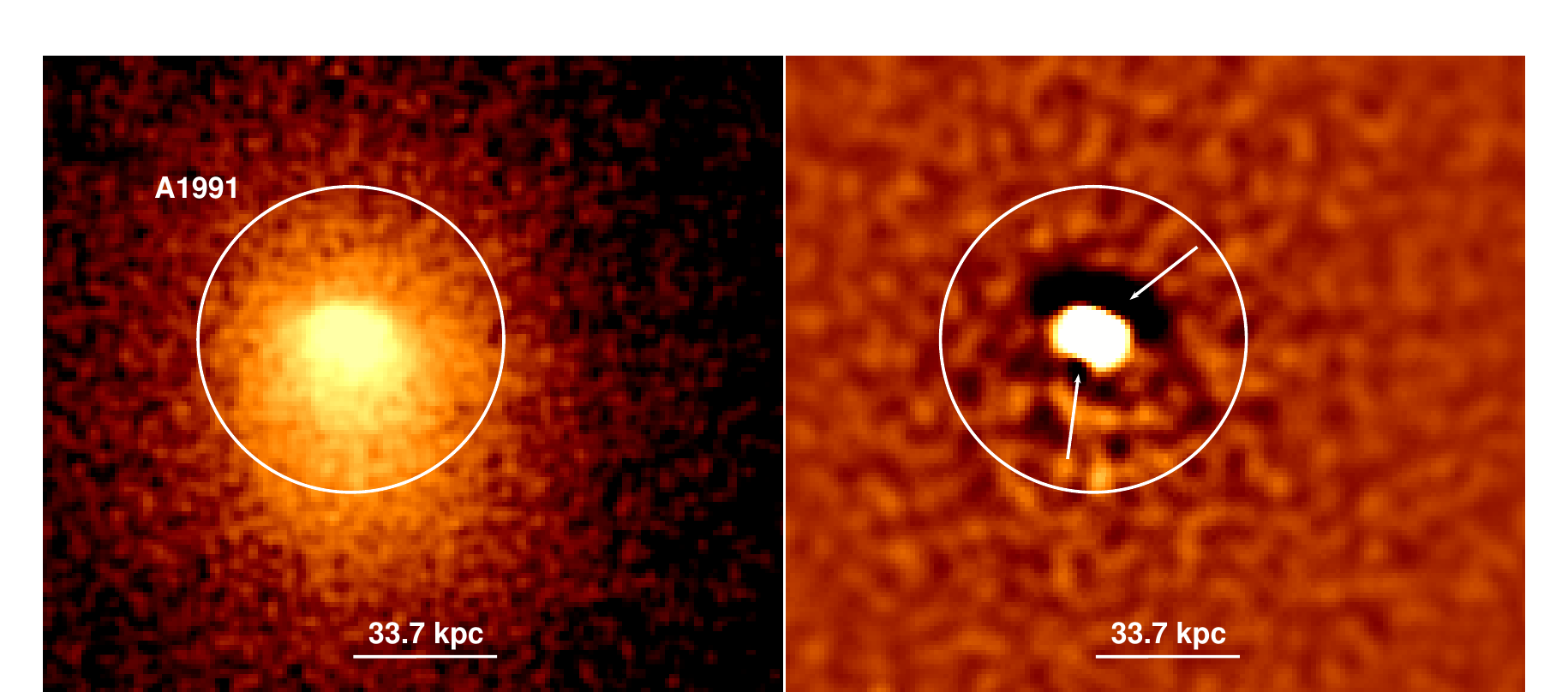}
\includegraphics[trim = 1.2cm 11.5cm 1.2cm 10cm, clip, height=4.0cm, width=8.8cm]{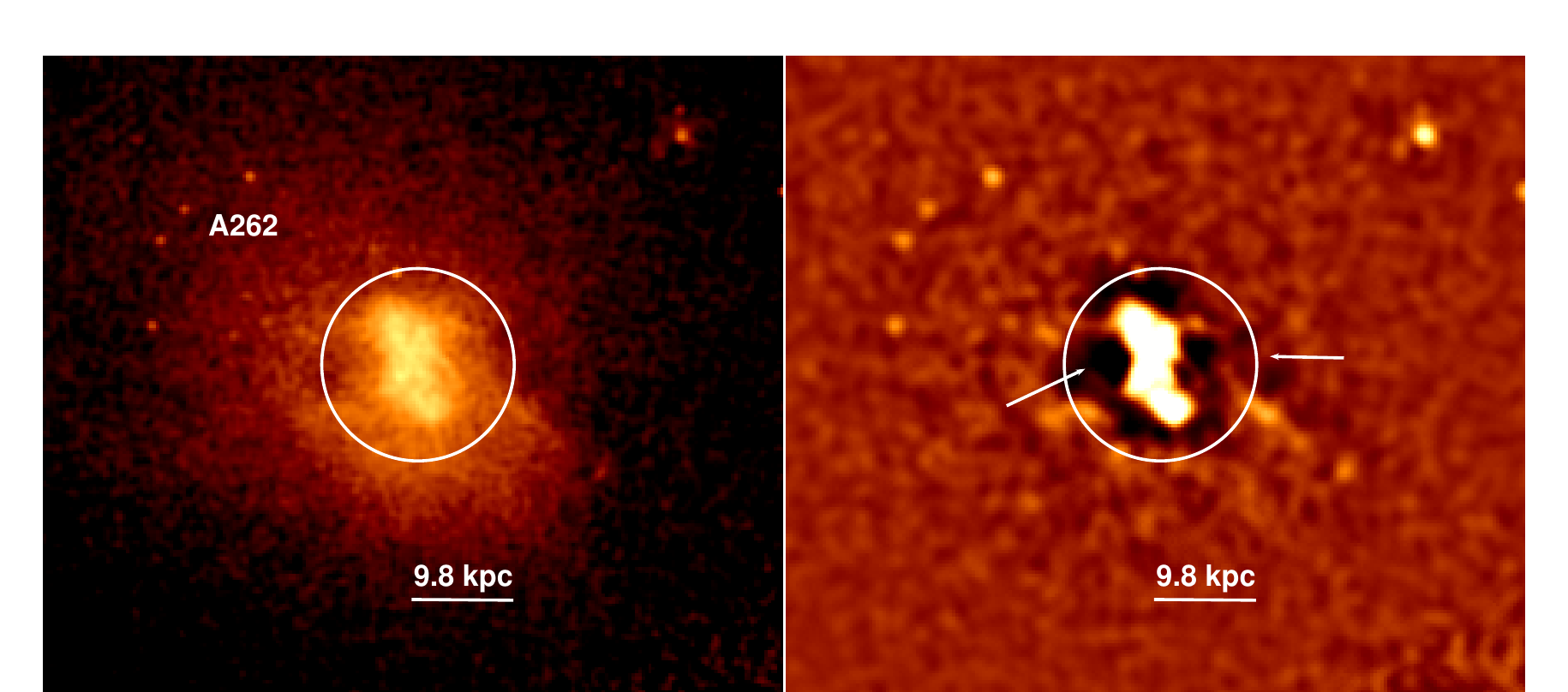}
\includegraphics[trim = 1.2cm 11.5cm 1.2cm 10cm, clip, height=4.0cm, width=8.8cm]{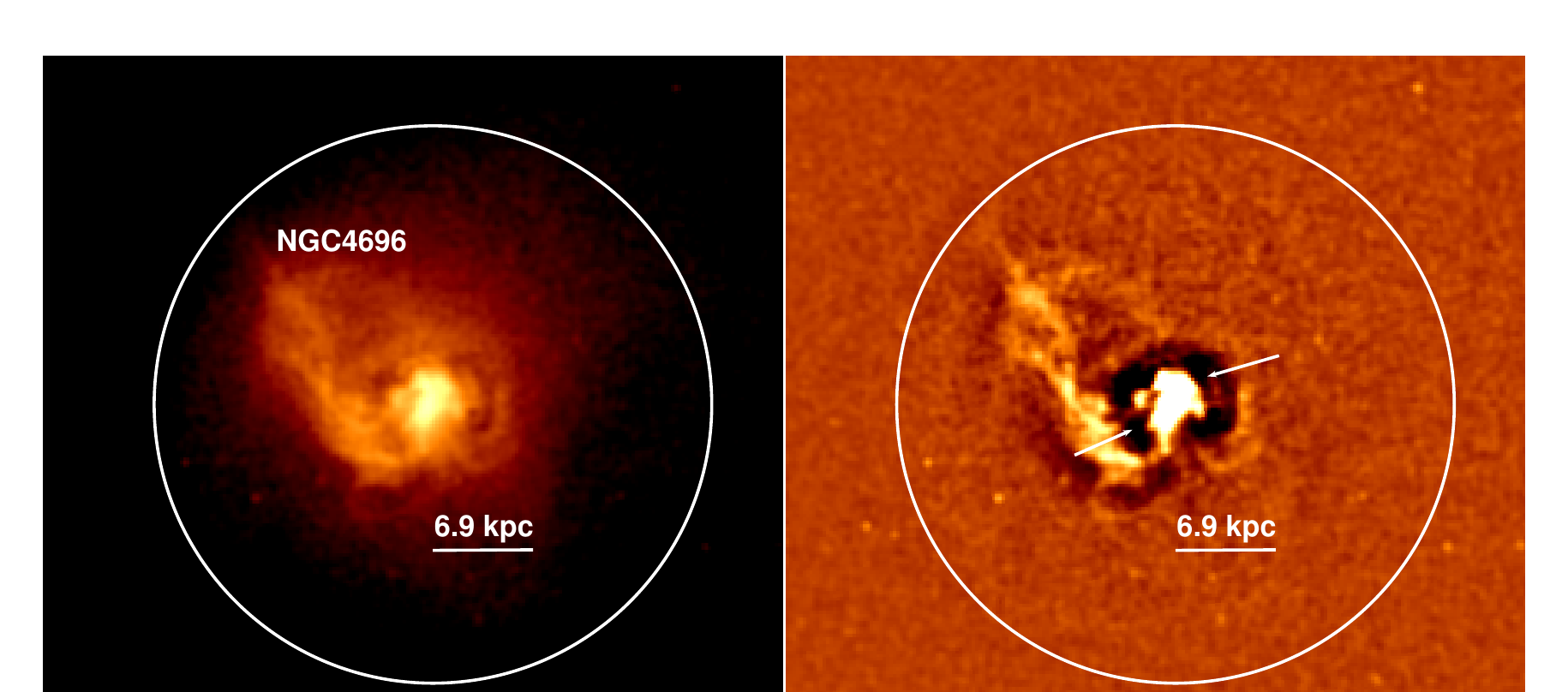}
\includegraphics[trim = 1.2cm 11.5cm 1.2cm 10cm, clip, height=4.0cm, width=8.8cm]{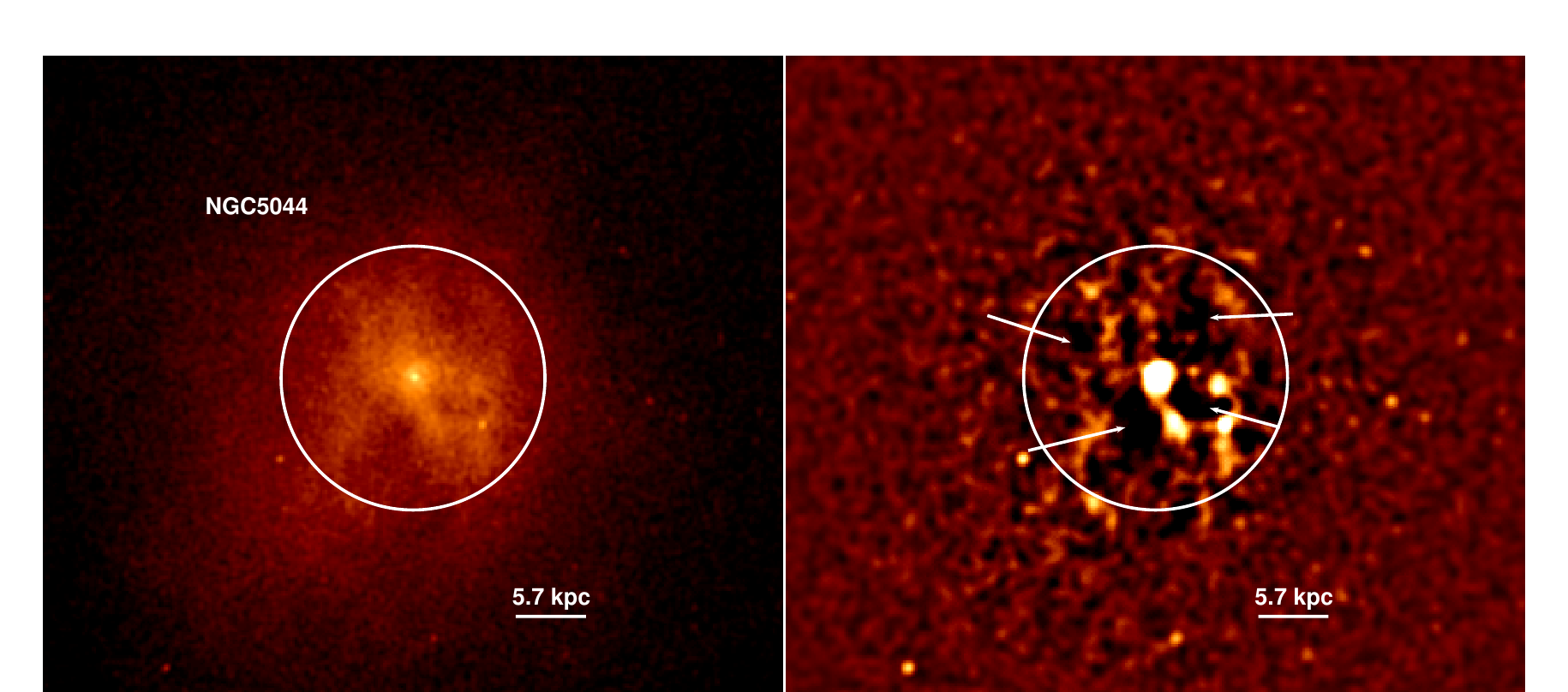}
\includegraphics[trim = 1.2cm 11.5cm 1.2cm 10cm, clip, height=4.0cm, width=8.8cm]{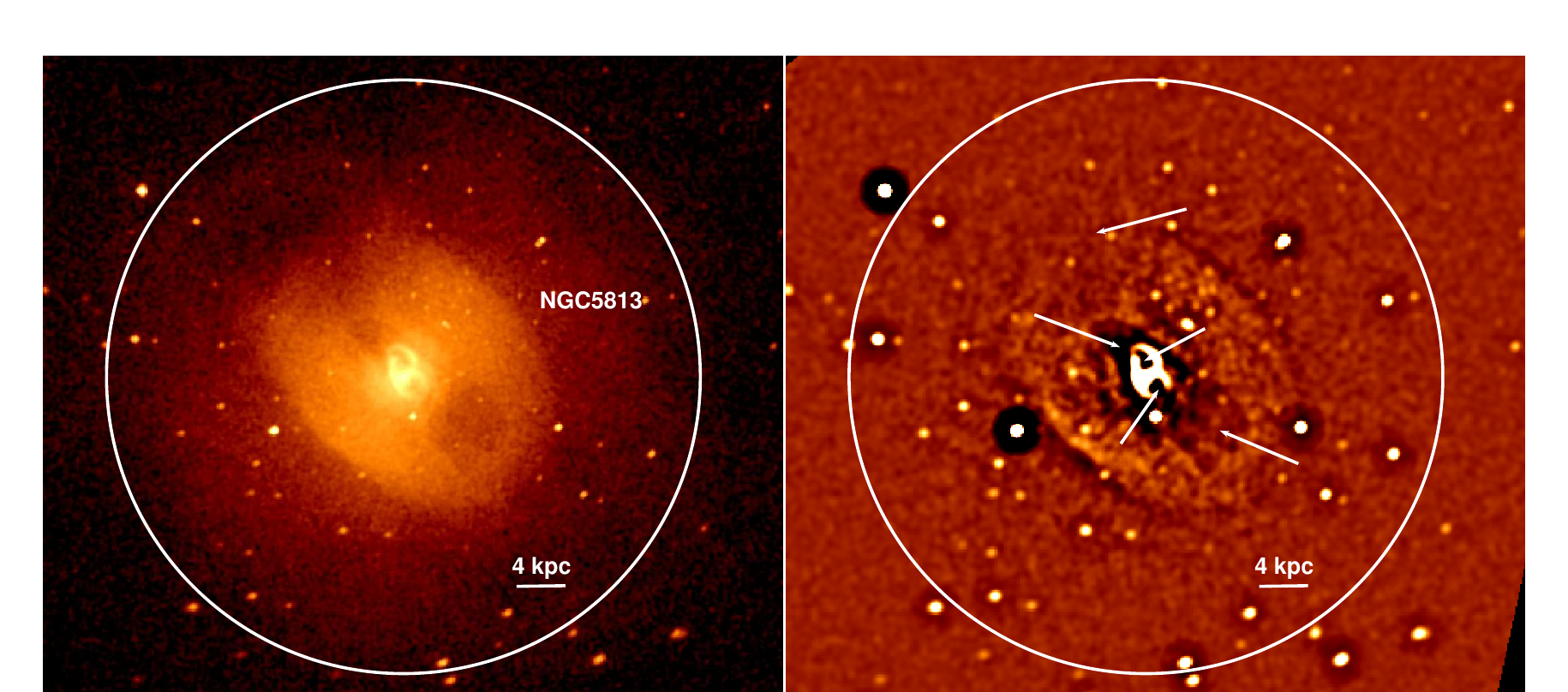}
\includegraphics[trim = 1.2cm 11.5cm 1.2cm 10cm, clip, height=4.0cm, width=8.8cm]{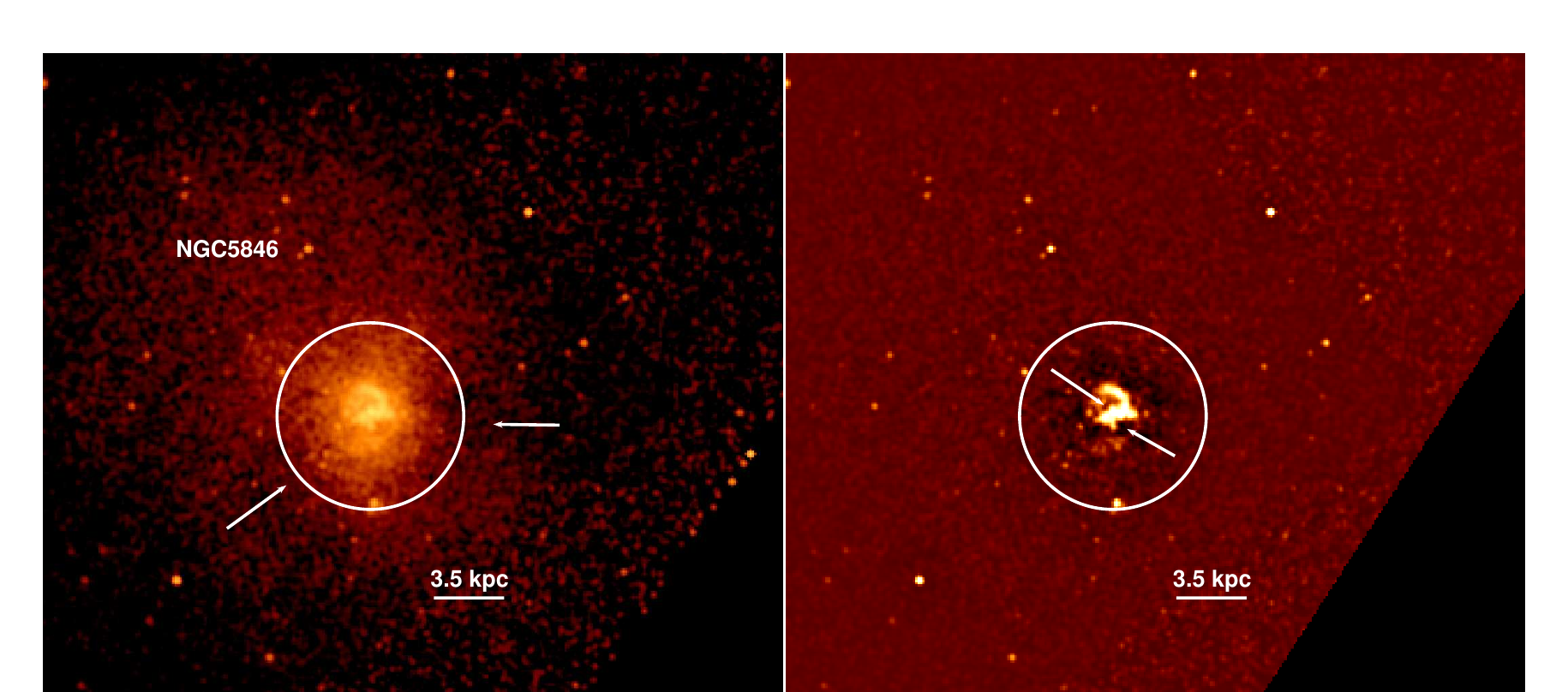}
\caption[]{0.5--7.0 keV background-subtracted, exposure-corrected images and the unsharp-masked images, adapted from Paper II, for all nine sources in our subsample with certain central abundance drops, with their X-ray cavities indicated by white arrows. The white circle in each image has a radius equal to the mean radius of the annulus within which the abundance peak is found, in the respective group or cluster. In each pair of images, the one on the left-hand side is the 0.5--7.0 keV exposure-corrected, background-subtracted image, while the one on the right is the unsharp masked image. All the background-subtracted, exposure-corrected images have been smoothed with a 2-pixel Gaussian. The bar is 0.5 arcmin long in all the images. }
\label{fig:imagescertain}
\end{center}
\end{figure*}

\begin{figure*}
\begin{center}
\includegraphics[trim = 1.2cm 11.5cm 1.2cm 10cm, clip, height=4.0cm, width=8.8cm]{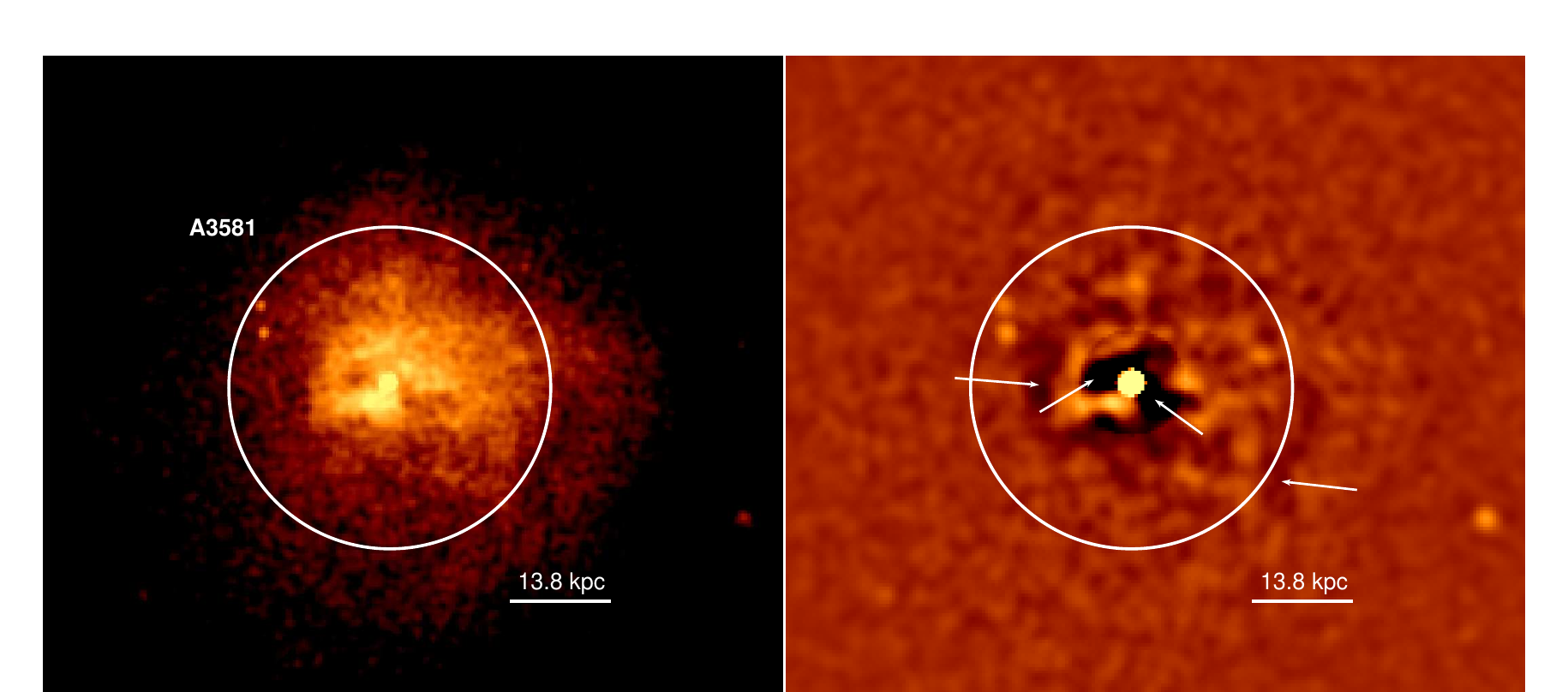}
\includegraphics[trim = 1.2cm 11.5cm 1.2cm 10cm, clip, height=4.0cm, width=8.8cm]{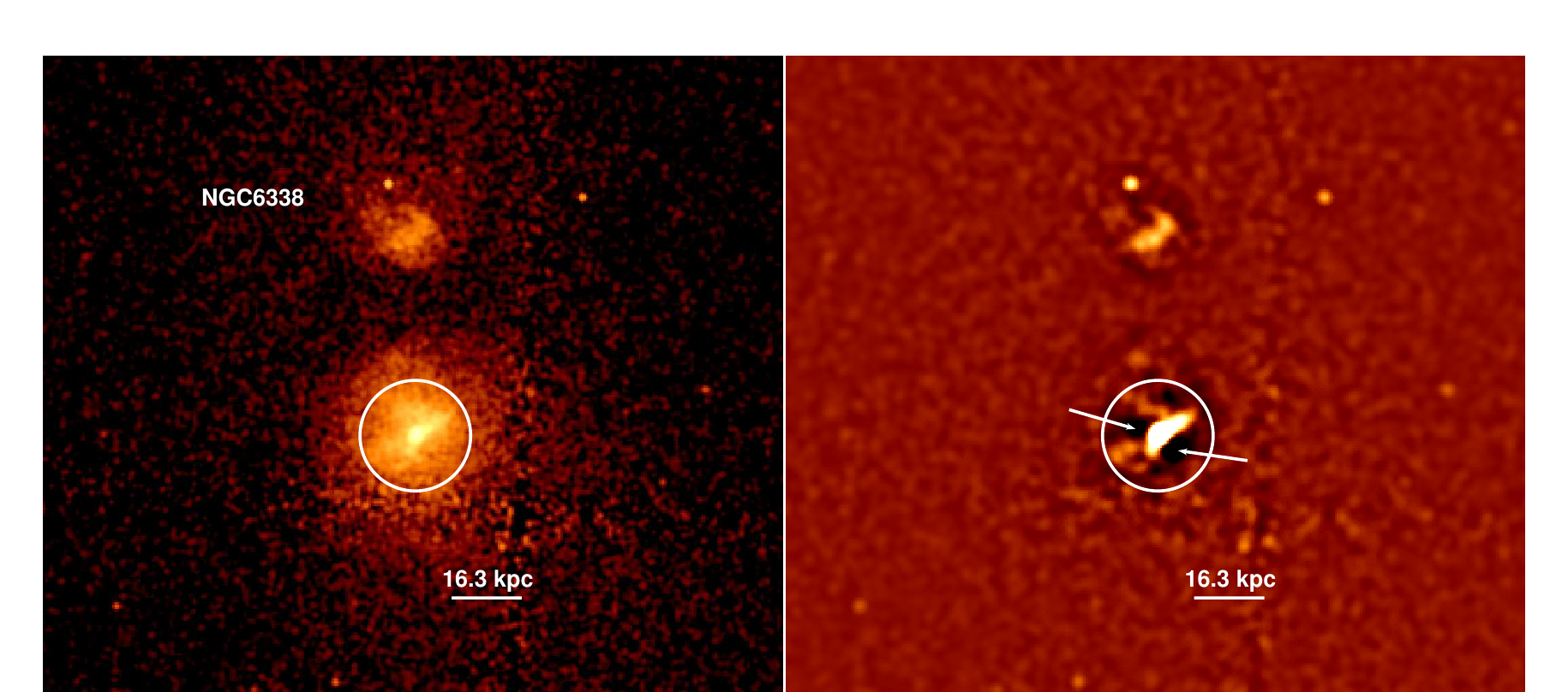}
\includegraphics[trim = 1.2cm 11.5cm 1.2cm 10cm, clip, height=4.0cm, width=8.8cm]{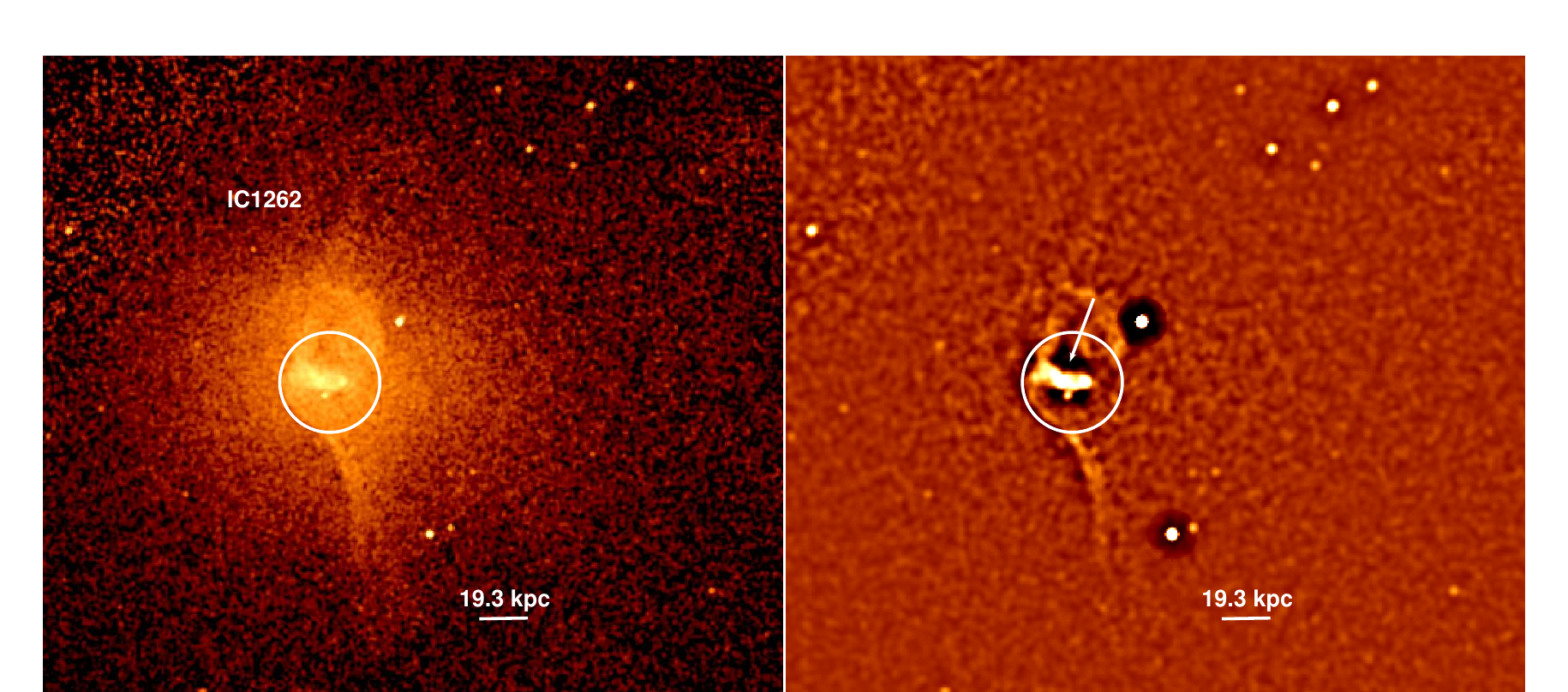}
\includegraphics[trim = 1.2cm 11.5cm 1.2cm 10cm, clip, height=4.0cm, width=8.8cm]{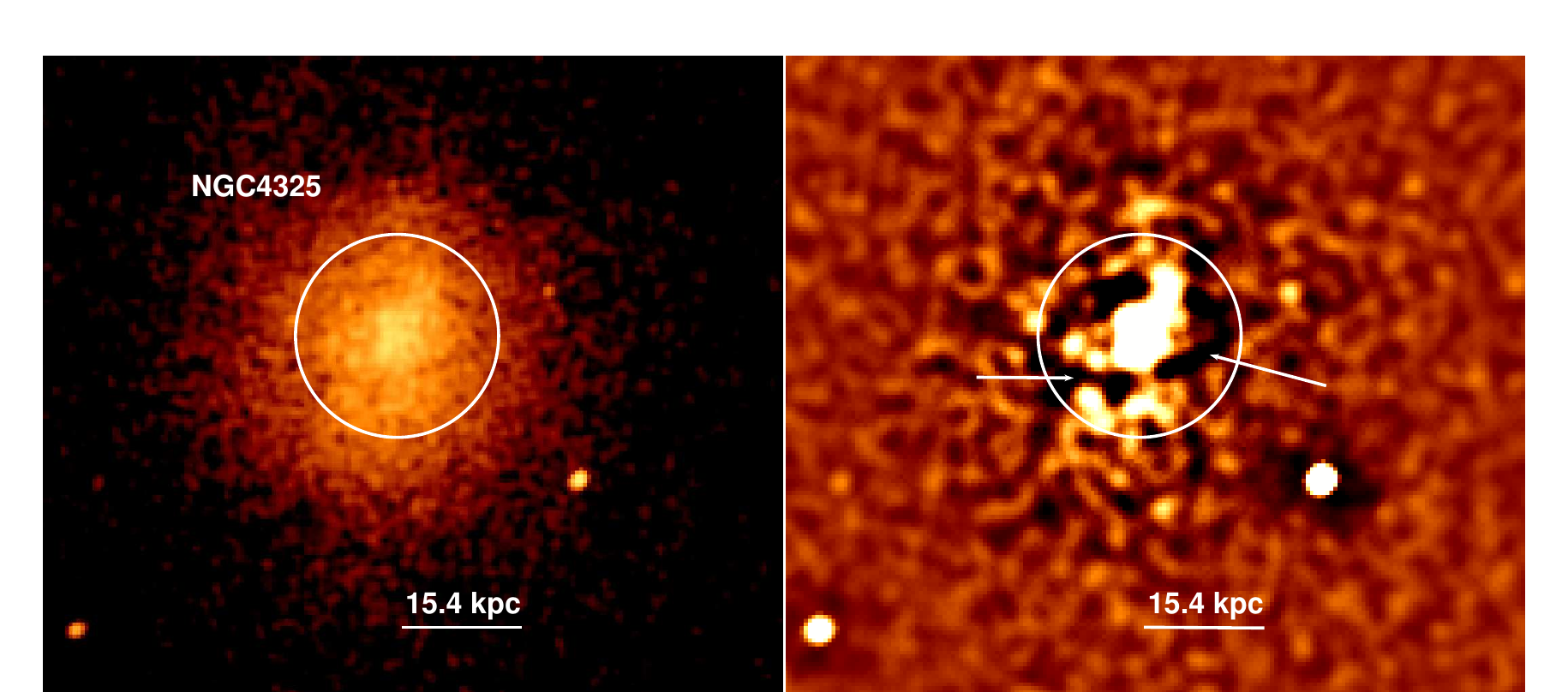}
\includegraphics[trim = 1.2cm 11.5cm 1.2cm 10cm, clip, height=4.0cm, width=8.8cm]{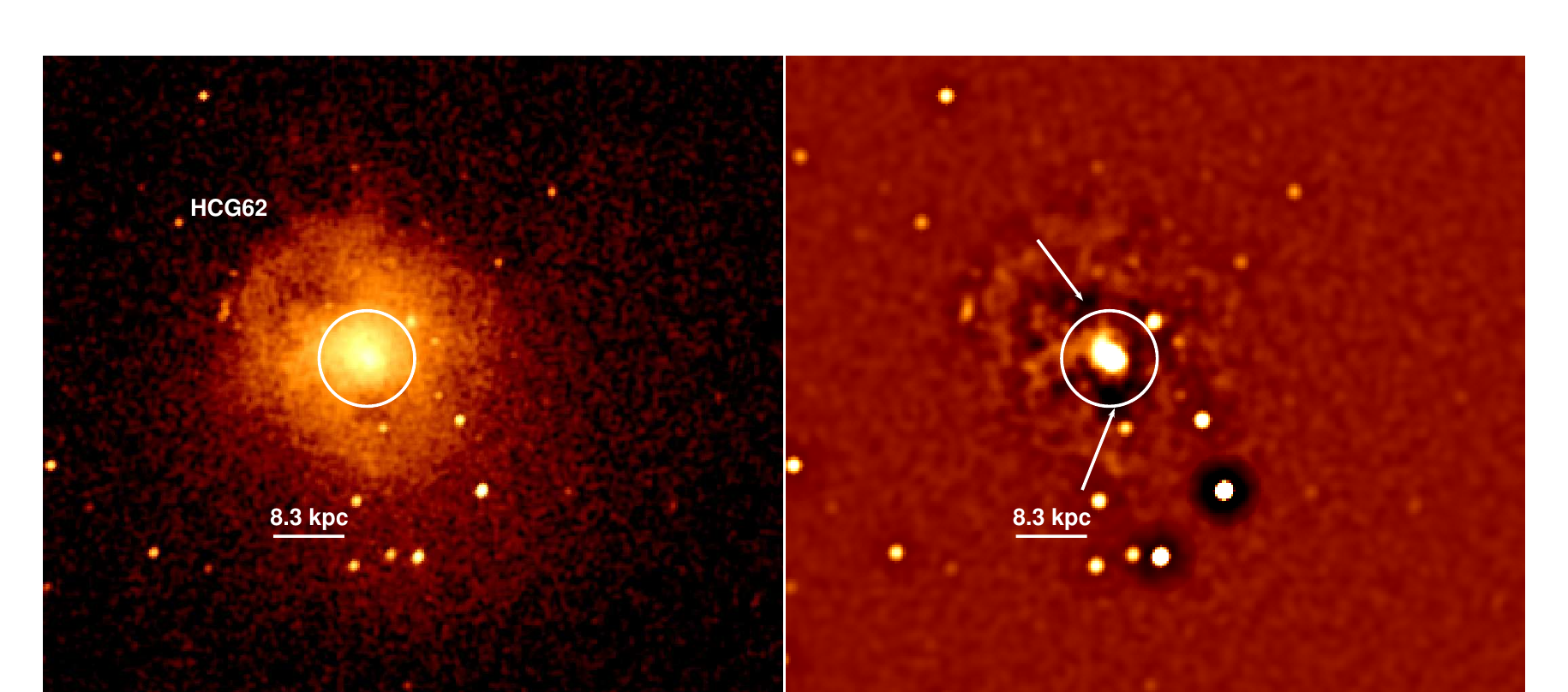}
\includegraphics[trim = 1.2cm 11.5cm 1.2cm 10cm, clip, height=4.0cm, width=8.8cm]{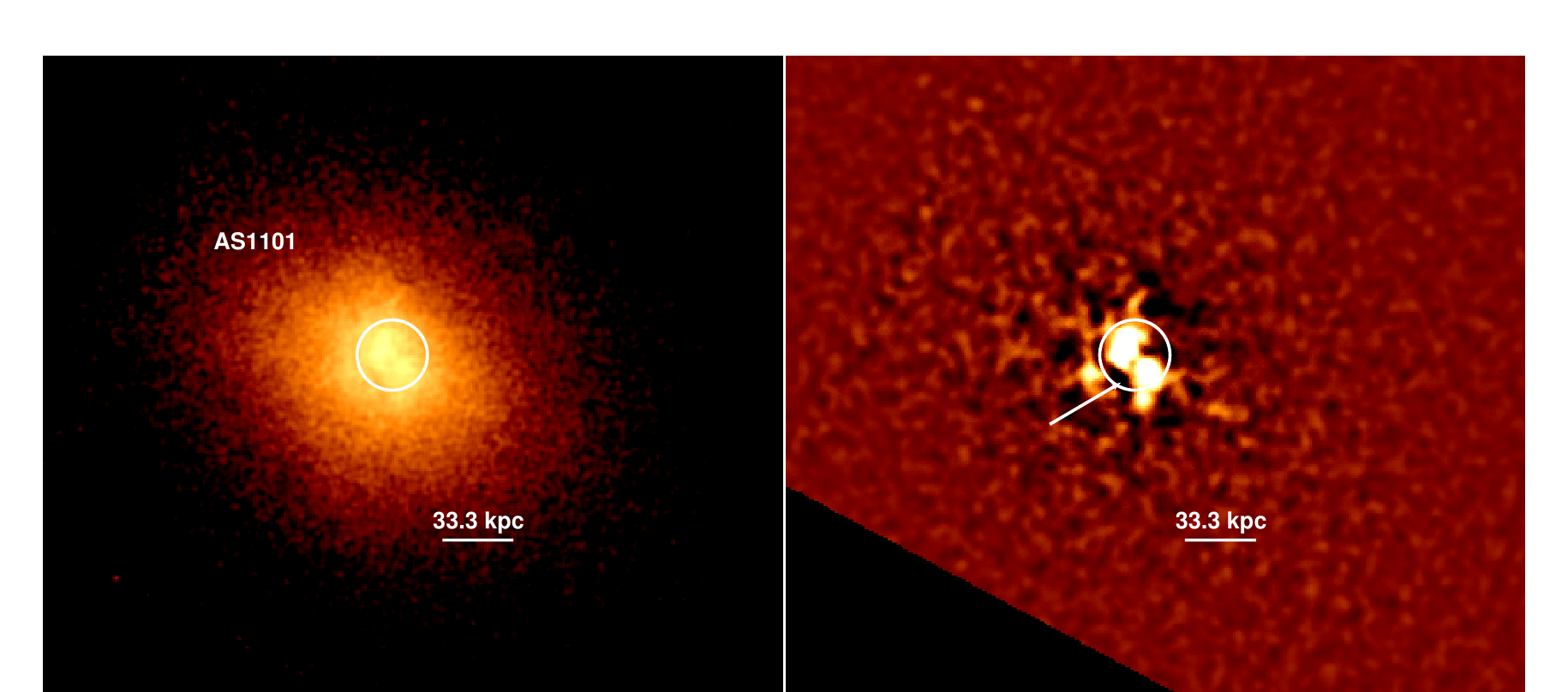}
\caption[]{As in Fig. \ref{fig:imagescertain}, but for the eight sources with possible central abundance drops.}
\label{fig:imagespossible}
\end{center}
\end{figure*}

The observation IDs and the clean exposure times for each source are listed in tables 2 and 4 in Paper I. All the sources in our subsample have been observed with {\it Chandra}, and, with the exception of Abell~1991, NGC~4325 and NGC~6338, have at least $\sim$50 ks of clean exposure time. We refer the reader to section 3 of Paper I for a detailed presentation of the data analysis. To summarise, the {\sc ciao acis\_reprocess\_events} pipeline was used to generate new Level 2 events files from the available Level 1 events files. In order to eliminate periods of background flaring from the data, we created and examined background lightcurves for each observation. The appropriate blank-sky observations were used to generate background images and spectra, after being adjusted to match the corresponding observations. When necessary, multiple datasets for one source were reprojected onto the same set of coordinates. Finally, 0.5--7.0 keV background-subtracted and exposure corrected images were made for each source, and were visually examined to single out contaminating point sources. These point sources were excised from future spectral analysis. For a detailed description of the data analysis of the NGC~4696 data, we refer the reader to \cite{Panagoulia13}.

We note that, in the case of NGC~4325, Abell~262, NGC~5813, NGC~5044 and NGC~5846, the data had to be re-analysed, in order to confirm the presence or absence of a central abundance drop. This warranted the use of larger spectral bins, to allow for more counts, and hence a larger signal-to-noise ratio, in each bin. A new version of the {\sc ciao} software was released between the first and second analyses of the data for these two sources, which meant that it had to be re-obtained and reprocessed with a different and more recent version of the {\sc ciao} software than that used for the data from the rest of the sources. We do not expect this to affect the validity, or the magnitude, of the abundance drops observed in these sources. 

\subsection{Spectral analysis}
As previously mentioned, this work focuses on the radial metallicity profiles of our sample of sources, and whether variations in them can be attributed to the presence of X-ray cavities. Again, we refer readers to Section 4.1 of Paper I for a detailed description of our spectral analysis, and to \cite{Panagoulia13} for the details of the analysis in the case of NGC~4696. To summmarise, the standard {\sc ciao} routines were applied to the data, in order to extract source and background spectra, ancillary region files (ARFs) and redistribution matrix files (RMFs). Projection effects were minimised by using the {\sc dsdeproj} routine \citep{Sanders07} to deproject the extracted source spectra. All spectral fitting was performed using {\sc xspec} \citep{Arnaud96} v12.7.1b. Specifically, all spectra were modelled using an absorbed optically thin thermal model, namely the {\sc wabs*apec} model in {\sc xspec}. Initially, just the iron abundance, temperature and normalisation were allowed to vary independently. The redshift was fixed to the value corresponding to each source, and the column density was frozed to the Galactic value \citep{Kalberla05}. The only exception to this was 2A0335+096, which has a measured column density of about 2.4$\times$10$^{+21}$ cm$^{-2}$, which is significantly higher than the galactic value of 1.8$\times$10$^{+21}$ cm$^{-2}$ \citep[e.g.][]{Sanders09a}. In our analysis, we used a column density of 2.34$\times$10$^{+21}$ cm$^{-2}$, obtained by fitting regions outside of the cluster core. If lines from elements other than iron, such as silicon or sulphur, were visible in the spectra, the spectral model was changed to {\sc wabs*vapec}, which allows the measurement of additional elemental abundances, and the corresponding abundances were allowed to vary. We consider the detection of an element, other than iron, to be real if allowing its metallicity to vary did indeed result in the visible fitting of a spectral line, rather than just influencing the continuum, and if it improved the statistical fit by $\Delta\chi^{2} \leq$ -2.7 (i.e. a detection was made at $\geq$ 90 per cent confidence). Where necessary, we added an extra thermal component to model the emission, and recorded the new abundance value(s). This was necessary for the innermost spectral bins of some of the 14 sources under study here, namely NGC~4696, Abell~262, NGC~5813 and Abell~3581, which have known multi-temperature structure in their core. This was done because the use of a single-temperature model to fit multi-temperature emission, biases the metallicity low \citep[e.g.][]{Buote03}. In these cases, the elemental abundances of the two thermal components were tied together, while the temperature and normalisation of each component were allowed to vary independently. The issue of multi-temperature models is discussed further in Section 5.4, and Appendix A.

\subsection{X-ray cavity detection}
We used unsharp-masking to determine the location, number, size and distance of the X-ray cavities from the centre of their host source. We smoothed the 0.5--7.0 keV background-subtracted, exposure-corrected images with 1-, 2-, 8- and 10-pixel Gaussians, and then subtracted each of the two more heavily smoothed images from the two less smoothed images. The corresponding images are shown in Figs. \ref{fig:imagescertain} and \ref{fig:imagespossible} for the sources with certain and possible central abundance drops respectively, adapted from Paper II. We only show the unsharp masked image in which the X-ray cavities are most obvious. In each set of two images, the left-hand panel shows the background-subtracted, exposure-corrected image, which has been smoothed with a 2-pixel Gaussian, while the right-hand panel shows the unsharp-masked image. The white bar in each image is 0.5 arcmin long. The white circles indicate the mean radius of the annulus in which the peak in the radial abundance profile is found in the corresponding source. 
%The unsharp-masked images shown here are the 2-8 pixel ones for Abell~1991, Abell~2052, NGC~4325 and Abell~262, the 1-8 pixel image for NGC~4696, the 2-10 pixel image for 2A0335+096, and the 2-6 pixel image for Abell~1991. 

\section{Results}
\subsection{Imaging}
Images of all fourteen sources in our sample that have a central abundance drop are shown in Figs. \ref{fig:imagescertain} and \ref{fig:imagespossible}, adapted from Paper II, with the X-ray cavities in each source pointed out by the white arrows. The white circle indicates the mean radius of the annulus within which the abundance peak in the respective group or cluster was found (see the following section). We point out that, in Paper II, all the cavities listed in the seventeen sources, with the exception of those of NGC~4325, the two outermost cavities in NGC~5846 and the outermost western cavity in Abell~3581, were listed as ``certain'' cavities, i.e. they are clearly visible in both the original and unsharp-masked image, or are unambiguously visible in the unsharp-masked image. The cavities in NGC~4325, the outermost western cavity in Abell~3581, and the two outermost cavities in NGC~5846 were listed as ``possible'', as they are visible only in the unsharp-masked image, and the data are too noisy to class the cavities as ``certain''. For more details on the X-ray cavity classification scheme, as well as the measured properties of each cavity, interested readers are referred to section 5.1 of Paper II.

%As listed in Table \ref{tab:bubbles}, all five of our sources display H$\alpha$ filaments in their core. The filamentary structure of NGC~4696 has been studied in detail by \cite{Crawford05a}, and closely follows the X-ray structure in the core of the cluster. The H$\alpha$ filaments in NGC~4636 \citep{Werner13} show a striking resemblance to its X-ray morphology, though on a much smaller scale. The filamentary structure seen in H$\alpha$ images of 2A0335+096 traces the X-ray structure in the core of the cluster, with the brightest knots of H$\alpha$ emission coinciding with the surface brightness peaks \citep{Donahue07, Romanishin88}. The H$\alpha$ emission of Abell~1991 has a peculiar spear shape, pointing to the north from the centre of the cluster, and is thought to be due to a bow shock \citep{McDonald11}. The spear head of the H$\alpha$ emission is cospatial with a bright area of soft X-ray emission, but the base of the spear, which is the brightest in H$\alpha$, shows no X-ray counterpart \citep{McDonald11}. Finally, the X-ray brightest regions in the core of Abell~2052 show a remarkable correlation with the H$\alpha$ contours \citep{Blanton11}. 

\subsection{Abundance profiles}

\begin{figure*}
\begin{center}
\includegraphics[trim=0cm 14cm 5cm 0cm, clip, width=2.9in, height=2.9in]{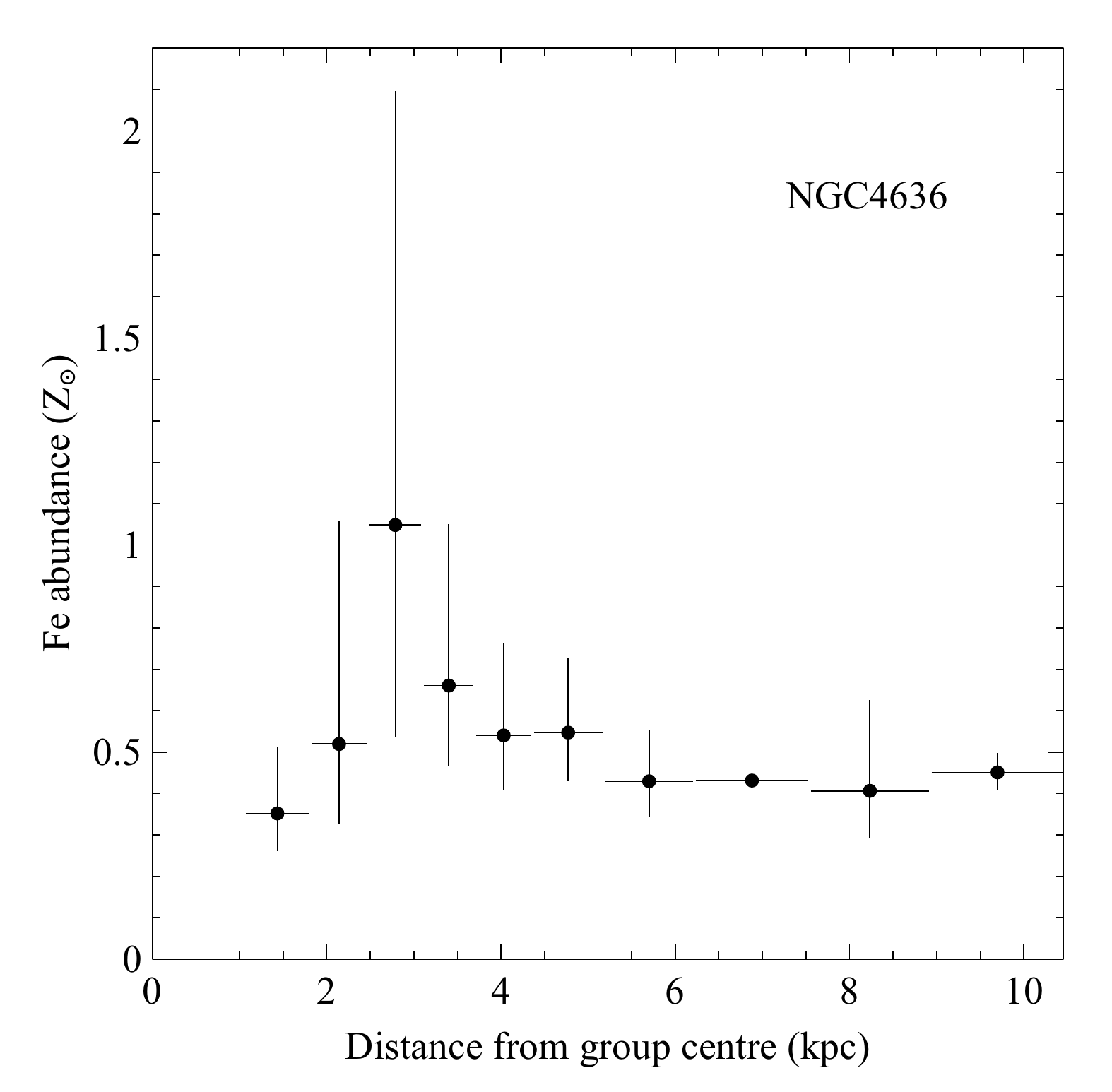}
\includegraphics[trim=0cm 14cm 5cm 0cm, clip, width=2.9in, height=2.9in]{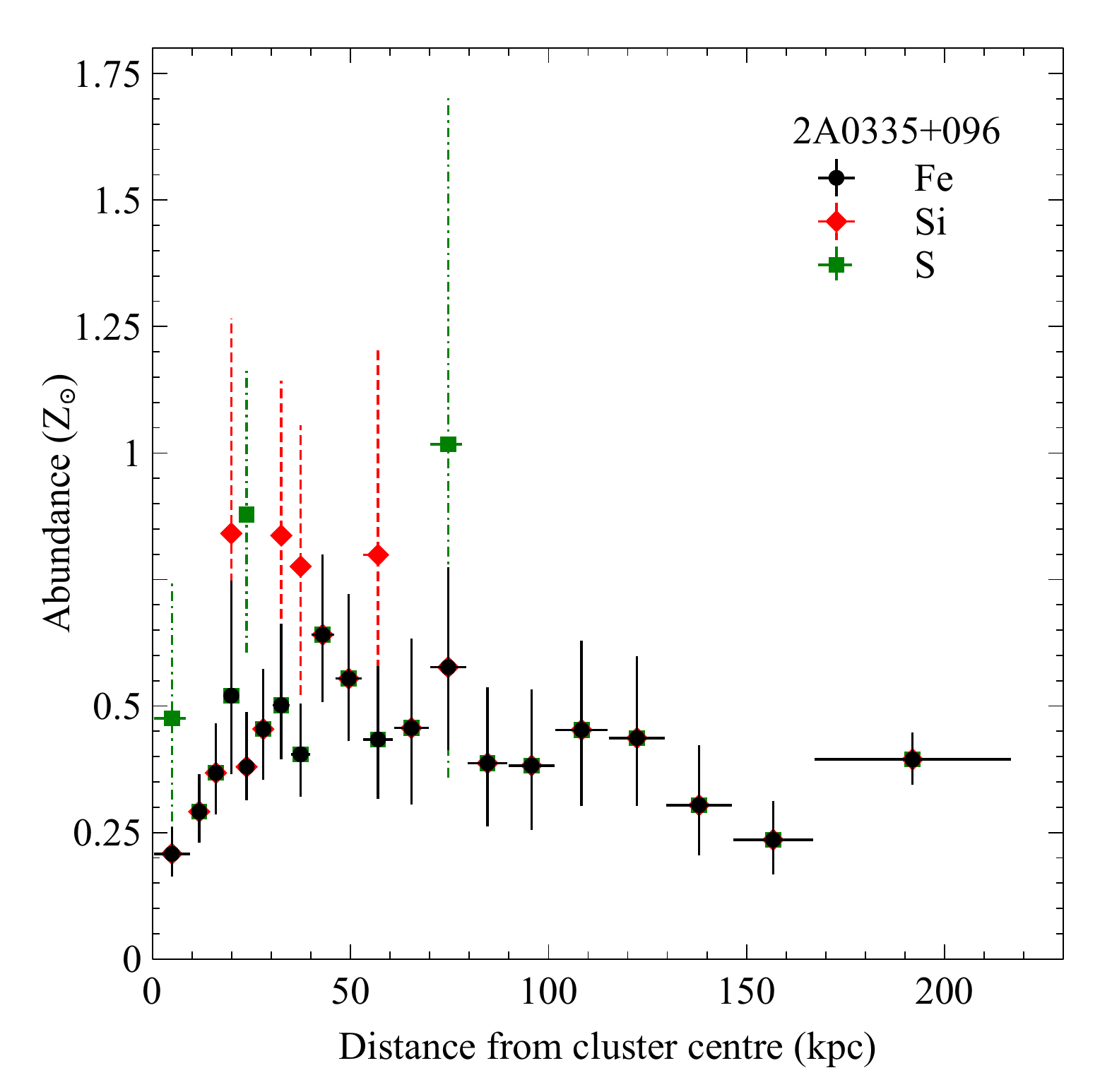}
\includegraphics[trim=0cm 14cm 5cm 0cm, clip, width=2.9in, height=2.9in]{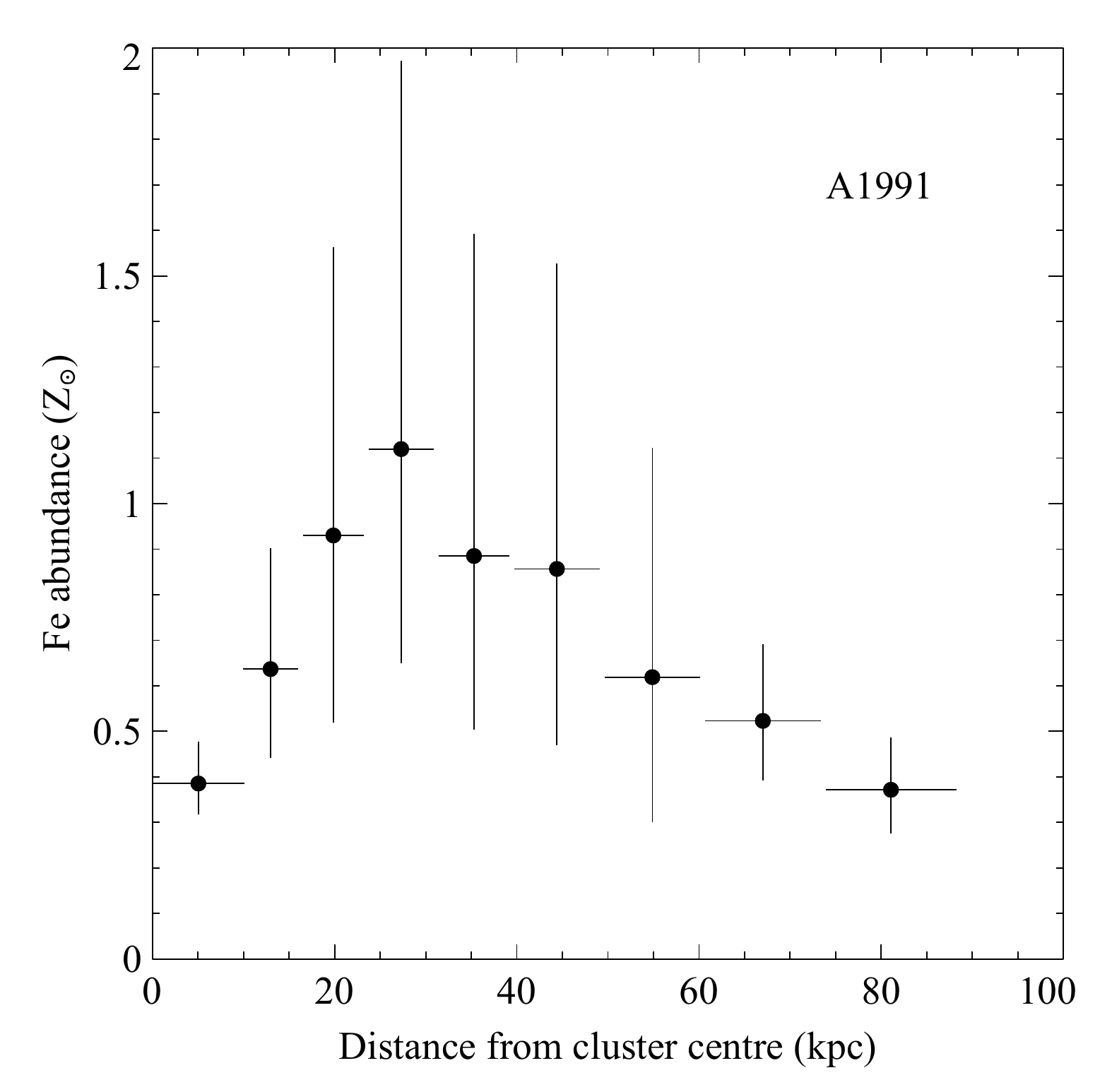}
\includegraphics[trim=0cm 14cm 5cm 0cm, clip, width=2.9in, height=2.9in]{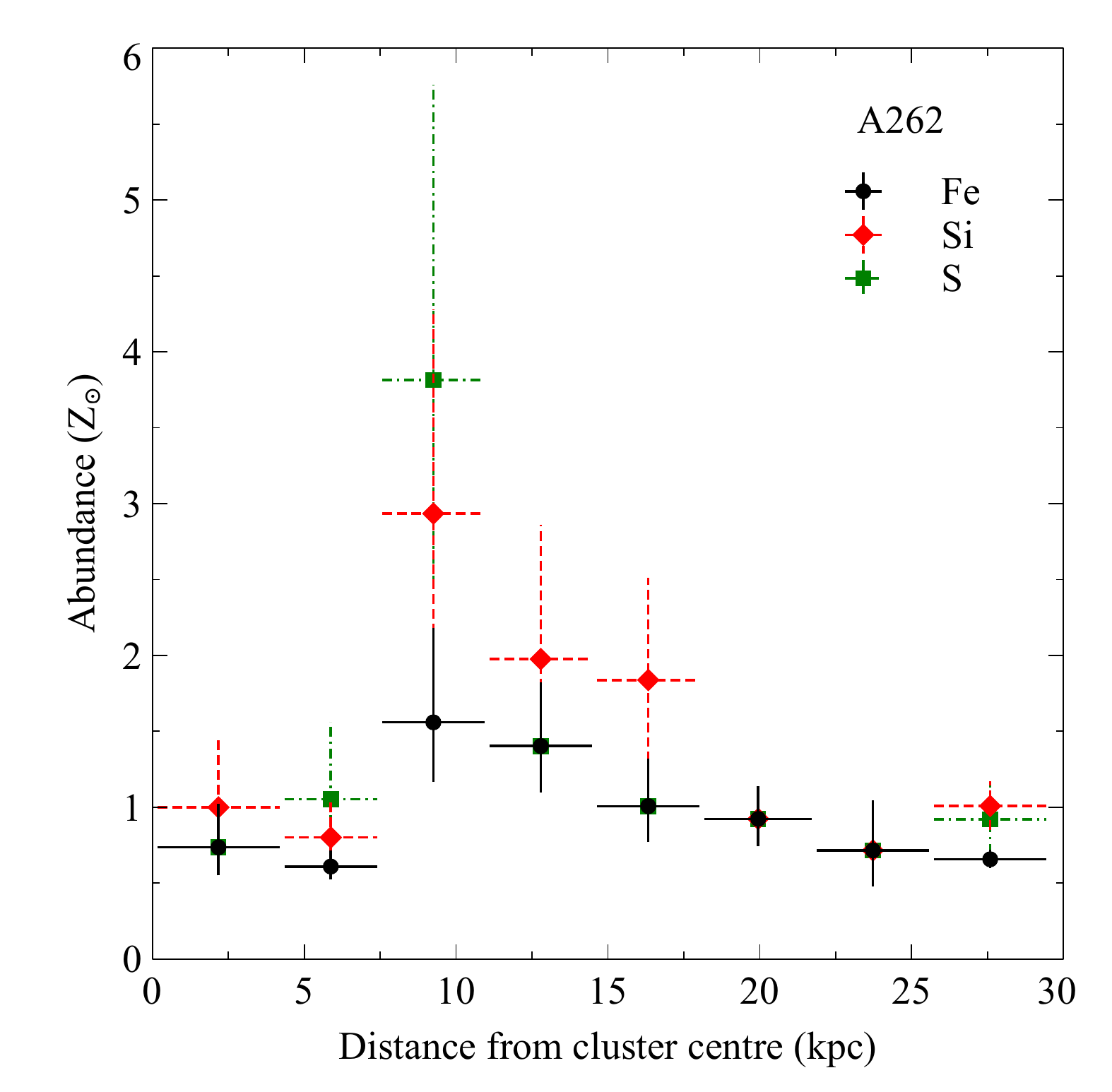}
\includegraphics[trim=0cm 14cm 5cm 0cm, clip, width=2.9in, height=2.9in]{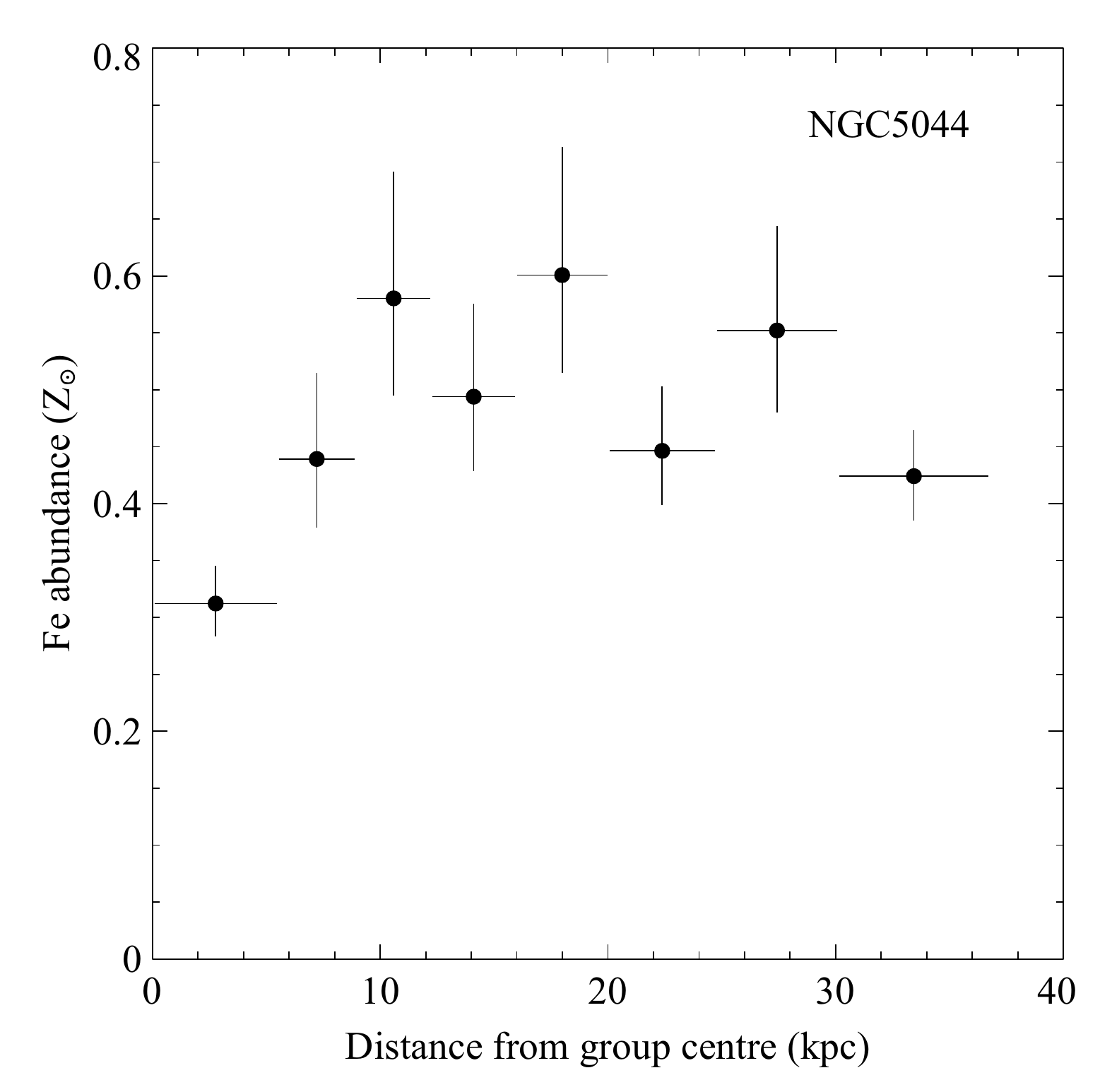}
\includegraphics[trim=0cm 14cm 5cm 0cm, clip, width=2.9in, height=2.9in]{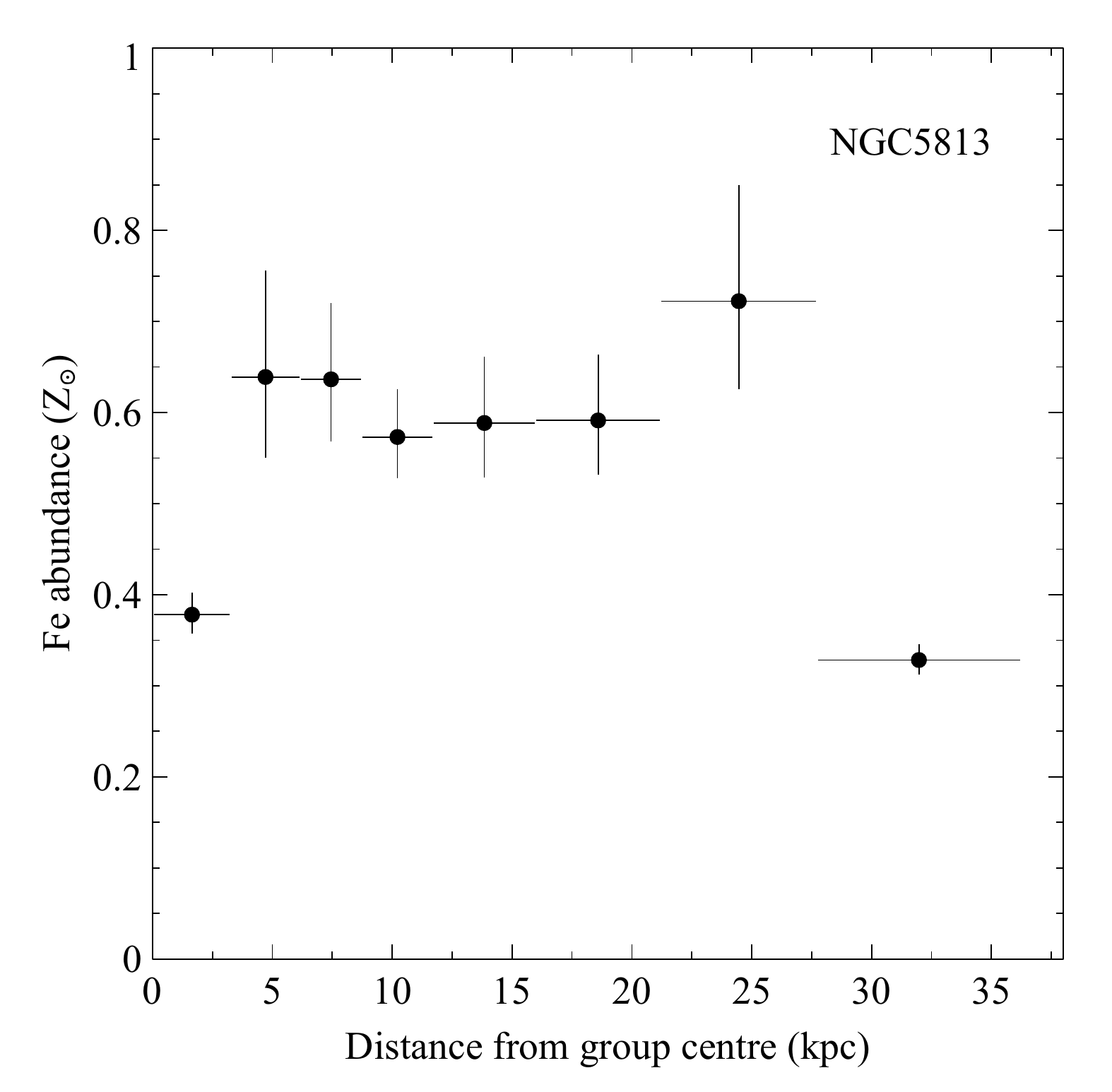}
\caption[]{Deprojected abundance profiles for the sources with certain central abundance drops. For NGC~4696 (the corresponding figure is adapted from \cite{Panagoulia13}), 2A0335+096 and Abell~262, we have plotted the silicon (red diamonds) and sulphur (green squares) abundance profiles, as well as the iron (black circles) abundance profile. In the latter two profiles, where the points of the silicon and sulphur profiles are not individually visible, it is due to their coincidence with the points of the iron profile (i.e. the statistical difference in the spectral fit when silicon and/or sulphur were allowed to vary was insignificant). There is a clear abundance drop in the innermost few kpc for each of the sources.}
\label{fig:abundprofscertain}
\end{center}
\end{figure*}

\begin{figure*}
\begin{center}
\contcaption{}
\includegraphics[trim=0cm 14cm 5cm 0cm, clip, width=2.9in, height=2.9in]{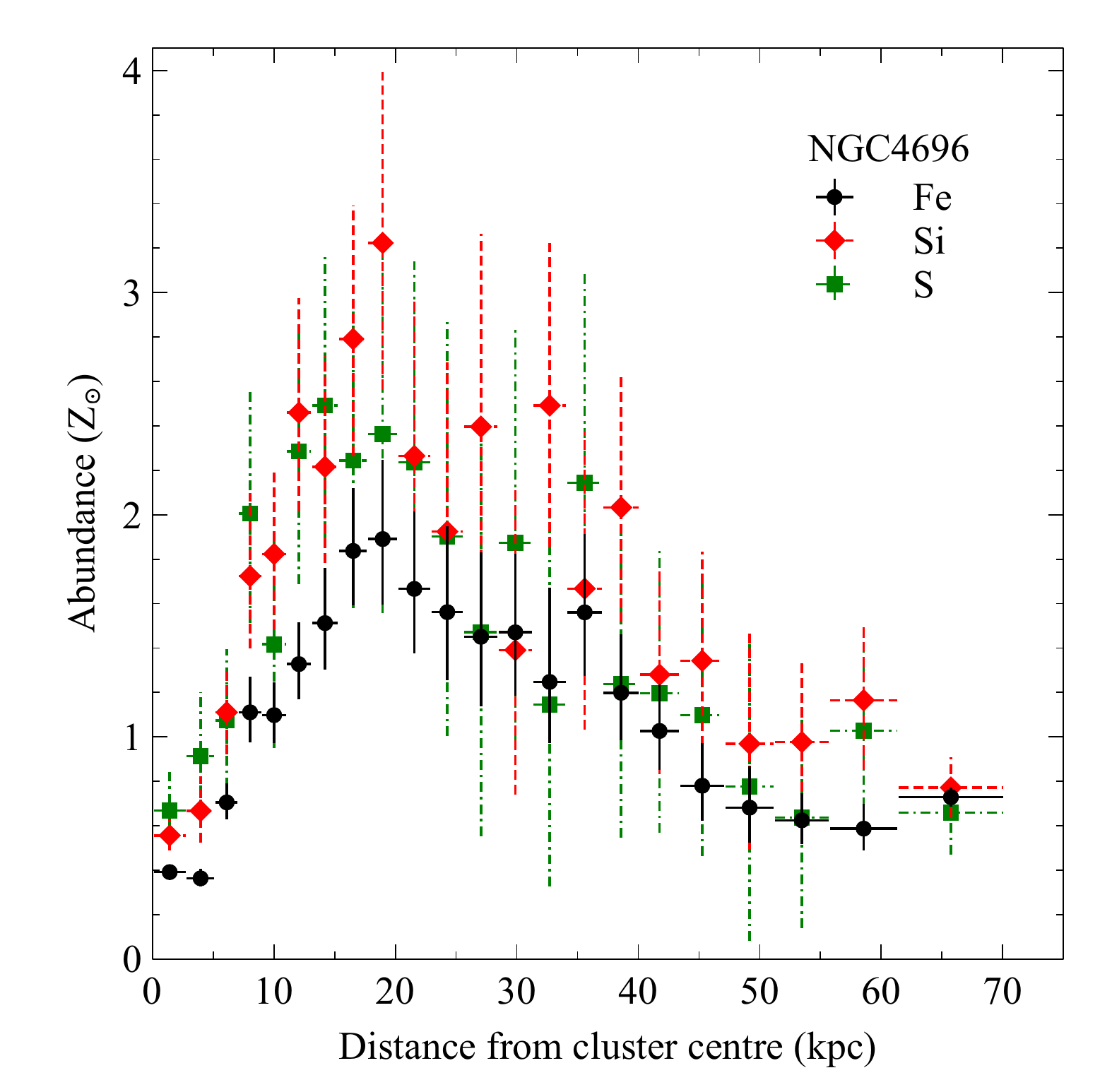}
\includegraphics[trim=0cm 14cm 5cm 0cm, clip, width=2.9in, height=2.9in]{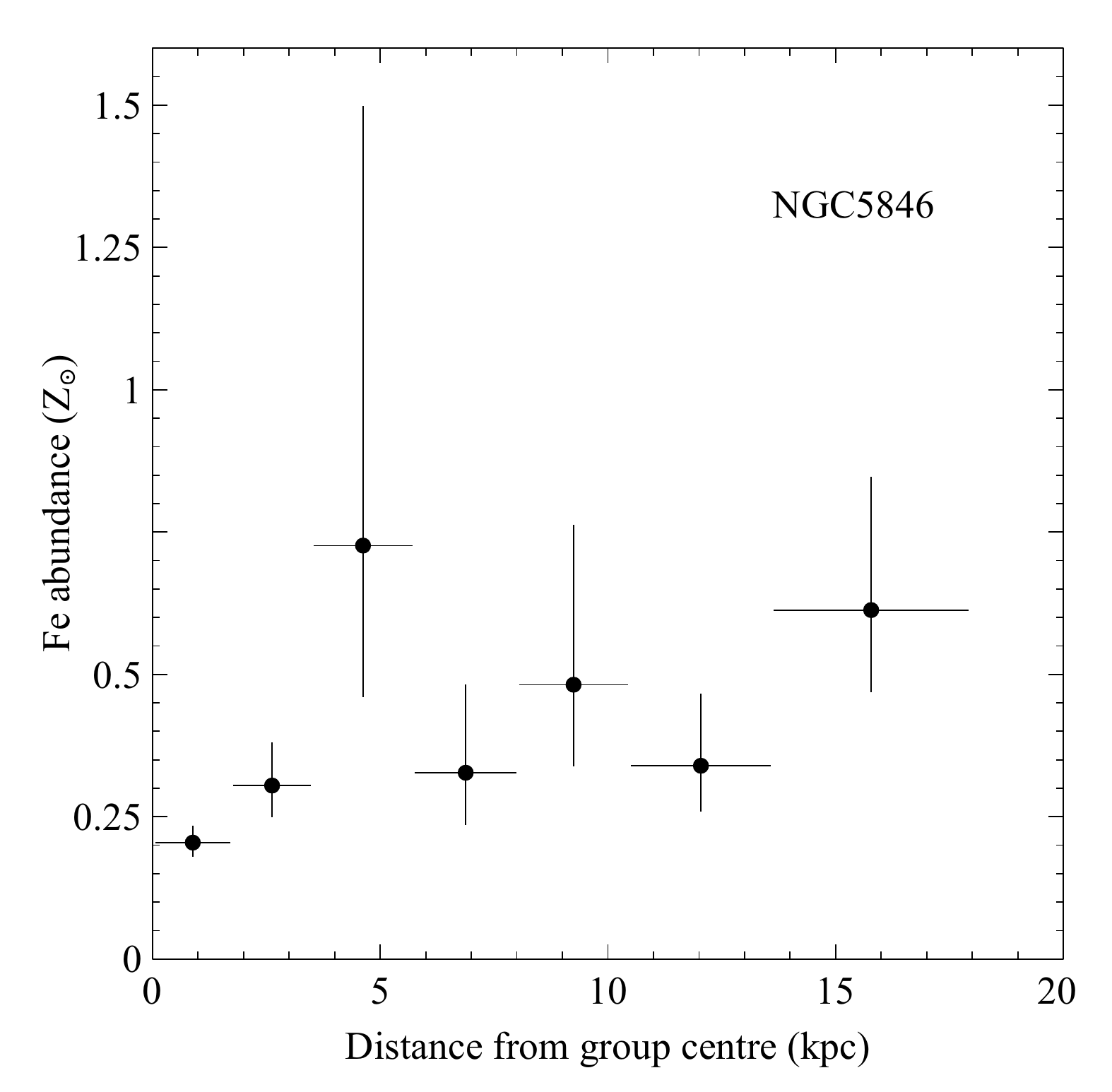}
\end{center}
\end{figure*}

\begin{figure*}
\begin{center}
\includegraphics[trim=0cm 14cm 5cm 0cm, clip, width=2.9in, height=2.9in]{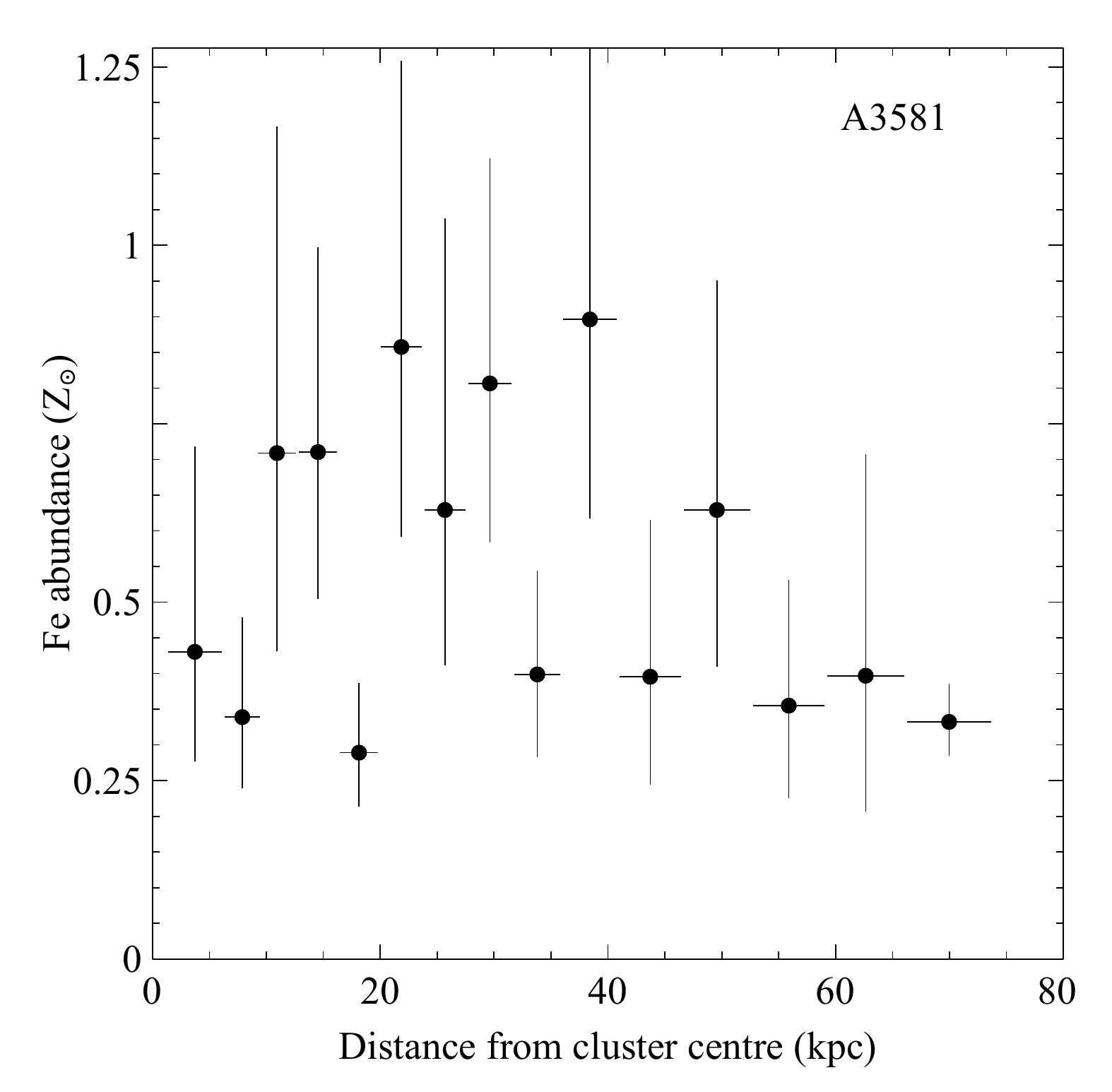}
\includegraphics[trim=0cm 14cm 5cm 0cm, clip, width=2.9in, height=2.9in]{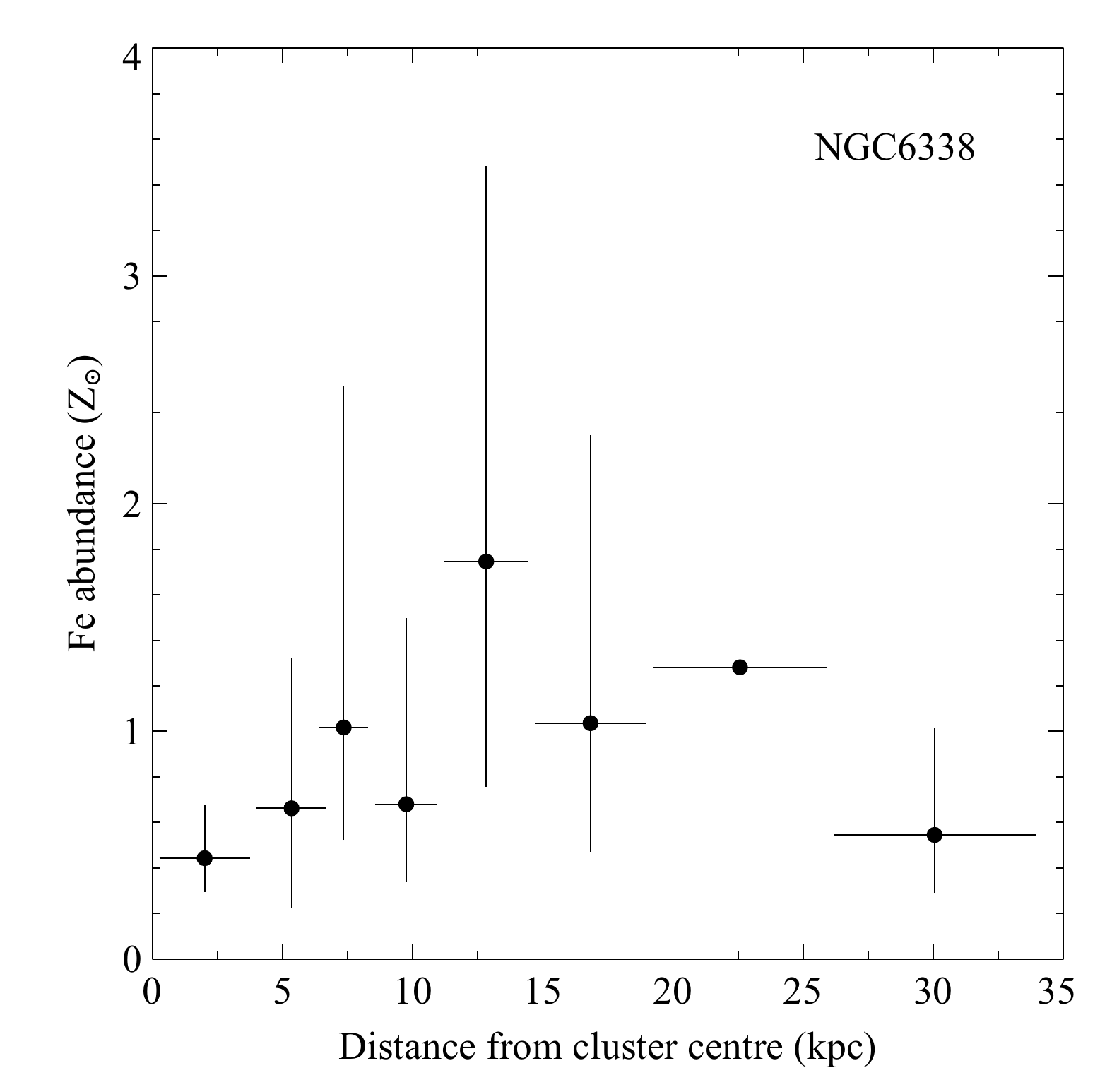}
\includegraphics[trim=0cm 14cm 5cm 0cm, clip, width=2.9in, height=2.9in]{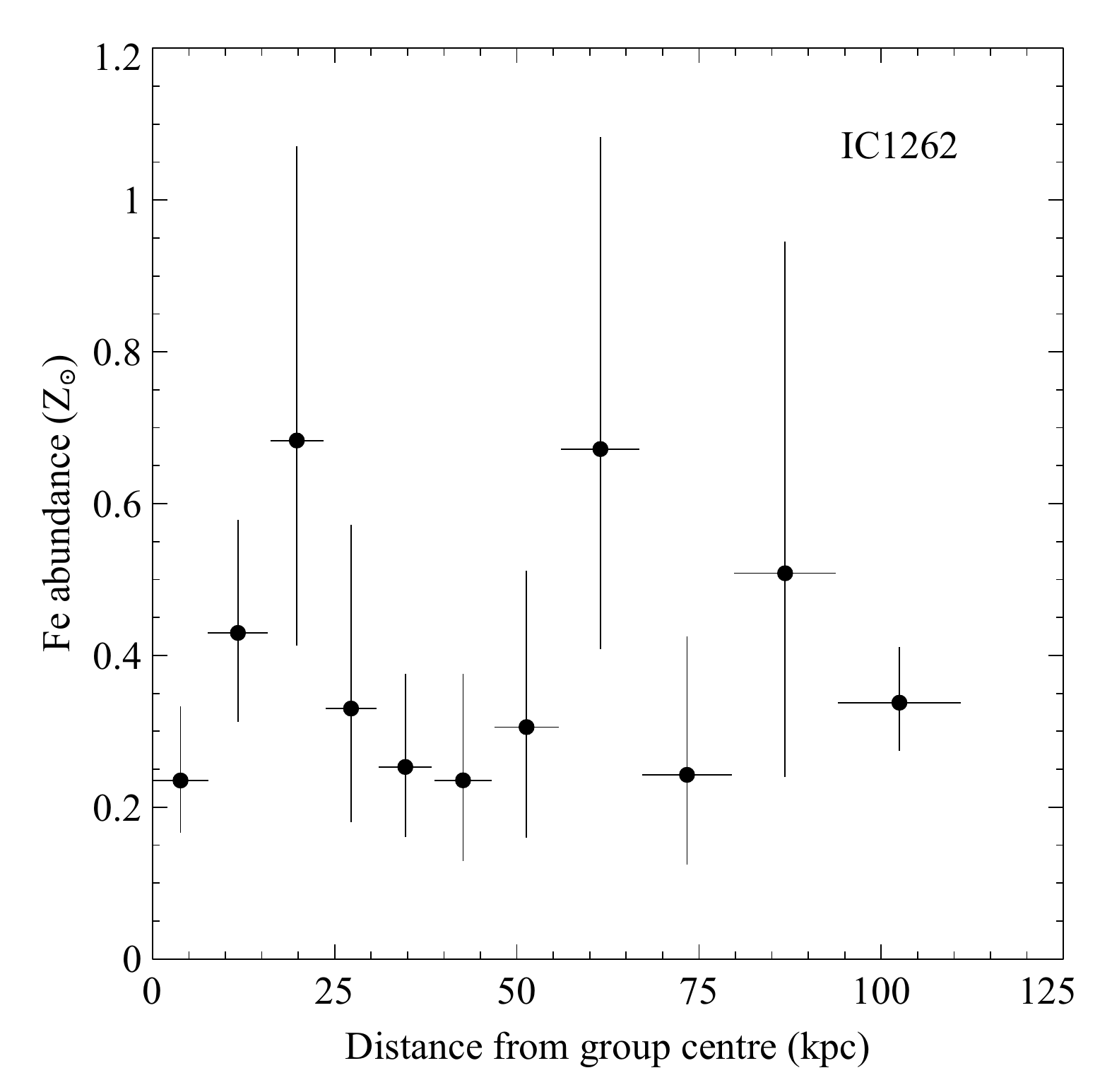}
\includegraphics[trim=0cm 14cm 5cm 0cm, clip, width=2.9in, height=2.9in]{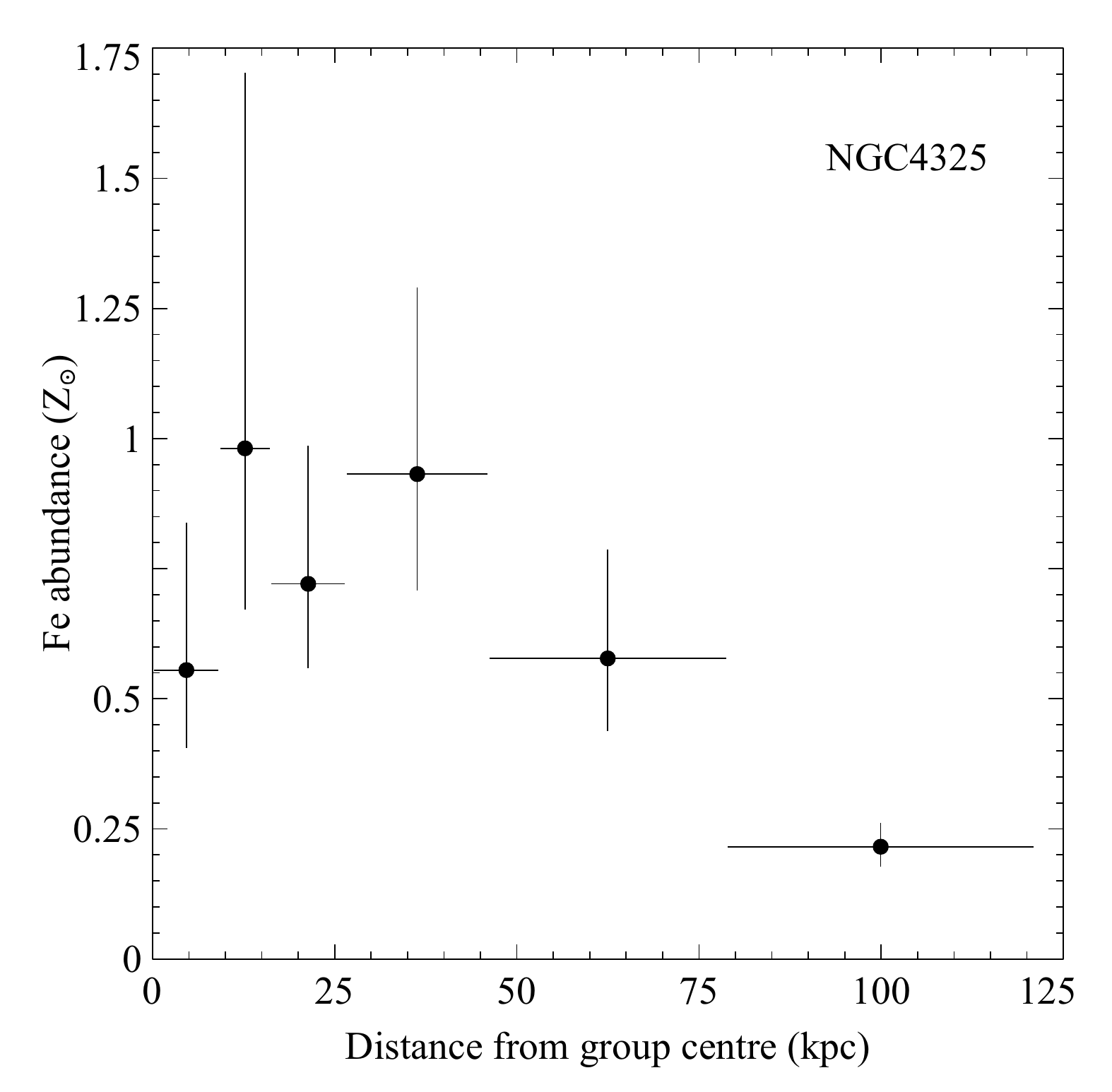}
\includegraphics[trim=0cm 14cm 5cm 0cm, clip, width=2.9in, height=2.9in]{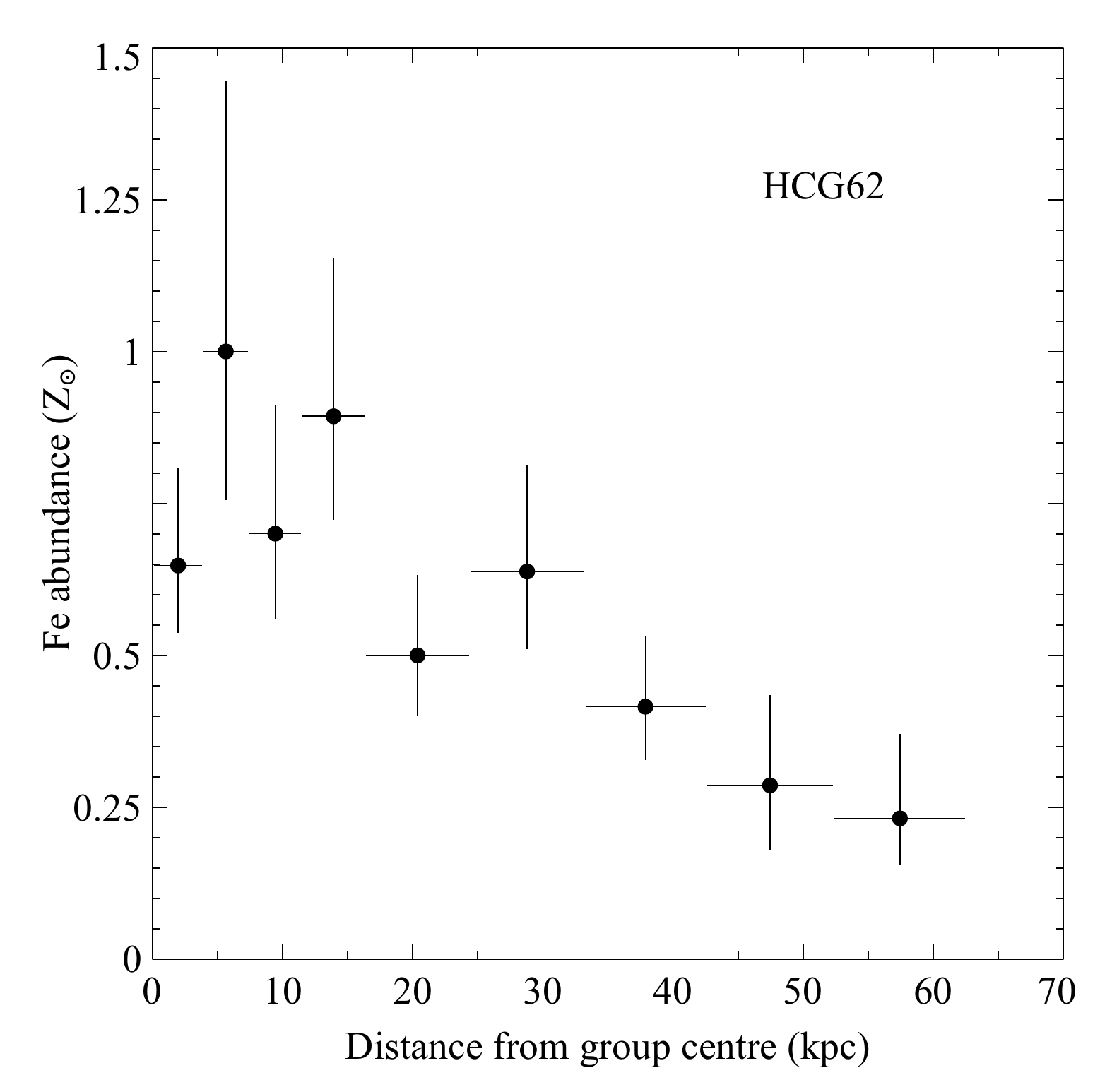}
  \includegraphics[trim=0cm 14cm 5cm 0cm, clip, width=2.9in, height=2.9in]{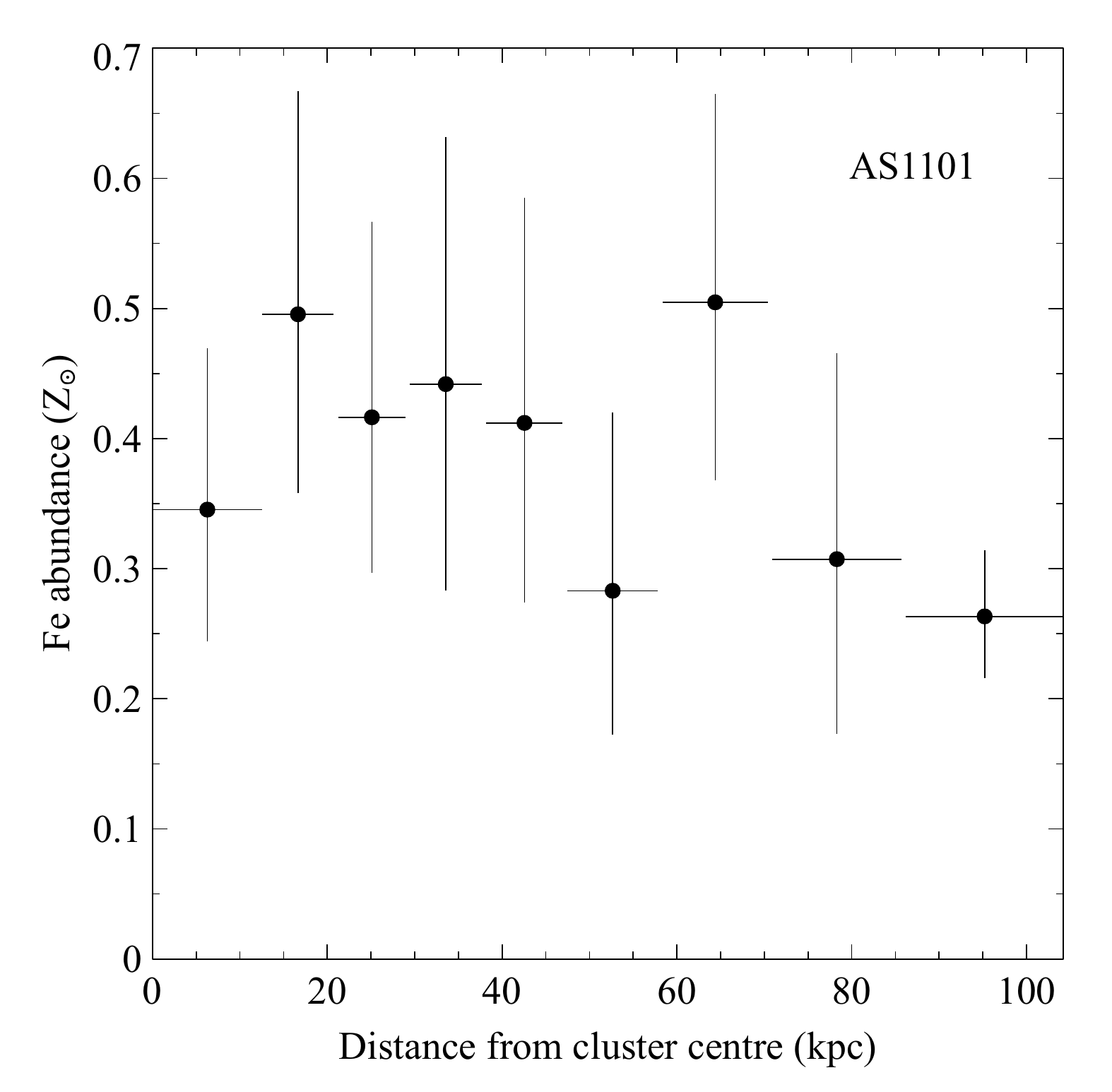}
\caption[]{As for Fig. \ref{fig:abundprofscertain}, but for the sources with possible central abundance drops. Although the metallicity drops in the inner few bins of each source, the drop is not as statistically significant as the drops in the sources with certain central abundance drops.}
\label{fig:abundprofspossible}
\end{center}
\end{figure*}

%\begin{figure*}
%\begin{center}
%\contcaption{}
%\includegraphics[trim=0cm 14cm 5cm 0cm, clip, width=2.9in, height=2.9in]{ironabundprof_hcg62.ps}
%  \includegraphics[trim=0cm 14cm 5cm 0cm, clip, width=2.9in, height=2.9in]{ironabundprof_as1101.ps}
  %   \caption{Deprojected iron, silicon and sulphur (bottom panel) abundance profiles for NGC~4696, adapted from \cite{Panagoulia13}, and iron abundance profiles for Abell~4059 and NGC~5846. The iron, silicon and sulphur abundance profiles are indicated by the black circles, red diamonds and green squares, respectively.}
%  \label{fig:abundprofs3}
%\end{center}
%\end{figure*}

The resulting deprojected abundance profiles are shown in Figs. \ref{fig:abundprofscertain} and \ref{fig:abundprofspossible}, for sources with certain and possible central abundance drops, respectively. We define a central abundance drop as ``certain'' when the drop in abundance is sharp and statistically significant, and there are no similar features (i.e. a sharp drop in abundance followed by an increase) further out in the radial abundance profile. Sources with ``possible'' central abundance drops have a central drop in metallicity, but the drop is not as statistically significant, and there are features of a similar statistical significance and shape at larger distances in the radial abundance profile. The difference in the statistical significance of the abundance drops between the certain and possible abundance drops is evident in these two figures. We note that, since we are interested in {\bf central} abundance drops, we focus on the behaviour of the radial abundance profile in the inner 50 kpc. We were able to recover detailed abundance profiles, for elements other than iron, only for NGC~4696, 2A0335+096 and Abell~262. In the abundance profiles of 2A0335+096 and Abell~262, some of the points of the silicon and sulphur profiles are coincident with the points of the iron abundance profile. This is because, for those spectral bins, allowing the silicon and/or sulphur abundances to vary independently in the spectral fit, had a negligible effect on the statistical ``goodness'' of the fit. We note that, in \cite{Panagoulia13}, we performed additional 2-temperature fits for the two innermost spectral bins of NGC~4696. We do not discuss the results of these fits here, as the abundance values for iron, silicon and sulphur were very similar to those of the single-temperature fits. We refer the interested reader to section 3.3 of \cite{Panagoulia13}, for a more detailed description of the spectral fitting procedure.  

As can be seen, there are abundance drops in the inner bins of all fourteen sources, meaning that up to nearly half of the sources that have X-ray cavities, might also have central abundance drops (14/30). For some of the sources, the abundance values in this region are, in fact, lower than the values in the outermost regions. The general shape of the abundance profile for all sources is a central drop, followed by a peak, after which the profile drops down to lower values. 

The central abundance drops in NGC~4696 have been previously observed \citep{Sanders02, Sanders08, Panagoulia13}, as has that in NGC~4636 \citep[e.g.][]{OSullivan03}. The abundance drop in the core of 2A0335+096 is reported in \cite{Mazzotta03}, though the authors make no comment on abundance profiles of individual elements. \cite{Pandge13} do not present a metallicity profile in their study of Abell~1991, and the metallicity profile shown in \cite{Sharma04} is consistent with a roughly constant metallicity for the inner 100 kpc of the same cluster. A moderate central abundance drop in NGC~4325 is reported in \cite{Rasmussen07}, and in the projected radial metallicity profile of \cite{Lagana14}. A central abundance drop in Abell~262 has been previously reported by \cite{Vikhlinin05}. A central abundance drop in Abell~3581 is evident in the metallicity map of \cite{Canning13}, as well as both the projected and deprojected radial profiles in \cite{Johnstone05}. A modest central abundance drop is seen in the projected radial profile of NGC~5044 in \cite{Rasmussen07}, and the deprojected and projected profiles of \cite{Buote03}. 
%\cite{Ghizzardi13} found metallicity jumps across the cold fronts of Abell~496, with moderate abundance drops in the more central regions, though radial profiles are not discussed. 
A central abundance drop for NGC~6338 is also reported in \cite{Rasmussen07}. 
%A significant central drop in metallicity in Abell~4059 is seen in the deprojected radial profile in \cite{Choi04}, though this disappears when a two-temperature model is used to model the cluster emission. \cite{Reynolds08} also find a central abundance drop, using a single-temperature model for their deprojection analysis. 
The central abundance drop in NGC~5846 has been reported previously in \cite{Rasmussen07}, who observe a drop in both the silicon and iron projected abundance profiles. The central abundance drop seen in HCG~62 has also been previously observed by \cite{Rasmussen07}, though their abundance values are generally higher. This may be due to the use of projected profiles by \cite{Rasmussen07}, in contrast to the deprojected profile shown here. \cite{Rafferty13} also report a central drop in the deprojected metallicity profile of HCG~62, using the same deprojection method as we have used here. Abell~S1101 has not been shown to have a central drop in radial abundance profiles, though the metallicity map of \cite{Werner11} shows that the metallicity in the central regions of the cluster is lower than at larger distances. Central metallicity drops have not been previously identified for NGC~5813 and IC~1262. 

We note that, in the cases of some of the sources with the possible central drops in their deprojected abundance profiles, e.g. Abell~3581, the central abundance drops seen in the deprojected profiles were not visible at all, or were far less pronounced, in the corresponding projected profiles. This could be due to noise introduced by means of the deprojection method used here. The implicit assumption of spherical symmetry made by {\sc dsdeproj}, may be a poorer approximation for some groups and clusters than for others, especially if the data are of poor quality. 
%In addition, the measured abundance in the centre of groups and clusters, depends on whether a single-temperature or two-temperature model is used to model the emission there; it is well-known that using a single-temperature model to represent multi-temperature thermal emission from groups and clusters, results in a low measured abundance \citep[e.g.][]{Werner08}. This particularly affects projected metallicity profiles, where inner bins are contaminated by the emission of neighbouring, and overlying, bins. We note that all the bins in our deprojected metallicity profiles, save the two inner ones in NGC~4696 \citep[see][for more details]{Panagoulia13}, were fit using a single-temperature model. This is because additional temperature components made a negligible improvement to the goodness of the spectral fit. 
Hence, we advise caution when considering the abundance drops seen in the ``noisier'' deprojected metallicity profiles presented here. 

\begin{table}
\begin{center}
\footnotesize{ 
\begin{tabular}{cccc}
  \multicolumn{1}{c}{Source name}&\multicolumn{1}{c}{$Z_{\rm {mag}}$}&\multicolumn{1}{c}{Missing iron mass}&\multicolumn{1}{c}{$D_{\rm {peak}}$}\\
  \multicolumn{1}{c}{}&\multicolumn{1}{c}{}&\multicolumn{1}{c}{(M$_{\odot}$)}&\multicolumn{1}{c}{(kpc)}\\
  \multicolumn{1}{c}{(1)}&\multicolumn{1}{c}{(2)}&\multicolumn{1}{c}{(3)}&\multicolumn{1}{c}{(4)}\\
    \hline
NGC~4696 & 3.7$^{+0.8}_{-0.6}$ & 1.4$\times$10$^{6}$ & 19.0$\pm$1.2 \\
NGC~5846 & 2.6$^{+2.7}_{-1.0}$ & 1.0$\times$10$^{5}$ & 4.6$\pm$1.1 \\
2A0335+096 & 2.1$^{+0.7}_{-0.6}$ & 1.4$\times$10$^{7}$ & 43.0$\pm$2.9 \\
NGC~4636 & 2.0$^{+2.2}_{-1.1}$ & 5.0$\times$10$^{5}$ & 2.8$\pm$0.3 \\
Abell~1991 & 1.9$^{+1.5}_{-0.9}$ & 3.3$\times$10$^{7}$ & 27.3$\pm$3.6 \\
%Abell~2052 & 1.7$^{+0.5}_{-0.4}$ & 1.3$\times$10$^{7}$ & 20.7$\pm$1.9 \\
Abell~262 & 1.1$^{+0.6}_{-0.3}$ & 1.9$\times$10$^{6}$ & 9.3$\pm$1.7 \\
NGC~5813 & 0.9$^{+0.2}_{-0.1}$ & 7.2$\times$10$^{5}$ & 24.5$\pm$3.2 \\
NGC~5044 & 0.9$^{+0.2}_{-0.1}$ & 3.1$\times$10$^{5}$ & 10.6$\pm$1.6 \\
\hline
IC~1262 & 1.9$^{+1.3}_{-0.9}$ & 9.4$\times$10$^{5}$ & 19.8$\pm$3.6 \\
NGC~6338 & 1.3$^{+2.0}_{-0.8}$ & 1.9$\times$10$^{6}$ & 7.3$\pm$0.9\\
Abell~3581 & 1.1$^{+0.9}_{-0.5}$ & 2.3$\times$10$^{6}$ & 21.9$\pm$1.8 \\
%Abell~4059 & 1.0$^{+0.4}_{-0.3}$ & 9.2$\times$10$^{5}$ & 21.7$\pm$5.0 \\
%Abell~496 & 0.8$\pm$0.3 & 5.7$\times$10$^{5}$ & 19.8$\pm$2.9 \\
NGC~4325 & 0.8$^{+0.7}_{-0.3}$ & 9.8$\times$10$^{4}$ & 12.7$\pm$3.4 \\
HCG~62 & 0.5$^{+0.3}_{-0.2}$ & 8.6$\times$10$^{4}$ & 5.6$\pm$1.7 \\
Abell~S1101 & 0.4$\pm$0.2 & 2.3$\times$10$^{5}$ & 16.6$\pm$4.1 \\
\hline
\end{tabular}
}
\end{center}
\caption{List of the sources that have a central abundance drop, and some of their derived properties, in order of descending $Z_{\rm {mag}}$. The sources above the line are the ones with a certain central abundance drop, while the ones below have a possible central abundance drop. The name of each source is given in column (1), while the magnitude of the central abundance drop, $Z_{\rm {mag}}$ as defined in Equation \ref{eqn:zmag}, and the calculated missing iron mass are given in columns (2) and (3), respectively. The extent of the spectral bin in which the peak of the abundance distribution is seen, $D_{\rm {peak}}$, is shown in column (4).}
\label{tab:properties}
\end{table}

\subsection{Single- vs. multi-temperature fits}
As has been mentioned, using a simple 1-T model to represent the emission from a more complex temperature structure can bias the measured metallicity low. This is particularly crucial for the cores of CC groups and clusters, because of the presence of multiphase ICM and cooling flows. For this reason, we have fit the central bins, in which an abundance drop is present or suspected, of each of the 14 sources that are under study here, with single-temperature and multi-temperature models, in order to determine which of the two better represents the ICM emission. The results of these fits are shown in Tables \ref{tab:multitfits} and \ref{tab:multitfits2}, in Appendix A.  

\section{Discussion}
\subsection{``Missing'' iron masses for individual sources}
Here, we calculate the ``missing'' iron mass of each of the fourteen groups and clusters we study in this paper. We have utilised the metallicity ratios of \cite{Anders89} to obtain the metallicity profiles of Figs. \ref{fig:abundprofscertain} and \ref{fig:abundprofspossible}. The number ratio relative to iron is then 4.68$\times$10$^{-5}$, which we use for our calculations of the ``missing'' iron mass in the seventeen sources. Here, we define the ``missing'' iron mass as the additional iron mass needed in order to obtain a flat central abundance profile. The value of the flat central abundance profile is different for each source, and values for individual sources are given in the relevant sections below. We also list the missing iron mass for each source in Table \ref{tab:properties}, along with the magnitude of the abundance drop, $Z_{\rm {mag}}$ (as defined in Section 5.4), and the mean radius of the spectral bin in which the peak in the metallicity distribution is seen, $D_{\rm {peak}}$. The line splits the sources into the ones with a certain (top 8 sources) and possible (bottom 6 sources) central abundance drop. 

\subsubsection{NGC~4636}
NGC~4636 is the central galaxy in a group that is on the outskirts of the Virgo cluster. In the X-ray, it shows a set of spectacular bright arm-like features, identified as the rim of ellipsoidal bubbles, likely produced through the shock heating of the gas around the bubbles as these propagate through the intragroup medium (IGM) \citep{OSullivan05, Baldi09}. The iron abundance in NGC~4636 drops in the central 2 spectral bins, which correspond to the innermost 2.5 kpc, as can be seen in Fig. \ref{fig:abundprofscertain}. Most of the H$\alpha$ filaments in this group are contained within the innermost 1 kpc \citep{Werner14}, and the measured dust mass is $\sim$5$\times$10$^{5}$ M$_{\odot}$, from observations with {\it Herschel} and {\it Spitzer} \citep{Smith12, Martini13}.\cite{Temi07b} used {\it Spitzer} to study the dust in NGC~4636, and find that the infrared emission from NGC~4636 is concentrated in the innermost 1--2 kpc, which coincides with the iron abundance drop. The same authors measure an excess dust mass of $\sim$10$^{5}$ M$_{\odot}$ within the central 4--5 kpc of NGC~4636. We estimate that the ``missing'' iron mass relative to 0.7 central Solar abundance is $\sim$5$\times$10$^{5}$ M$_{\odot}$. 

\subsubsection{NGC~4696}
NGC~4696 is the brightest cluster galaxy (BCG) of the Centaurus cluster, and is the second closest galaxy cluster after the Virgo cluster (NGC~4696 has a redshift of $\sim$0.01). It has a pair of well-defined X-ray cavities, and a plume of X-ray emitting material that stretches to the north east of the cluster \citep[e.g.][]{Fabian05}. We find that the ``missing'' iron mass with respect to unit Solar abundance is 1.4$\times$10$^{6}$ M$_{\odot}$, in the central 5 kpc of this cluster \citep[Fig. \ref{fig:abundprofscertain}; see also][]{Panagoulia13}. The central 5 kpc of NGC~4696 contain most of the optical filamentary structure present in the galaxy \citep{Crawford05a}. {\it Herschel} images, created by \cite{Mittal11}, show that the vast majority of the infrared emission comes from the inner 5 kpc of the cluster. In addition, the same authors detected emission from cold (19 K) and warm (46 K) dust, with respective masses of 1.6$\pm$0.3$\times$10$^{6}$ M$_{\odot}$ and 4.0$\pm$2.8$\times$10$^{3}$ M$_{\odot}$. Emission was detected across the infrared spectrum by \cite{Kaneda05, Kaneda07}, using {\it Spitzer} data. We note that the innermost two spectral bins of NGC~4696, in which we see the drop, were fit using a two-temperature model \citep[for more details of the spectral fitting, see][]{Panagoulia13}. 

\subsubsection{Abell~1991}
Abell~1991 hosts a moderate cooling flow \citep{Stewart84}, and shows moderate amounts of star formation \citep{McNamara92} and accreted H$_{\rm {I}}$ gas \citep{McNamara90}. In this cluster, the drop in iron abundance occurs in the innermost 15 kpc, corresponding to the two inner spectral bins (Fig. \ref{fig:abundprofscertain}). We calculate that the missing iron mass, with respect to 0.9 Solar abundance, is 3.3$\times$10$^{7}$ M$_{\odot}$. The majority of the H$\alpha$ emission is also contained within the inner 15 kpc \citep{McDonald11}. \cite{Cox95} found that Abell~1991 contains $>$1.6$\times$10$^{6}$ M$_{\odot}$ of cold dust, and $>$6.6$\times$10$^{4}$ M$_{\odot}$ of warm dust. 

%\subsubsection{Abell~2052}
% Abell~2052 is a cool core cluster, and has a complex system of X-ray shocks and cavities \citep{Blanton11}. It also shows enhanced H$\alpha$ and N$_{\rm {II}}$ emission, which coincides with the location of the lower entropy X-ray gas \citep{Blanton01, dePlaa10}. Abell~2052 hosts a spectacular system of H$\alpha$ filaments, located within 20 kpc of its core \citep{McDonald12}, which closely matches the brightest X-ray structures. It contains $\leq$9.2$\times$10$^{7}$ M$_{\odot}$ of dust \citep{Edge01}. There is a very steep drop in iron abundance in the central $\sim$17 kpc of the cluster (Fig. \ref{fig:abundprofscertain}), which corresponds to 1.3$\times$10$^{7}$ M$_{\odot}$ of iron, relative to 0.7 Solar abundance.

\subsubsection{2A0335+096}
2A0335+096 is a nearby X-ray luminous cluster, and harbours a cool core. It has a companion that lies to its northwest, with which it has interacted in the past, and has a ``knotty'' appearance in its core \citep{Donahue07, Sanders09a}. The H$\alpha$ filament system in 2A0335+096 is concentrated in the inner 25 kpc of the cluster \citep{Romanishin88}. The dust mass for this cluster was calculated to be about 7.3$\times$10$^{5}$ M$_{\odot}$ by \cite{Edge01}. \cite{Donahue11} detected emission from polycyclic aromatic hydrocarbons (PAHs) in 2A0335+096 using {\it Spitzer}. In addition, \cite{Wilman11} detected a large amount of molecular H$_{2}$ gas in the central galaxy of this cluster, which appears to form an accretion disk and so fuels the central galaxy's AGN. As can be seen in Fig. \ref{fig:abundprofscertain}, there is a sharp drop in the iron abundance in the inner $\sim$20 kpc of the cluster, with indications that a similar trend is followed by the silicon and sulphur abundance profiles. We calculate the missing iron mass to be 9.2$\times$10$^{6}$ M$_{\odot}$, with respect to 0.5 Solar abundance. We note that there are multiple temperature components in the core of 2A0335+096, as detected by the Reflection Grating Spectrometer (RGS) on board {\it XMM-Newton} \citep[e.g.][]{Sanders09a}, and hinted at by our multi-temperature fit results (see Appendix A). Due to the much lower effective area of the {\it Chandra} ACIS instruments at low temperatures relative to the {\it XMM-Newton} RGS, these lower-temperature components cannot be resolved using the currently available {\it Chandra} data. However, these lower-temperature components are not expected to have a significant impact on the existence of a central abundance drop in 2A0335+096, as suggested by the multi-temperature fits shown in Appendix A.

\subsubsection{Abell~262}
Abell~262 is a member of the Perseus supercluster, with a complex central structure in the X-rays, which includes a pair of X-ray cavities \citep{Blanton04}. As can be seen in Fig. \ref{fig:abundprofscertain}, silicon, sulphur and iron all experience an abundance drop in the two innermost spectral bins of Abell~262. This is similar to the abundance profiles seen in NGC~4696 and 2A0335+096, and points to a common origin for these drops. We estimate that the ``missing'' iron mass, relative to 1.3 Solar abundance, is about 1.9$\times$10$^{6}$ M$_{\odot}$. We point out that the emission from the innermost spectral bin, was fit using a two-temperature thermal model, as the use of an extra thermal component statistically improved the fit. Using data from the 30m IRAM telescope, \cite{Salome03} calculated that the mass of dust with a temperature of 35 K is 8.7$\times$10$^{6}$ M$_{\odot}$. The N$_{\rm {II}}$ emission seen in the core of this cluster \citep{Plana98} traces the X-ray structure at the same location, as does the optical emission in its centre \citep{Blanton04}.

\subsubsection{NGC~5044}
NGC~5044 shows filamentary H$\alpha$ structure in its core \citep{Werner14}, and has several small cavities \citep{David09}. In addition, NGC~5044 has been studied in the infrared by \cite{Temi07a} and \cite{Panuzzo11}, who used {\it Spitzer} to observe this and other clusters. Both authors also find emission from PAHs. In a later paper, \cite{Temi07b} studied NGC~5044 more thoroughly, and calculated an excess dust mass of 1.5$\times$10$^{5}$ M$_{\odot}$, with a physical extent of 4--5 kpc. This is also where the steep central abundance drop is seen in this group. As can be seen from the top right panel of NGC~5044, there is a steep drop in abundance in the inner spectral bin, which corresponds to a missing iron mass of 3.1$\times$10$^{5}$ M$_{\odot}$ with respect to 0.55 Solar metallicity.

\subsubsection{NGC~5813}
NGC~5813 has a spectacular well-studied system of X-ray cavities, produced in three separate AGN outbursts, and shocks \citep{Randall11}. It also has H$\alpha$ filaments in its inner $\sim$5 kpc, which lie along the same direction as the X-ray cavities \citep{Werner14}. \cite{Tran01} found a dusty circumnuclear disk in NGC~5813, using WFPC2 on board the {\it Hubble Space Telescope}. Using the {\it Infrared Space Observatory} ({\it ISO}), \cite{Temi04} calculated a dust mass of $\leq$7$\times$10$^{5}$ M$_{\odot}$, for a dust temperature of 20 K. From the abundance profile of NGC~5813 in Fig. \ref{fig:abundprofscertain}, there is a steep abundance drop in the innermost spectral bin. With respect to 0.6 Solar metallicity, the missing iron mass was estimated to be $\sim$7.2$\times$10$^{5}$ M$_{\odot}$. 

\subsubsection{NGC~5846}
NGC~5846 has two X-ray cavities, that are roughly 0.6 kpc from its core, and a ghost cavity at 5.2 kpc to the west of its centre \citep{Machacek11}. In Paper II, we found an additional ghost cavity to the east of the group centre (see also the relevant image in Fig. \ref{fig:imagescertain}). In addition, \cite{Machacek11} find that the rims of the inner two bubbles have a bright, knot-like structure, that are proposed to be either extended filamentary structures, or gas clouds that were shock heated by the most recent AGN outburst. In fact, \cite{Trinchieri02} find that the H$\alpha$ emission traces the bright knotty X-ray filaments to excellent precision. As can be seen from the corresponding abundance profile, there is a steep iron abundance drop in the innermost 2 spectral bins in NGC~5846. Assuming a flat abundance profile value of 0.6 times the Solar value, we estimate that the missing iron mass for the two innermost bins is about 10$^{5}$ M$_{\odot}$, in total. \cite{Tran01} used data from the {\it Hubble Space Telescope} to study the dust features in NGC~5846, and find that they are filamentary in nature. The same authors derive a dust mass of 2.6$\times$10$^{3}$ M$_{\odot}$ for NGC~5846, from the visual extinction of the visible dust features. Finally, \cite{Temi07a} detected NGC~5846 in the 24 $\mu$m, 70$\mu$m and 160$\mu$m energy bands with {\it Spitzer}. \\ 

\subsubsection{NGC~4325}
NGC~4325 is a group with a disturbed core, and has two small X-ray cavities \citep{Russell07}. There are also optical filaments in the core of NGC~4325, which appear to be extended in the same direction as the X-ray emission and are contained within the innermost 15 kpc \citep{McDonald12}. An iron abundance drop is evident in the central spectral bin of this group (Fig. \ref{fig:abundprofspossible}). We calculate the mass of ``missing'' iron to be 9.8$\times$10$^{4}$ M$_{\odot}$, with respect to 0.8 times the Solar abundance. Unfortunately, measurements for the dust mass contained in the core of this group appear to not exist in the literature to date. 

\subsubsection{Abell~3581}
Abell~3581 hosts the most luminous cool core of any nearby group/poor cluster, as well as two pairs of cavities \citep{Canning13}. It also shows signs of sloshing in its core, as well as H$\alpha$ filaments and dust lanes that are cospatial with the central X-ray structure \citep{Canning13}. As can be seen in Fig. \ref{fig:abundprofspossible}, there is a steep central abundance drop in the innermost two spectral bins of Abell~3581, corresponding to the innermost 15 kpc. We estimate that the total missing iron mass, for the two innermost spectral bins, is about 2.3$\times$10$^{6}$ M$_{\odot}$ with respect to 0.8 Solar abundance. We note that the spectrum of the innermost spectral bin was modelled using a two-temperature thermal model, as the addition of an extra thermal component improved the statistical ``goodness'' of the fit. Unfortunately, we were unable to obtain a dust mass estimate from the available literature. We point out that the dust emission is brightest where the abundance drop is steepest in this cluster \citep{Canning13}.

%\subsubsection{Abell~496}
%As can be seen from the corresponding abundance profile in Fig. \ref{fig:abundprofspossible}, the innermost spectral bin of Abell~496 shows a sharp abundance drop. With respect to unit Solar abundance, we estimate that the missing iron mass is equivalent to about 5.7$\times$10$^{5}$ M$_{\odot}$. \cite{McDonald10} observed filamentary H$\alpha$ structure in the central 5 kpc of the core of this cluster, and \cite{Bregman90} calculated a dust mass of 1.1$\times$10$^{7}$ M$_{\odot}$, assuming a dust temperature of 30 K. In addition, the cluster was detected by {\it Spitzer} at 3.6$\mu$m, 4.5$\mu$m, 5.8$\mu$m and 8.0$\mu$m \citep{Hoffer12}. In terms of X-ray structure, Abell~496 displays a sloshing-induced spiral and cold fronts \citep{Roediger12}.  

\subsubsection{NGC~6338}
NGC~6338 has two, maybe three, X-ray cavities within just 6 kpc of its core, and luminous X-ray filaments which coincide with the H$\alpha$ filaments \citep{Pandge12}. It forms a pair with MCG+10-24-117, which lies to its north \citep{Pandge12}. \cite{Martel04} show that the dust distribution in the innermost $\sim$5 kpc of NGC~6338 consists of knots, while there are also H$\alpha$ filaments in the same region. As can be seen from the radial abundance profile (Fig. \ref{fig:abundprofspossible}), there is an abundance drop in the inntermost $\sim$7 kpc, which correspond to the innermost two spectral bins. We estimate that the missing iron mass, with respect to 1.3 times the Solar metallicity, for these two bins is 1.9$\times$10$^{6}$ M$_{\odot}$. We were unable to obtain a dust mass measurement from the literature. 

\subsubsection{IC~1262}
In the X-rays, the core of IC~1262 is host to a complex environment. Its main features are a bright arm stretching to the south of the cluster, a sharp discontinuity to the east, and a loop/arm structure to the north \citep{Trinchieri07}. \cite{Crawford99} detected optical H$\alpha$ line emission from IC~1262. From the corresponding radial abundance profile (Fig. \ref{fig:abundprofspossible}), it is evident that there is a steep abundance drop in the central 10 kpc of the group, which correspond to the inner spectral bin. With respect to 0.6 times the Solar abundance, the missing iron mass is calculated to be 9.4$\times$10$^{5}$ M$_{\odot}$. We were unfortunately unable to obtain a value for the dust mass from the literature. 

%\subsubsection{Abell~4059}
%The most prominent features of the X-ray structure of the core of Abell~4059 are a sharp discontinuity to the south west, and a clearly-defined cavity to the north west with a weak shock that encircles it in the north east \citep{Reynolds08}. There is a clear central drop in the radial abundance profile of Abell~4059, shown in Fig. \ref{fig:abundprofspossible}. We estimate the missing iron mass in the innermost spectral bin to be 9.2$\times$10$^{5}$ M$_{\odot}$, with respect to unit Solar abundance. The H$\alpha$ emission in Abell~4059 is centered on its core, but is only slightly extended \citep{McDonald10}. No dust mass measurements were available for Abell~4059 in the literature. \\

\subsubsection{HCG~62}
The calculated missing iron mass for HCG~62 is 8.6$\times$10$^{4}$ M$_{\odot}$, with respect to 0.85 times the Solar metallicity. We were unable to find a dust mass measurement for HCG~62 in the currently available literature. We note that \cite{Gallagher08} detect no excess dust emission from this group. \cite{Rafferty13} report the detection of 2 sets of cavities in HCG~62, with the innermost set at just a couple of kpc from the core. No H$\alpha$ filaments have been detected in this group \cite{Valluri96}. \\

\subsubsection{Abell~S1101}
In Abell~S1101, we estimate the missing iron mass to be 2.3$\times$10$^{5}$ M$_{\odot}$, with respect to 0.45 times the Solar metallicity. \cite{Hansen00} observed Abell~S1101 using {\it ISO}, and detected dust with a temperature of 40 K, and mass of 1.4$\times$10$^{6}$ M$_{\odot}$. In addition, \cite{Jaffe05} report heavy extinction of 1.6--1.8 magnitudes in this source, in the H$\alpha$ band. The same authors show H$\alpha$ images of Abell~S1101, which reveal a H$\alpha$ filament that stretches out to the north of the cluster centre. \\
\vspace{16pt}

\indent 2A0335+096, Abell~1991, Abell~2052, Abell~262, NGC~4325, NGC~6338, IC~1262 and Abell~4059 were all imaged by \cite{Quillen08}, using {\it Spitzer} in the 3.6$\rm {\mu}$m, 8$\rm {\mu}$m and 24$\rm {\mu}$m bands. In these images, the emission coincides with the innermost 1--2 spectral bins of each source, which is where the abundance drops are steepest. In addition, the same authors detected all five sources in the 3.6$\rm {\mu}$m, 4.5$\rm {\mu}$m, 5.8$\rm {\mu}$m, 8$\rm {\mu}$m and 24$\rm {\mu}$m bands. In addition, \cite{Hu92} used the ratio of the Ly$\alpha$ and H$\alpha$ fluxes to estimate the intrinsic extinction in 10 clusters, amongst which were Abell~262, Abell~1991 and Abell~2052. The intrinsic extinction calculated by \cite{Hu92} for all three sources was 0.14 mags or higher, indicating a large amount of intrinsic dust. 

\subsection{Comparison with previous studies}
Out of the sources in our parent sample, that we examine for a central abundance drop and that have been studied previously, some have central abundance drops that we do not detect in our analysis. Two such sources are NGC~533 and MKW~4, both of which have been studied in \cite{Rasmussen07}. The same authors find central abundance drops, for both iron and silicon, in both sources. We examined both sources for a central abundance drop, adjusting the number of counts in each bin, and hence the bin size, as necessary. We did not find convincing evidence in support of a central abundance drop in these two sources. 

In MKW~4, the resulting profile we obtain is consistent with the iron abundance being flat in the inner few bins, with the measured iron abundance decreasing gradually at larger radii. On the other hand, the projected iron and silicon metallicity profiles, shown in \cite{Rasmussen07}, show a sharp peak in the second innermost spectral bin. However, the value of both the iron and silicon abundance in this bin is about 7 times the solar value, which is an unnaturally high value for galaxy groups. Therefore, it is not certain whether this particular central abundance drop, reported in \cite{Rasmussen07}, is real or not. 

In the case of NGC~533, we do indeed see a central abundance drop of a factor of about 2; the innermost bin and second innermost bin, wherein would be the abundance peak, have respective abundance values of 0.52$^{+0.12}_{-0.09}$ Z$_{\odot}$ and 0.96$^{+0.27}_{-0.19}$ Z$_{\odot}$. However, the available data are not sufficient in quality for us to be certain of this, as we were restricted to just four spectral bins, in order to have enough counts in each one to get a reliable abundance measurement. 

We note that \cite{Rasmussen07} use projected profiles in their analysis, which means that each of their spectral bins will be contaminated with emission from neighbouring bins. This will particularly affect the central bins, in which abundance drops are most likely to occur, as they will be the most contaminated by the layers of overlying emission. In fact, in section 4.1 of the same paper, the authors discuss the potential effect that deprojection would have on their data, and show the resulting temperature and abundance profile for NGC~4325 in figure 5 of the same paper. The general effect on the temperature profile is the decrease of the temperature value in individual spectral bins, without changing the overall shape of the temperature profile. However, the impact on the abundance profile is more profound. In some spectral bins the abundance value increases, while in others it decreases, resulting in a significant change in the abundance profile shape. Finally, we point out that, where necessary, \cite{Rasmussen07} added a power law of a fixed photon index to compensate for unresolved emission from low-mass X-ray binaries. These two factors, i.e. the use of an additional power law component and projected spectra, may explain, at least in part, the disagreement between the abundance profiles in \cite{Rasmussen07}, and those resulting from our analysis.

We point out that \cite{Rasmussen07} also use the abundance table of \cite{Grevesse98} in their spectral analysis, while we use that of \cite{Anders89}. Although this difference in the use of abundance tables will have an impact on measured abundance values, we do not expect it to affect the shape of individual abundance profiles. 

%\begin{figure}
%  \includegraphics[trim=0cm 14cm 5cm 0cm, clip, width=8.5cm, height=8.5cm]{abundprof_ngc5846.ps}
%  \caption{Deprojected abundance profile for NGC~5846. As can be seen, although the two inner spectral bins do show a metallicity drop with respect to the third inner bin (the abundance peak), it is not clear whether this feature is a central abundance drop, or part of a general positive abundance gradient.}
%  \label{fig:ngc5846abund}
%\end{figure}

\subsection{Where is the missing iron?}
Assuming now that the abundance drops in the cores of the sources studied here, are caused by a real absence of iron, we examine where the iron might be, or might have gone. 

As previously mentioned, the inner few kpc of each group and cluster studied here contain most of the observed optical filaments, dust emission and infrared emission. This points to a complex multiphase environment, in which cold gas and dust ($\sim$30 K), warm gas (10$^{3}$--10$^{4}$ K) and hot X-ray emitting gas (10$^{7}$ K) all co-exist. Finding a mechanism through which all these phases of the ICM can occupy the same region of a group or cluster is, therefore, a tricky and somewhat speculative process process. 

Iron is commonly depleted on to grains in cold dust \citep[e.g.][]{Draine09}, and could constitute a significant fraction of its mass. The cold gas may have a higher iron abundance than the hot gas in a group or cluster, and so a sizeable fraction of the missing iron could be in the grains \citep[see also][]{Canning11}. As is visible in Figs. \ref{fig:imagescertain} and \ref{fig:imagespossible}, all the sources studied in this paper display AGN feedback activity in the form of bubbles filled with radio-emitting plasma, which are also mostly located within the central 15--20 kpc of their respective host source. These bubbles are buoyant in the ICM gas, and they will drag the optical filaments outwards as they rise to larger distances from the source centre, as has been discussed for the filaments in e.g. NGC~4696 \citep{Fabian03, Crawford05a, Panagoulia13}. As a result, the quasi-continous bubbling process, which offsets cooling in group and cluster cores \citep[see][for a review]{Fabian12}, drags grains outwards. It is possible that both grains and filaments are destroyed beyond a certain distance in each of the seventeen sources studied here, through e.g. sputtering in the hot gas. Thus, the iron would be returned to the hot X-ray emitting ICM at that location, enriching the hot ICM and contributing to the peak in the iron abundance profiles in Figs. \ref{fig:abundprofscertain} and \ref{fig:abundprofspossible}. It is possible for silicon and sulphur to also be embedded in dust grains \citep[see the discussion section in][and references therein]{Calura09}, which would explain their abundance profile shapes, in the cases of 2A0335+096, Abell~262 and NGC~4696. 

The calculation of the timescales involved in such a scenario in the case of NGC~4696 are given in section 4 of \cite{Panagoulia13}. As all fourteen groups and clusters studied in this paper are relatively similar in age and physical characteristics, we expect that roughly the same timescales will apply to the additional thirteen sources studied here. 

Alternatively, a central abundance drop can come about if the iron-enriched ejecta from Type Ia supernovae (SNe) remain clumped, instead of mixing with the surrounding ICM gas. Metal-rich gas clouds cool radiatively on short timescales, thus having weak overall iron emission. This arises since, while in a high metallicity gas clump, the iron ions only need to cool themselves, in a low metallicity gas cloud the iron ions need to cool themselves as well as the surrounding hydrogen, helium etc. As a result, the iron emission will be stronger \citep{Fabian01, Morris03}. However, if this were the case, we would not expect the central abundance drops we observe to coincide so well with the extent of the dust emission. Rather, we would expect to see a more extended drop. Even so, this mechanism may well play a lesser role in defining the overall shape of the abundance profiles.  

\subsubsection{The role of sedimentation and thermal diffusion}
Measuring elemental abundances in the ICM is based on the line emission of individual elements, which falls in the X-ray band, due to the high temperatures typically found in the ICM. However, no information is available on the abundance of helium in the ICM, as it and hydrogen are both fully ionised at ICM temperatures. Thus, it is often assumed that the ICM abundance of helium is the same as its primordial value. This assumption may, however, be flawed, as sedimentaion of helium and metals may take place in and affect cluster cores \citep{Fabian77, Gilfanov84, Chuzhoy04, Ettori06, Medvedev13}. Unless transport processes in the ICM are heavily suppressed, the helium abundance in a cluster centre could be a factor of 2 higher than assumed. This has an impact on emission measure and metal abundance estimates \citep{Ettori06}. 

Although sedimentation has been extensively studied in the literature, little attention has been paid to the effects of thermal diffusion, which arises because of temperature gradients in the ICM. Simulations have shown that, in cool core clusters, thermal diffusion is strong enough to counteract gravitational sedimentation, and reverse the inflow of heavier particles into the cluster core, bringing about the removal of helium and metals from the innermost regions of clusters \citep{Shtykovskiy10}. \cite{Medvedev13} show that underestimating the helium abundance in the ICM leads to an underestimation of the abundance of metals. Using Abell~2029 as an example of a cool core cluster, the same authors demonstrate that thermal diffusion removes all the iron from the cluster core within 7 Gyr of diffusion being active. In addition, \cite{Medvedev13} show that the combined effect of thermal diffusion and gravitational sedimentation leads to a peak in the abundances of both helium and iron about 200 kpc from the core of Abell~2029. We note that \cite{Medvedev13} did not include magnetic fields and large-scale turbulence in their model, which are the two processes that may significantly decrease the impact of thermal diffusion. 

It is possible that thermal diffusion in the core of the sources considered in this study does indeed help shape their metal abundance profiles. However, since all fourteen sources display signs of AGN feedback in the form of X-ray cavities, we expect that the relative importance of thermal diffusion to be small compared to that of buoyantly rising bubbles, in terms of removing metals from cluster and group cores and distributing them further out in the ICM. This is supported by the fact that the diffusion velocity is much lower than the expected turbulent velocity generated by the AGN bubbling process \citep{Medvedev13}. 

\subsubsection{The role of resonance scattering}
It is possible for the strength of resonance lines to be affected by resonance scattering \citep{Gilfanov87, Sanders06b, Werner09, Zhuravleva13}. Many of the principal iron emission lines in the X-ray are resonance lines. If a group or cluster of galaxies is optically thick at the energies of strong resonance lines along a line of sight, the emission from these lines that would normally travel directly to an observer along that line of sight could then be scattered away from that line of sight. This could then lead to an underestimate of the central abundance, and, conversely, the abundance in outer regions could be overestimated. This phenomenon could strongly affect CC groups and clusters, many of which host systems of optical filaments \citep[e.g. see table 2 in][]{Panagoulia14b}. 

\cite{Sanders06b} considered the effect of resonance scattering on two clusters that show central abundance dips, namely NGC~4696 (the Centaurus cluster) and Abell~2199, using the {\it Chandra} observations that were available for each cluster at the time. As previously mentioned, we have confirmed the presence of a central drop in the deprojected radial abundance profile of NGC~4696 \citep{Panagoulia13}, though we do not detect one in Abell~2199, due to the lower quality of the data resulting in ``noisy'' deprojected spectra. The spectral model used by \cite{Sanders06b} to determine whether resonance scattering does result in off-centre peaks in abundance, takes effects such as projection and photoelectric absorption into account, as well as resonance scattering of many spectral lines. In the case of NGC~4696, \cite{Sanders06b} found that accounting for resonance scattering makes only a small improvement to the spectral fits, and does not change the measured abundance noticeably ($<$ 10 per cent) or remove the central abundance dip. We note that the same authors did not find a resonance scattering-induced change in the measured abundance of Abell~2199, either. 

Although resonance scattering has a negligible impact on the measured abundance in NGC~4696 when relatively low quality CCD spectra are used, such as those from the {\it Chandra} ACIS detectors, \cite{Sanders06b} point out that resonance scattering should be included in the spectral models when higher resolution X-ray spectra are considered, such as those obtained from the {\it XMM-Newton} RGS or a future X-ray bolometer mission, such as {\it ASTRO-H}. In the core of a source such as the Centaurus cluster, assuming that turbulence is quite low, resonance scattering increases the strength of the resonant Fe-K lines by up to 30 per cent (by scattering photons produced in outer regions of the cluster inwards), and decreases that of the resonance Fe-L lines by up to 10 per cent. 

For all the clusters and groups studied in this paper, as previously mentioned, we have used {\it Chandra} ACIS spectra. We therefore estimate that the effect of resonance scattering is negligible for our estimates of the abundance profiles for each group and cluster, since morphologically they are very similar to the Centaurus cluster (existence of X-ray cavities, short central cooling times and presence of optical filaments their core).  

\subsection{A closer look at 2A0335+096}

\begin{figure*}
\begin{center}
\includegraphics[trim=0cm 14cm 5cm 0cm, clip, width=8.5cm, height=8.5cm]{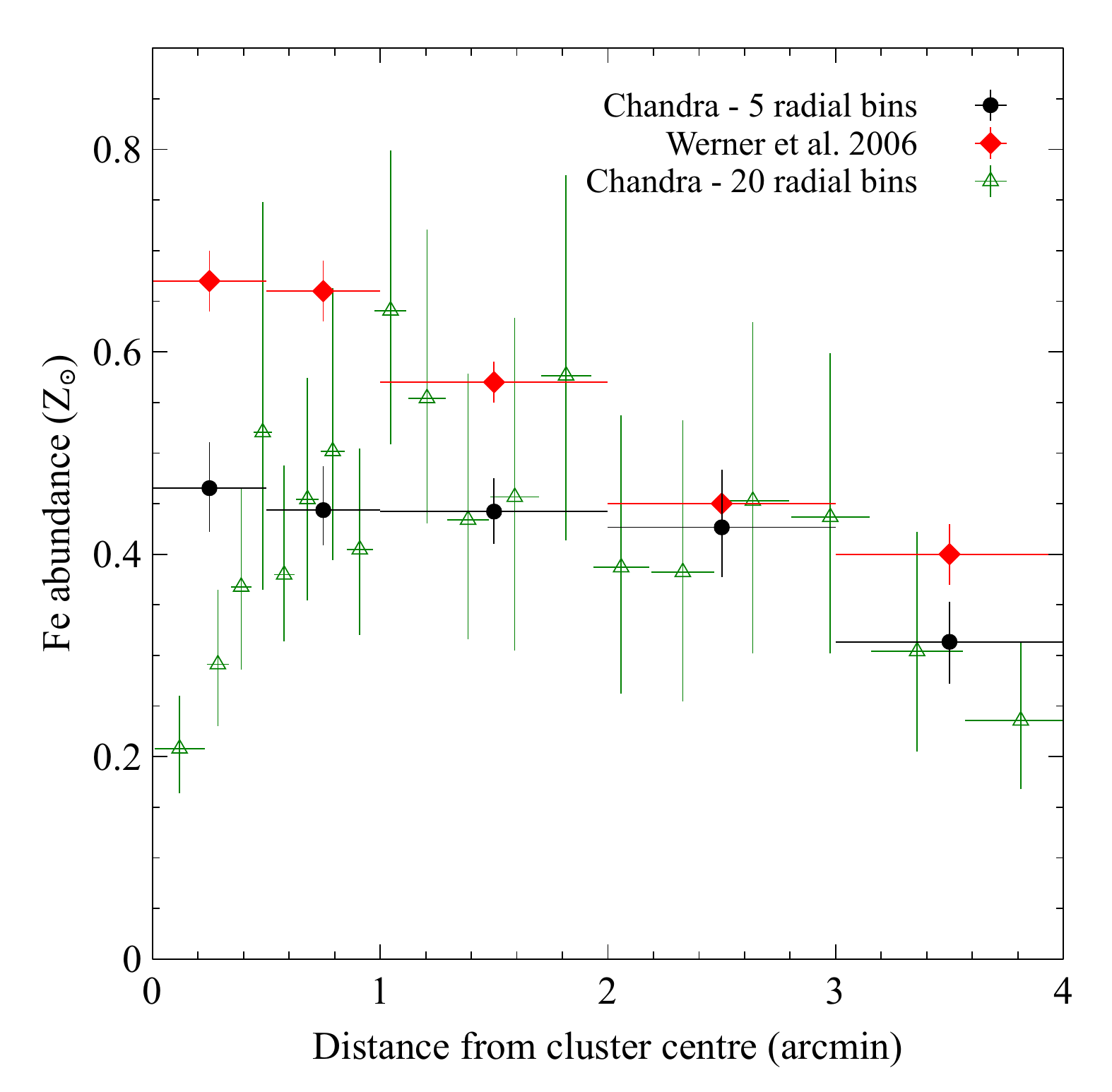}
\includegraphics[trim=0cm 14cm 5cm 0cm, clip, width=8.5cm, height=8.5cm]{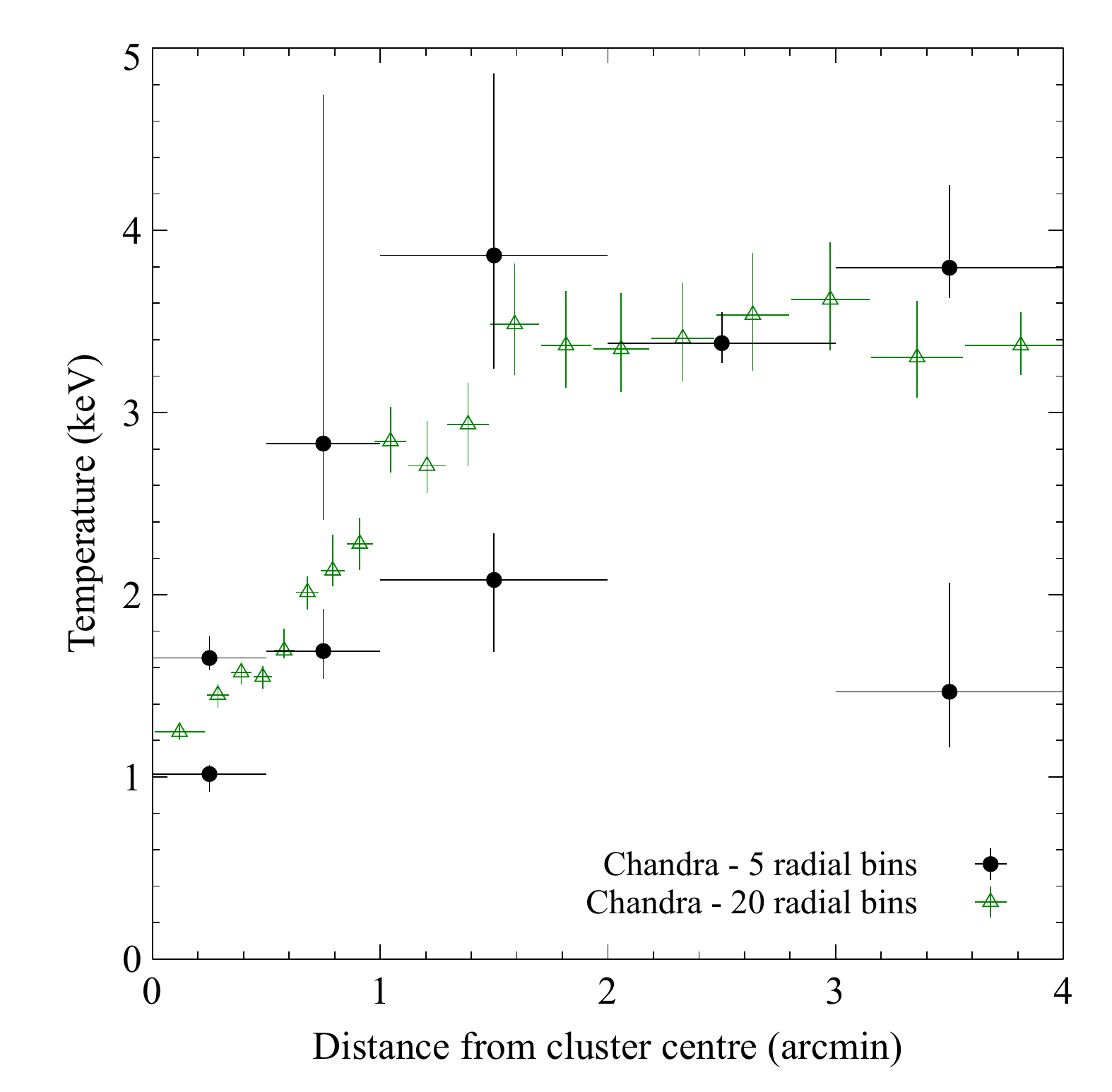}
\caption{{\it Left}: Iron metallicity profiles obtained using {\it Chandra} and {\it XMM-Newton} data. The black circles are spectra extracted from the {\it Chandra} data from the 5 radial {\it XMM-Newton} bins as defined by \protect\cite{Werner06}, the red diamonds are the iron metallicity profile calculated using a {\sc wdem} model from \protect\cite{Werner06} (see table 6 in the same paper), and the green triangles are the iron metallicity profile shown in Fig. \ref{fig:abundprofscertain}. The two profiles made using wider radial bins show no sign of a central metallicity drop. {\it Right}: Deprojected temperature profiles made using {\it Chandra} data. The black circles are the profile using the same radial bins as \protect\cite{Werner06}, while the green diamonds are the profile made using 20 radial bins. The 20-radial bin profile shows a very steep decrease in the inner $\sim$1.5 arcmin or so, which include the region that shows an abundance drop.}
\label{fig:wernerabund}
\end{center}
\end{figure*}

As previously mentioned, cool core groups and clusters harbour a multiphase ICM, from X-ray-emitting hot gas to cold dust. In the X-ray, this can lead to a complex temperature structure in the core. Incorrectly modelling this complex structure can give rise to a well-known metallicity bias: if a multi-temperature structure is modelled with a single-temperature model, the calculated metallicity can be biased low \citep[see e.g.][]{Buote00, Buote03}. We have, as mentioned in Section 3.2, added extra thermal components to the models used to represent the ICM emission from the sources under study, where this was necessary. This was the case with just a few of the sources, for which the addition of an extra thermal component significantly improved the statistical ``goodness'' of the fit, and whose data were of a high enough quality so that the temperature of the additional component could be accurately constrained. However, previous studies of some of these sources, have indicated that there is in fact no central abundance drop in them, particularly if different data were used and/or the metallicity profiles were made using larger bins and over larger scales. To investigate whether our analysis methods produce results consistent with those of previous studies, we take a closer look at the metallicity profile of 2A0335+096. We choose to further study this cluster because it has $\sim$102 ks of clean exposure time, is bright and has a well-studied multiphase ICM. In addition, \cite{Mazzotta03}, using {\it Chandra} data, detected a central abundance drop and a peak at $\sim$55'' ($\simeq$38 kpc) in both the projected and deprojected metallicity profiles of 2A0335+096, which is in good agreement with our results presented in Fig. \ref{fig:abundprofscertain}. Furthermore, \cite{Sanders09a} generated a projected metallicity map, in which an off-centre metallicity peak is seen south of the cluster centre. In the same map, the metallicity in the core of the cluster appears lower. On the other hand, \cite{Werner06} used {\it XMM-Newton} to study 2A0335+096. Using much wider bins than those used in this paper and by \cite{Mazzotta03}, the authors found no central drop in the abundance profile of 2A0335+096; rather, the deprojected metallicity profile seems to have a central peak and then it smoothly decreases to lower values with increasing radius, when a two-temperature model is used. 

Since our results agree with those presented in \cite{Mazzotta03}, in this section we try to replicate the deprojected metallicity profile shown in \cite{Werner06}. For this reason, we extract deprojected spectra from the three {\it Chandra} observations (ObsIDs 919, 7939 and 9792), in the same annuli used in \cite{Werner06} for the deprojected metallicity profiles: 0--0.5', 0.5--1.0', 1.0--2.0', 2.0--3.0 and 3.0--4.0'. Our spectral extraction follows the method outlined in Section 3.2, with few differences. Specifically, we use a {\sc vapec} model for the emission from each annulus, and we allow Mg, Si, S, Ar, Ca and Ni to vary freely in addition to Fe, independent of the change to the statistical ``goodness'' of the fit, in order to be consistent with the results presented in \cite{Werner06}. We note that, even when all the aforementioned elements were allowed to vary freely, the change to the measured Fe abundance in each annulus was minimal ($\leq$3 per cent, or 0.01 Z$_{\odot}$), with the exception of the second annulus (0.5--1.0' radius), where the change was 0.04 Z$_{\odot}$ ($\sim$8 percent). In addition, we allowed the column density, $N_{\rm {H}}$, to vary freely, in line with the method used in \cite{Werner06}. The 3 innermost annuli and the outermost annulus were best-fit using a two-temperature model, while a single-temperature model was used for the second but last outer annulus (2.0--3.0' radius), since the value of the second temperature could not be constrained. The addition of an extra thermal component to the spectral model used for the second outermost annulus, did not change the measured iron abundance significantly. 

The resulting Fe metallicity profile is shown in the left-hand panel of Fig. \ref{fig:wernerabund}, where the black circles are the deprojected iron profile extracted from the {\it Chandra} data using the radial bin sizes of \cite{Werner06}, the red diamonds are the deprojected iron metallicity profile calculated using a {\sc wdem} model from \cite{Werner06} (see table 6 in the same paper), and the green triangles are the deprojected iron metallicity profile shown in Fig. \ref{fig:abundprofscertain}. We note that, in the time between the creation of the abundance profile shown in Fig. \ref{fig:abundprofscertain} and the 5 radial bin profile presented in this section, the {\sc ciao} software package had been updated to a new version (version 4.6). For this reason, the data used to generate the 5 radial bin profile had to be reprocessed using a different {\sc ciao} version to that used to reprocess the 20 radial bin profile data. Although there is an offset between the 5 radial bin profile calculated in this paper and the one shown in \cite{Werner06}, in both cases there is no central drop in abundance, although the \cite{Werner06} profile shows a central peak that the equivalent {\it Chandra} profile does not. Therefore, as far as the central drop in iron abundance is concerned, our results are broadly consistent with those of \cite{Werner06}. The left-hand panel of Fig. \ref{fig:wernerabund} also demonstrates the need for high-quality data in order to detect central metallicity drops (see the next section for a more detailed discussion), since the central metallicity drop is undetected by both {\it Chandra} and {\it XMM-Newton} when larger radial bins are used. 

We note that the difference in the metallicity profiles extracted from the {\it Chandra} data, using 5 radial bins and 20 radial bins (black circles and green triangles in Fig. \ref{fig:wernerabund}, respectively), is likely due to the fact that the 20-bin temperature profile has a very steep gradient inwards from the 12th radial bin ($\sim$1.6'), which includes the region in which an abundance drop is seen. In fact, the temperature drops from $\sim$3.4 keV in the 12th radial bin, down to just $\sim$1.2 keV in the innermost bin, as shown in the right-hand panel of Fig. \ref{fig:wernerabund}. In this figure, the black circles represent the profile made using the radial bins of \cite{Werner06}, while the green triangles are the profile made using 20 radial bins. Therefore, using just 3 radial bins to model the inner 2' of 2A0335+096, results in the temperature gradient being smoothed over, which in turn affects the metallicity measurements. Again, we point out the existence of previously detected low-temperature components in the core of 2A0335+096 \citep[][see also Appendix A]{Sanders09a}, though these are not expected to significantly affect the existence of a central metallicity drop. 

There is a well-known discrepancy between the temperatures measured, for the same object, by the EPIC and ACIS instruments, which can in turn affect the corresponding metallicities \citep{Schellenberger14}. This may be one of the more significant underlying factors contributing to the difference between the two 5 radial bin profiles, presented in Fig. \ref{fig:wernerabund}. In addition, the metallicity values measured by {\it Chandra} could be lower than those measured by {\it XMM-Newton} because of the use of the {\sc vapec} model in this analysis, vs. the use of the {\sc mekal} model in \cite{Werner06}. Generally speaking, the values calculated using a {\sc vapec} model are lower than those obtained using a {\sc mekal} model \citep{Sanders06a}. The use of a {\sc wdem} model by \cite{Werner06}, with {\sc mekal} thermal components, as opposed to a single- or two-temperature {\sc vapec} model in our analysis, will further differentiate the obtained metallicities. In addition, different deprojection methods may play a role in the discrepancy between the two profiles. 

Overall, it is clear from both the analysis presented above, and from previous studies, that 2A0335+096 is a cluster with a highly complicated core structure \citep[e.g.][]{Sanders09a}. For this reason, accurately deprojecting spectra from this region is quite hard, particularly because of the high column density observed there ($\sim$2$\times$10$^{21}$ cm$^{-2}$). Therefore, care must be taken when modelling the core of 2A0335+096, and when interpreting the results of the spectral fitting. One way to disentangle the relative metallicity contributions from multiple temperature components, would be to measure the metallicity of each individual temperature component. Unfortunately, this is not possible with the currently available data on the clusters under study in this paper.  

\subsection{Data quality effects}

\begin{figure}
\begin{center}
\includegraphics[trim=0cm 14cm 5cm 0cm, clip, width=8.5cm, height=8.5cm]{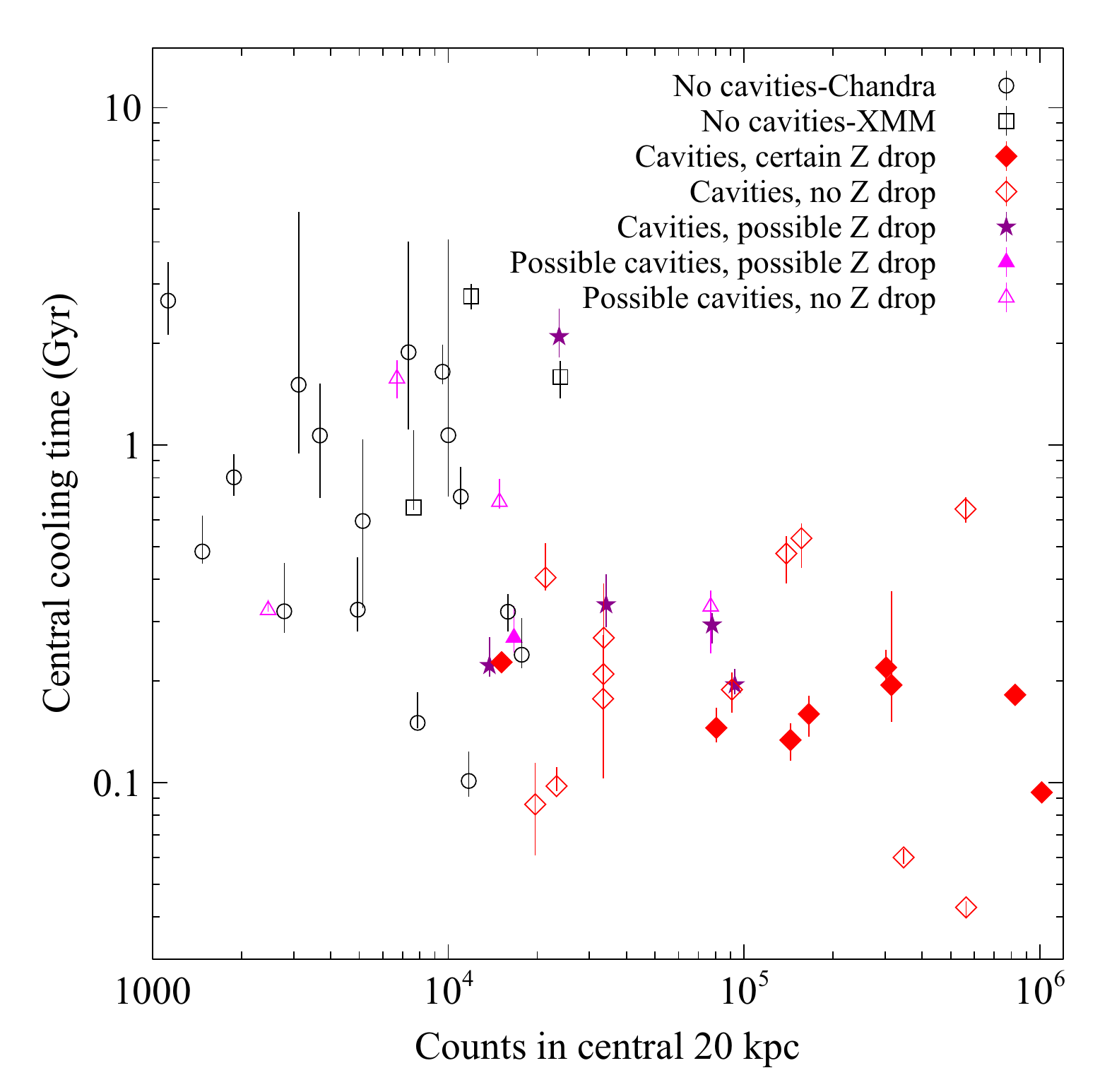}
\caption{Central cooling time vs. number of counts within a 20 kpc-radius circle, centered on the source, for the 49 sources of our parent sample, which have a central cooling time $\leq$3 Gyr, adapted from figure 4 of Paper II. Empty and filled symbols represent sources without and with certain or possible central abundance drops, respectively. For a more detailed explanation of the individual symbols, please see text.}
\label{fig:cavsanddrops}
\end{center}
\end{figure}

In Paper II, we examined whether the detection rate of X-ray cavities was affected by the quality of the available data. To this end, we calculated the central cooling time, and the number of counts within a circle of a 20 kpc radius centered on the source, for the 49 sources in our short central cooling time subsample (i.e. for sources that have a central cooling time of $\leq$3 Gyr), and plotted the former against the latter. We found that there is indeed a strong correlation between the detection of cavities, and the number of counts within the central 20 kpc of each group and cluster (see figure 4 of Paper II). Motivated by this work, we examine whether there is a similar trend for the detection of central abundance drops. To this end, we make use of figure 4 from Paper II, adjusting the symbols so that they additionally reflect the presence or absence of a central abundance drop. The resulting plot is shown in Fig. \ref{fig:cavsanddrops}, where we have plotted the central cooling time against the number of counts in the 20 kpc-radius circle centered on the corresponding  source. We note that no central abundance drops were found in sources without X-ray cavities. The empty and filled symbols represent sources without and with certain or possible central abundance drops, respectively. The empty black circles and squares indicate sources that do not have X-ray cavities or a central abundance drop, and for which {\it Chandra} and {\it XMM-Newton} data were used in the analysis, respectively. The filled (empty) pink triangles represent sources that have possible X-ray cavities, and do (do not) have a possible central abundance drop. The filled (empty) red diamonds are sources that have certain X-ray cavities, and do (do not) have a certain central abundance drop. Finally, the purple stars indicate sources that have certain cavities, but only a possible central abundance drop. For more details on the cavity classification scheme, we refer the interested reader to section 5.1 of Paper II. We note that no central abundance drops were found in sources without X-ray cavities. 
%We note that the two sources with possible cavities and the sources with certain cavities that have a central abundance drop (indicated by the filled pink triangles and the filled red diamonds, respectively), all have a steep central abundance drop, hence the relevant key descriptor in Fig. \ref{fig:cavsanddrops}. 

\begin{figure}
\begin{center}
\includegraphics[trim=0cm 14cm 5cm 0cm, clip, width=8.5cm, height=8.5cm]{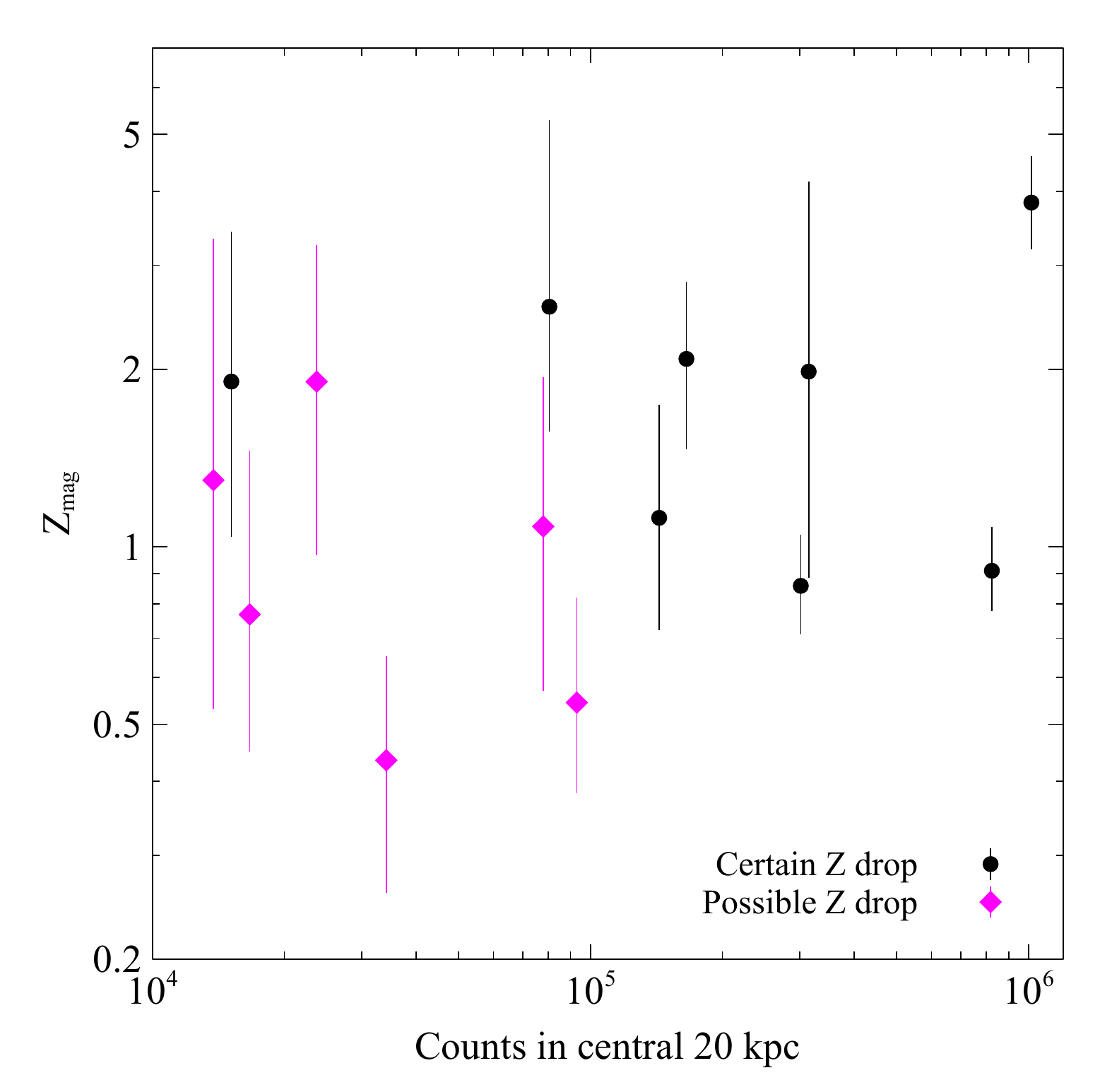}
\caption{$Z_{\rm {mag}}$ vs. number of counts within a 20 kpc-radius circle, centered on the core of the source, for the 14 sources in which we have detected a central abundance drop. Black circles indicate sources that have a certain central abundance drop, pink diamonds represent sources with a possible central abundance drop. As can be seen, no correlation exists between the magnitude of the abundance drop and the number of counts in the central 20 kpc. However, there does seem to be a threshold in the number of counts, above which all central abundance drops are certain.}
\label{fig:zmagvscounts}
\end{center}
\end{figure}

As can be seen from Fig. \ref{fig:cavsanddrops}, central abundance drops are generally more likely to occur in sources with cavities and a central cooling time $\leq$1 Gyr. These drops are detected when the number of counts within the 20 kpc radius circle is $\geq$13000. Half of the sources (1/20, or $\sim$65 percent) with $\geq$30000 counts in their central 20 kpc, have a certain or possible central abundance drop. Out of these sources, 7/20 have a certain central abundance drop (35 percent), while the remaining three, have possible drops. If we consider the 16 sources with more than 50000 counts in their central 20 kpc, then the fractions increase to 9/16 ($\sim$56 percent) for sources with a certain or possible central abundance drop, and 7/16 ($\sim$44 percent) for sources with a certain central abundance drop. Although there is no cut-off number of counts, above which all sources have a central abundance drop, as was the case with X-ray cavity detection in Paper II, there is a clear trend in as much as a higher number of central counts (in this case, $\geq$13000) means there is a greater possibility of detecting a central abundance drop. 

In addition, we examine if there is a correlation between the number of counts in the central 20 kpc of a group or cluster, and the magnitude of the central abundance drop, $Z_{\rm {mag}}$. We define $Z_{\rm {mag}}$ as 
\begin{equation}
Z_{\rm {mag}} = \left|\frac{Z_{\rm {in}} - Z_{\rm {peak}}}{Z_{\rm {in}}}\right| = \left|1 - \frac{Z_{\rm {peak}}}{Z_{\rm {in}}}\right|,
\label{eqn:zmag}
\end{equation}
where $Z_{\rm {in}}$ is the abundance measured in the innermost spectral bin of a source, and $Z_{\rm {peak}}$ is the abundance measured at the peak of the radial abundance profile. $Z_{\rm {mag}}$ is the value of the abundance calculated in the single spectral bin with the highest abundance value, rather than the average abundance value of the few spectral bins that might comprise the abundance peak ``plateau''. Here, we used standard error propagation formulae to calculate the errors on $Z_{\rm {mag}}$. We list the value of $Z_{\rm {mag}}$ for each source in column (2) of Table \ref{tab:properties}. We isolate the 14 sources in which we have found central abundance drops, and plot $Z_{\rm {mag}}$ against the number of counts within the 20 kpc radius circle, which is centered on their core. The resulting plot is shown in Fig. \ref{fig:zmagvscounts}, where the black circles and pink diamonds indicate sources with certain and possible central abundance drops, respectively. There is clearly no correlation between the two plotted quantities, indicating that the strength of the observed central abundance drop does not depend on the quality of the available data. However, there is a dependence of the statistical significance of the detected drop on the data quality. There is a threshold number of counts, $\sim$94000 counts, above which all the detected central abundance drops are ``certain''. Below this, only with a couple of exceptions, all central abundance drops are ``possible''.

We point out that the threshold above which central abundance drops are detected, from Fig. \ref{fig:cavsanddrops}, and that above which cavities are detected, in figure 4 of Paper II, are quite similar. Generally speaking, $\sim$13000 counts within a 20 kpc-radius circle, concentric with the source centre, are needed to detect either a central abundance drop, or X-ray cavities. This number of counts is so high as, when searching for either X-ray cavities or an abundance drop, a drop in surface brightness or metallicity relative to adjacent regions needs to be detected. In other words, in both cases, relative differences between different regions need to be detected, rather than absolute values in a certain region.

\subsection{Cavity location vs. location of abundance peak}

\begin{figure}
\begin{center}
\includegraphics[trim=0cm 14cm 5cm 0cm, clip, width=8.5cm, height=8.5cm]{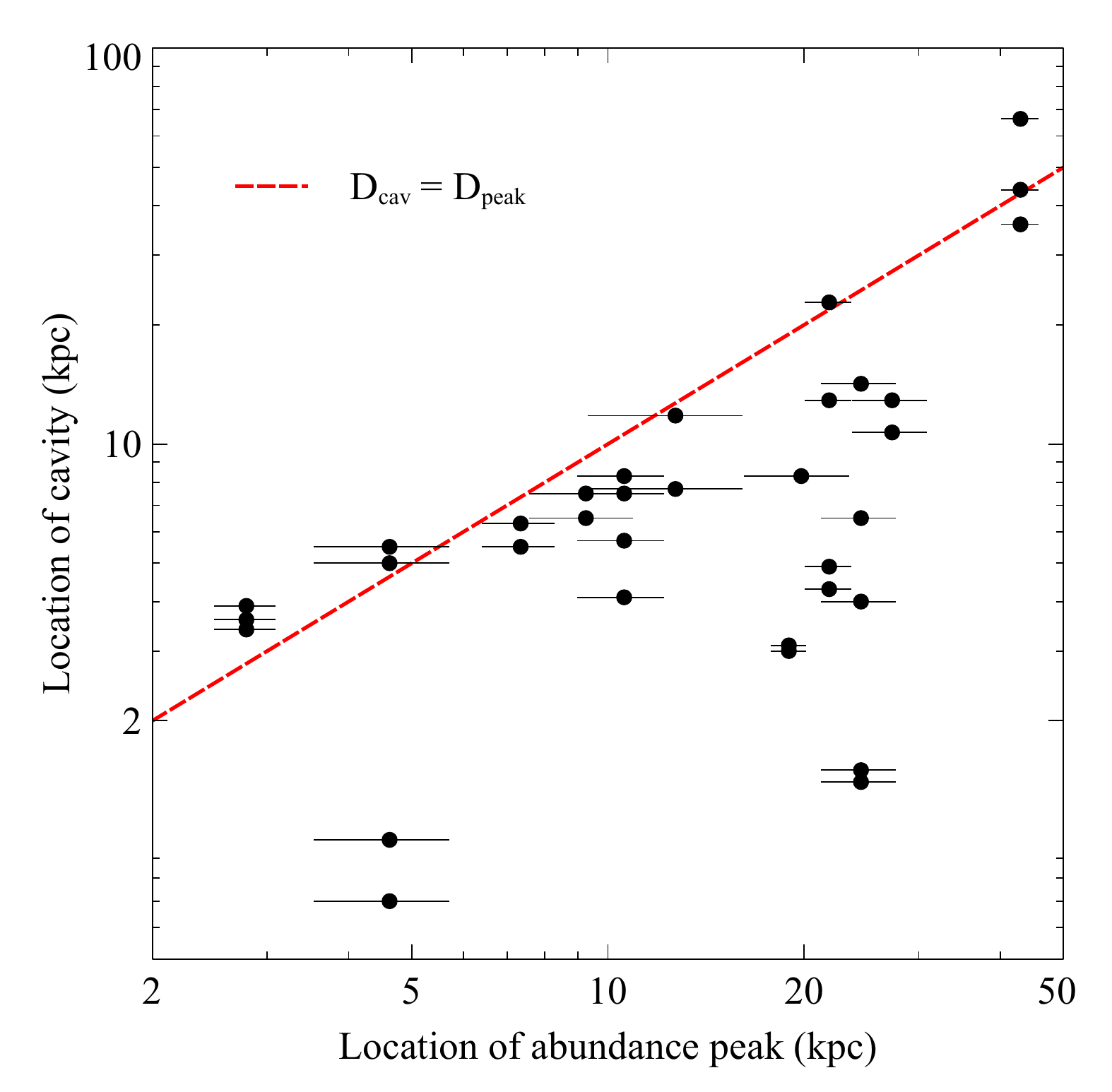}
\caption{Cavity location vs. abundance peak location, for all the cavities of the fourteen groups and clusters that exhibit central abundance drops. As can be seen, for most sources the abundance peak lies further away than the cavities themselves. The red dashed line represents $D_{\rm {cav}} = D_{\rm {peak}}$.}
\label{fig:abundpeaks}
\end{center}
\end{figure}

In Fig. \ref{fig:abundpeaks}, we plot the location of the cavities, $D_{\rm {cav}}$, of each of the sources that has a central abundance drop, against the mean radius of the spectral bin in which the abundance peak lies, $D_{\rm {peak}}$. The values for $D_{\rm {cav}}$ are taken from table 2 of Paper II, while the value of $D_{\rm {peak}}$ for each source is listed in column (4) of Table \ref{tab:properties}. The red line in the same figure indicates $D_{\rm {cav}} = D_{\rm {peak}}$. We see that, for most sources, the abundance peaks lie closer to the source centre than the cavities, as has been found previously by e.g. \cite{Kirkpatrick11}. This is in line with the theory that bubbles drag up metal-rich material from group and cluster cores, and dump it at larger distances. As such, we would expect the abundance peak location to be no further than the location of the cavities; in other words, $D_{\rm {cav}}$ acts as an upper limit to $D_{\rm {peak}}$. The fact that some abundance peaks lie further out than the respective cavities may be due to the presence of ghost cavities, that can no longer be detected. 

Steep abundance peaks such as some of the ones seen here, are expected to be relatively long-lived structures in their host sources, as it would take a significant amount of time to transport enough material outwards to create them. As such, it is possible that the metallicity peaks in groups and clusters are tracing the AGN outburst history of their host source. 

\subsection{Cavity power vs. magnitude of central abundance drop}
As has been mentioned in previous sections, all groups and clusters found to have a central abundance drop, also have X-ray cavities. One question that naturally arises from this fact, is whether cavity power influences the magnitude of the central abundance drop, $Z_{\rm {mag}}$, defined in Equation \ref{eqn:zmag}. We list the value of $Z_{\rm {mag}}$ for each source in column (2) of Table \ref{tab:properties}. 
%Na\"{\i}vely, one might expect more powerful cavities, and hence AGN outbursts, to produce larger central abundance drops. 

\begin{figure}
\begin{center}
\includegraphics[trim=14cm 4cm 0cm 0cm, clip, width=8.5cm, height=8.5cm, angle=90]{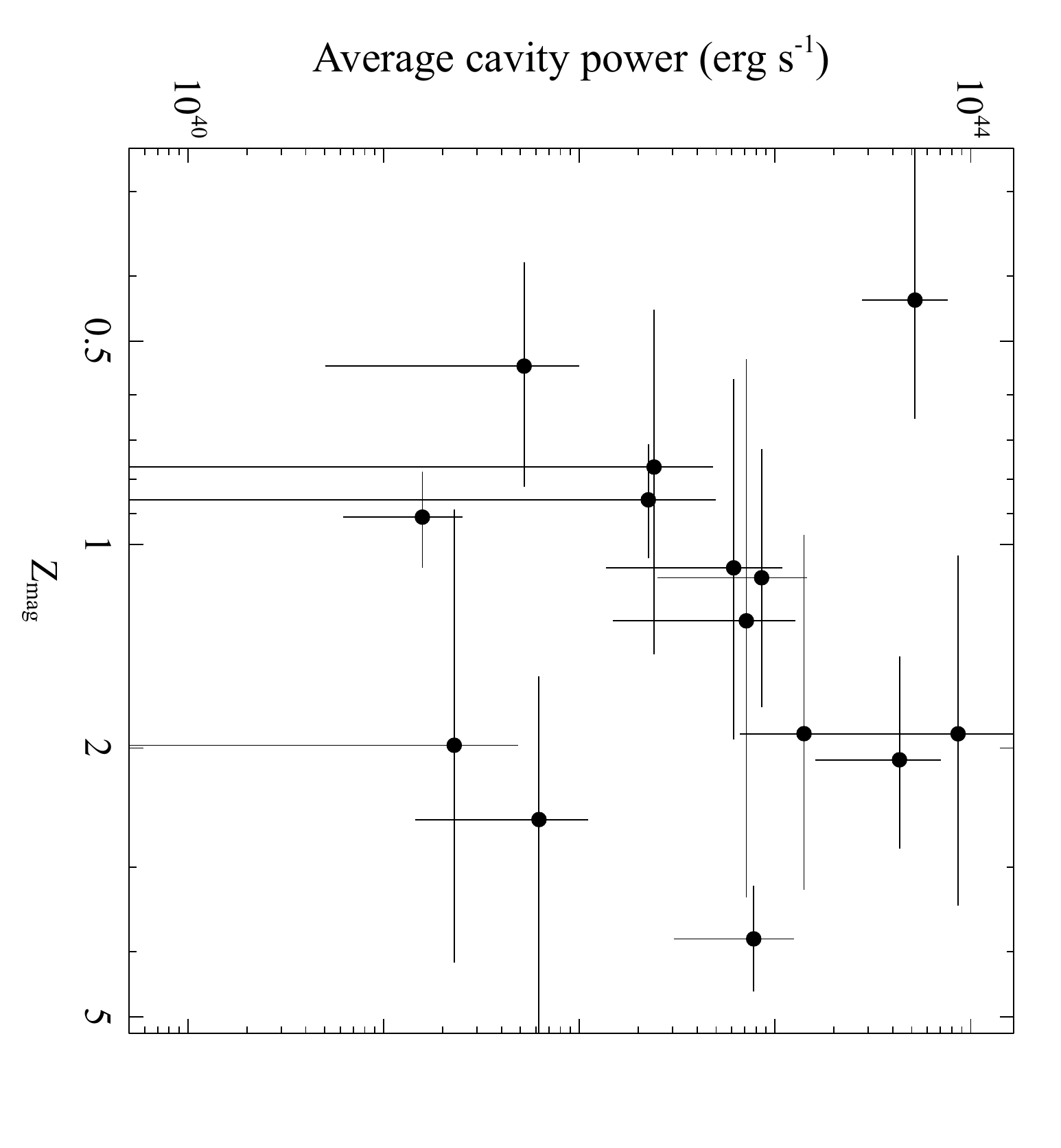}
\caption{Average power of the inner set of cavities of each source, vs. the magnitude of the corresponding central abundance drop, for all 17 sources with a central abundance drop. No correlation is visible between the two quantities, hinting at a complex relation between the measured central abundance drop, and the source's AGN outburst history.}
\label{fig:cavpowervszmag}
\end{center}
\end{figure}

For the sake of clarity, we plot the average cavity power of the inner set of X-ray cavities in a source, against the corresponding magnitude of the central abundance drop, in Fig. \ref{fig:cavpowervszmag}. For NGC~5044 and NGC~4636, since all the cavities in each of these sources are at very similar distances from their respective group core (see Fig. \ref{fig:imagescertain}), we averaged over the power of all the cavities in each source. In the case of 2A0335+096, we used the average power of the cavities to the northwest of the core. As can be seen from Fig. \ref{fig:cavpowervszmag}, there is no correlation between the two plotted quantities. This is supported by the spread of values in $Z_{{\rm mag}}$, for a certain average cavity power. This implies that the magnitude of the central abundance drop depends on factors other than the power of the cavities produced by an AGN outburst. It is likely that $Z_{\rm {mag}}$ depends on several different factors, all of which comprise at least parts of the AGN outburst history of the cluster. 

\subsection{Testing the dust model}
As mentioned previously, a likely cause for the central abundance drops in all the sources studied here, is the embedding of iron, and possibly silicon and sulphur, in dust grains. This assumes that some of the stellar mass loss material in the core of a group or cluster, can remain cool and separate from the surrounding ICM, preserving its embedded dust, as proposed by \cite{Voit11}. As a result, we can test our proposed theory by measuring the radial abundance profiles of elements that are not expected to partake in the creation of dust grains. Two of these elements, whose abundance is measurable using {\it Chandra}, are neon and argon. We would expect to obtain a relatively flat abundance profile for both these elements, if they are indeed not depleted in dust grains. 

Recently, the first detection of a noble gas molecular ion, $^{36}$ArH$^{+}$, was reported by \cite{Barlow13}, who used {\it Herschel} to observe the Crab Nebula. The same authors suggest that the detected emission lines are produced by electron collisions in partially ionised regions, where the electron densities are of the order of a few 100 cm$^{-3}$. Based on the work of {\cite{Barlow13}, \cite{Schilke14} then assigned a previously unidentified line, detected by {\it Herschel} in the Galactic sources Sagittarius B2(M) and (N), to $^{36}$ArH$^{+}$. It is therefore possible that some of the argon we would expect to see in the cores of galaxy groups and clusters, is actually locked up in such molecular ions, resulting in a central drop in the argon radial abundance profile. However, $^{36}$ArH$^{+}$ has yet to be detected in a galaxy group or cluster, and the amount of argon that can be depleted in such a manner is unclear. 

%The currently available NGC~4696 data are insufficient in quality to measure reliable radial profiles for both the neon and argon abundance. However, we submitted a proposal for a 500 ks {\it Chandra} observation during the Cycle 16 Call for Proposals, which was approved. We expect all the relevant observations to be completed in June 2014. As a result, the available exposure time for NGC~4696 will be more than trebled, which will in turn allow us to measure deprojected abundance profiles for argon and possibly neon.   

\section{Summary and conclusions}
In this paper, we searched for central abundance drops in the sources of our parent sample, for which we have reliable radial abundance profiles. As it consists of nearby groups and clusters which have well-resolved cores, our sample is ideal for the identification of central abundance drops, which cannot be detected with certainty in more distant sources. We find central iron abundance drops in fourteen sources: NGC~4636, NGC~4696, 2A0335+096, Abell~1991, Abell~262, NGC~5813, NGC~5044, NGC~5846, Abell~3581, NGC~6338, IC~1262, Abell~S1101, HCG~62 and NGC~4325. The first eight sources have certain central abundance drops, while the rest have possible central abundance drops. All of the aforementioned sources are also part of a subsample of 30 sources, all of which have X-ray cavities and a central cooling time $\leq$3 Gyr. This indicates that up to $\sim$47 percent of the sources that have X-ray cavities, also have central abundance drops (14/30). In NGC~4696, Abell~262 and 2A0335+096, we find evidence for a similar trend in the abundance profiles of silicon and sulphur, indicating a common origin for these metallicity drops. Following our earlier work on NGC~4696, we propose that these metallicity drops occur because of AGN feedback activity in these groups and clusters. Stellar mass loss processes eject metal-rich material into the core of a group or cluster, which can, at least in part, remain in  locked up in dust grains \citep{Voit11}. The ``missing'' iron is then embedded in dust grains, which are in turn embedded in cold gaseous filaments. These are detected in the optical and infrared, and are spatially coincident with the abundance drops seen in the X-ray. The peak in the observed metallicity profiles could be due to these dusty filaments being dragged outwards by the bubbling feedback process, and then being destroyed beyond a certain distance. Then, the metals will be returned to the hot X-ray emitting ICM, contributing to the creation of a local abundance peak. A lesser role in the creation of these abundance drops and peaks may be played by thermal diffusion. 

We point out that care must be taken when studying the spectra, whether projected or deprojected, from the cores of groups and clusters, especially when the metallicity is the property under study. This is particularly pertinent for clusters with a high column density towards their core, which can be mistaken for an extra thermal component in the modelling process. Furthermore, Fig. \ref{fig:cavsanddrops} and the spectral fitting results in Tables \ref{tab:multitfits} and \ref{tab:multitfits2}, highlight the need for high quality data to verify the existence of the central abundance drops presented in this paper. 

In terms of the available data quality influencing the detection of central abundance drops, we find that, generally, the possibility of detecting a central abundance drop increases with the number of counts within the central 20 kpc of a source. There is not, however, a cut-off number of counts, above which all sources have a central abundance drop. In addition, the magnitude of the central abundance drop, $Z_{\rm {mag}}$, is not affected by the number of counts in the central 20 kpc of a group or cluster. On the other hand, the statistical significance of the central abundance drop does depend on the number of these counts, as nearly all sources with more than $\sim$94000 counts in their central 20 kpc have a certain central abundance drop, while those with fewer counts have possible central abundance drops, with only a couple of exceptions. As expected from the scenario for the generation of central abundance drops, outlined in the previous paragraph, the distance at which the cavities lie, $D_{\rm {cav}}$, acts as an upper limit to the distance at which the abundance peak is observed, $D_{\rm {peak}}$, for most sources. Finally, no relation is found between the power of a cavity, $P_{\rm {cav}}$, and $Z_{\rm {mag}}$.  

Our proposed explanation for the origin of the observed abundance drops, that the ``missing'' iron, silicon and sulphur are locked up in dust grains, can be tested by measuring the radial abundance profiles of elements such as neon and argon. The recent discovery of the molecular ion $^{36}$ArH$^{+}$ may affect the shape of the argon abundance profile in the inner regions of a group or cluster, however the extent to which argon is depleted into this ion is unclear. 

{\bf Overall, we find that central abundance drops all occur in sources with X-ray cavities, all bar one exception having a central cooling time of $\leq$1 Gyr. From our analysis in this paper, it is evident that AGN feedback transports metals as well as energy into the surrounding ICM. Assuming that AGN feedback is common and long-lived, our results have implications for the origin and interpretation of abundance gradients in massive galaxies.}

One question that naturally arises from the here presented analysis, is whether central abundance drops exist in more distant clusters. There is evidence of large amounts of molecular gas residing in the cores of distant clusters \citep{Edge01, Odea08}, and even of AGN feedback-driven outflows of molecular gas \citep{McNamara14}. However, central abundance drops have yet to be discovered in distant clusters, such as Abell~2204 \citep{Sanders09}, Abell~2597 \citep{Hicks02} and PKS~0745-191 \citep{Sanders14}. Whether this is due to low spatial resolution, or whether distant clusters actually do not have a central abundance drop, remains unclear. More detailed work, focusing on the central regions of distant clusters, is needed to resolve this matter. 

\section*{Acknowledgements}
EKP acknowledges the support of a STFC studentship. We thank the anonymous referee for comments that improved the contect of this paper. EKP would like to thank Helen Russell and Stephen Walker for helpful discussions and suggestions. 

The plots in this paper were created using {\sc veusz}.\footnote{http://home.gna.org/veusz/}
This research has made use of the NASA/IPAC Extragalactic Database (NED)\footnote{http://ned.ipac.caltech.edu/} which is operated by the Jet Propulsion Laboratory, California Institute of Technology, under contract with the National Aeronautics and Space Administration.

\bibliographystyle{mn2e}
\bibliography{references}

\appendix
\section{Single- and multi-temperature fits of source inner regions}
In this Appendix, we present the single-temperature (1-T) and two-temperature (2-T) fits of the innermost regions of the sources under study in this paper, with the exception of NGC~4696. We refer the reader to \cite{Panagoulia13} for a detailed analysis of the innermost regions of NGC~4696. Only the results from the annuli of each group or cluster in which we consider there to be an abundance drop (see Sections 5.1.1--5.1.14), are shown in the table below. The annuli are labelled 1st, 2nd, etc with increasing radius, i.e. the 1st annulus in a source is the one closest to the centre (the one closest to the x-axis zero point in the individual metallicity profiles in Figs. \ref{fig:abundprofscertain} and \ref{fig:abundprofspossible}), and so on. In Table \ref{tab:multitfits}, columns (1) and (2) give the name of the source and the annulus number, respectively. Columns (3)--(5) show the results of the 1-T fits, and columns (6)--(9) those from the 2-T fits, respectively. The temperatures in columns (3), (6) and (7) are given in keV. The iron abundance, $Z_{\rm {Fe}}$, in columns (4) and (8) is given with respect to Solar metallicity. We note that in the cases of NGC~5813 (1st annulus), Abell~3581 (1st annulus) and Abell~262 (1st annulus), in the metallicity profiles presented in Fig. \ref{fig:abundprofscertain}, the relevant metallicity values were obtained using the results from the 2-T fits.   

In previous studies of the inner regions of CC groups and clusters, the complex temperature structure there has been modelled using multi-temperature models, where one temperature component is allowed to vary freely, while the rest are set to a fixed fraction of this temperature. In our case, we have named this freely varying temperature $T_{\rm {high}}$, and the temperatures were fixed at 0.5 times the value of $T_{\rm {high}}$ for the 2-T fits, and at 0.5 and 0.25 times the value of $T_{\rm {high}}$ in the 3-T fits. The results of these fits for each of the 14 sources are shown in Table \ref{tab:multitfits2}. As for Table \ref{tab:multitfits}, columns (1) and (2) indicate of the source and the annulus number, respectively. Columns (3)--(5) show the results of the 2-T fits with one fixed temperature component, while columns (6)--(8) show the results of the 3-T fits with two fixed temperature components. As can be seen from this table, in some cases even when freezing the additional temperature components at fixed ratios to each other, the metallicity cannot be constrained. 

As can be seen from the two tables in this appendix, there are definite abundance drops in the cores of some of the sources, such as 2A0335+096, that persist independent of the number of thermal components used to model the ICM emission. For some of the sources that have data of a poorer quality, the addition of extra thermal components does not significantly change the statistical ``goodness'' of the fit and results in large error bars, although it does alter the measured metallicity. As also indicated by Fig. \ref{fig:cavsanddrops}, better quality data are needed to confirm or disprove the existence of these abundance drops.

We note that in all the multi-temperature fits, the metallicities of the individual thermal components were tied, while their normalisations were allowed to vary independently.  

To create the metallicity profiles in Figs. \ref{fig:abundprofscertain} and \ref{fig:abundprofspossible},  we used the 2-T metallicity measurements when the two temperature values and the metallicity could be well constrained. This was the case with NGC~5813 (1st annulus), Abell~3581 (1st annulus) and Abell~262 (1st annulus). We used the results of the 1-T fits, whenever constraining the temperatures and/or metallicity using a 2-T fit was not feasible. As can be seen in Table \ref{tab:multitfits}, in a few cases the addition of an extra thermal component does not significantly improve the statistical ``goodness'' of the fit, and in fact constraining the value of the temperature of the additional component, and/or the abundance, is not possible (e.g. 2nd annulus of NGC~4636 and HCG~62). For several groups and clusters, the addition of a second thermal component resulted in much larger errorbars (than in the 1-T fits) on either one of the two temperature values, and/or the metallicity, while the change in the statistical ``goodness'' of the fit was, albeit statistically significant, relatively small. Some extreme examples of this are NGC~5044 and NGC~4325.

In general, we used the 2-T metallicity measurements when the two temperature values and the metallicity could be well constrained. In the case of unconstrained temperature values and/or metallicity values, we use the 1-T results. When the addition of an extra thermal component led to the generation of much larger errorbars for the measured metallicity and/or temperatures, than in the 1-T fits, we again use the 1-T fit results. 

\begin{table*}
\begin{center}
\footnotesize{ 
\begin{tabular}{ccccccccc}
  \multicolumn{1}{c}{Source name}&\multicolumn{1}{c}{Annulus number}&\multicolumn{3}{c}{1-T fit}&\multicolumn{4}{c}{2-T fit}\\
  \hline
  \multicolumn{1}{c}{}&\multicolumn{1}{c}{}&\multicolumn{1}{c}{$k_{\rm {B}}$T}&\multicolumn{1}{c}{$Z_{\rm {Fe}}$} &\multicolumn{1}{c}{$\chi^{2}$/d.o.f.}&\multicolumn{1}{c}{$k_{\rm {B}}$T$_{1}$}&\multicolumn{1}{c}{$k_{\rm {B}}$T$_{2}$}&\multicolumn{1}{c}{$Z_{\rm {Fe}}$}&\multicolumn{1}{c}{$\chi^{2}$/d.o.f.}\\
  \multicolumn{1}{c}{}&\multicolumn{1}{c}{}&\multicolumn{1}{c}{(keV)}&\multicolumn{1}{c}{(Z$_{\odot}$)}&\multicolumn{1}{c}{}&\multicolumn{1}{c}{(keV)}&\multicolumn{1}{c}{(keV)}&\multicolumn{1}{c}{(Z$_{\odot}$)}&\multicolumn{1}{c}{}\\
  \multicolumn{1}{c}{(1)}&\multicolumn{1}{c}{(2)}&\multicolumn{1}{c}{(3)}&\multicolumn{1}{c}{(4)}&\multicolumn{1}{c}{(5)}&\multicolumn{1}{c}{(6)}&\multicolumn{1}{c}{(7)}&\multicolumn{1}{c}{(8)}&\multicolumn{1}{c}{(9)}\\
    \hline
NGC~4636 & 1st & 0.52$^{+0.04}_{-0.03}$ & 0.38$^{+0.18}_{-0.10}$ & 211.62/204 & 0.21$^{+0.19}_{-0.08}$ & 0.60$^{+0.23}_{-0.03}$ & 1.30$^{+0.87}_{-0.83}$ & 194.75/202\\
 & 2nd & 0.57$\pm$0.03 & 0.55$^{+0.60}_{-0.21}$ & 242.58/204 & 0.56$\pm$0.03 & $<$63.2 & $<$0.92 & 228.28/202\\
\hline
Abell~1991 & 1st & 1.06$\pm$0.02 & 0.39$^{+0.09}_{-0.07}$ & 110.15/95 & 1.01$^{+0.03}_{-0.07}$ & 1.99$^{+0.96}_{-1.06}$ & 0.99$^{+0.87}_{-0.38}$ & 91.34/93\\
 & 2nd & 1.64$^{+0.07}_{-0.09}$ & 0.64$^{+0.27}_{-0.19}$ & 95.55/95 & $<$1.06 & $<$1.89 & 0.92$^{+1.04}_{-0.45}$ & 94.17/93 \\
\hline
%Abell~2052 & 1st & 1.22$\pm$0.04 & 0.27$^{+0.06}_{-0.05}$ & 1421.96/1425 & 1.01$^{+0.04}_{-0.06}$ & 2.11$^{+0.42}_{-0.33}$ & 1.05$^{+0.72}_{-0.64}$ & 1387.15/1423\\
% & 2nd & 1.34$\pm$0.01 & 0.42$^{+0.05}_{-0.04}$ & 1476.53/1425 & 1.03$^{+0.76}_{-0.24}$ & 1.55$^{+0.23}_{-0.13}$ & 0.64$^{+0.17}_{-0.13}$ & 1463.32/1423\\
% & 3rd & 1.45$\pm$0.03 & 0.32$\pm$0.04 & 1569.41/1425 & 1.03$\pm$0.03 & 1.99$^{+0.11}_{-0.12}$ & 0.81$^{+0.17}_{-0.14}$ & 1432.69/1423\\
%\hline 
2A0335+096 & 1st & 1.25$^{+0.03}_{-0.04}$ & 0.21$^{+0.05}_{-0.04}$ & 471.96/482 & $<$0.99 & 1.39$^{+0.32}_{-0.20}$ & 0.27$^{+0.16}_{-0.08}$ & 469.54/480 \\
 & 2nd & 1.45$^{+0.06}_{-0.07}$ & 0.29$^{+0.07}_{-0.06}$ & 502.25/483 & 1.19$^{+0.20}_{-0.14}$ & 1.99$^{+0.67}_{-0.28}$ & 0.54$^{+0.25}_{-0.18}$ & 487.24/481\\
 & 3rd & 1.57$\pm$0.06 & 0.37$^{+0.10}_{-0.08}$ & 506.47/483 & $<$0.27 & 1.61$^{+0.06}_{-0.08}$ & 0.43$^{+0.14}_{-0.11}$ & 504.39/481\\
% & 4th & 1.65$^{+0.04}_{-0.05}$ & 0.33$^{+0.07}_{-0.06}$ & 514.90/483 & 1.21$^{+0.38}_{-0.20}$ & 2.00$^{+1.10}_{-0.19}$ & 0.50$^{+0.21}_{-0.16}$ & 507.78/481 \\
\hline
Abell~262 & 1st & 0.98$\pm$0.01 & 0.42$^{+0.08}_{-0.06}$ & 283.44/223 & 0.94$^{+0.03}_{-0.14}$ & 1.81$^{+0.57}_{-0.65}$ & 0.78$^{+0.33}_{-0.20}$ & 264.44/221\\
 & 2nd & 1.24$\pm$0.02 & 0.61$^{+0.11}_{-0.09}$ & 299.59/223 & 1.01$^{+0.04}_{-0.12}$  & 1.66$^{+0.16}_{-0.18}$ & 1.50$^{+0.65}_{-0.42}$ & 260.57/223\\
\hline
NGC~5044 & 1st & 0.84$\pm$0.01 & 0.31$\pm$0.03 & 324.31/289 & 0.59$^{+0.66}_{-0.09}$ & 0.92$^{+0.34}_{-0.03}$ & 0.43$^{+0.08}_{-0.06}$ & 297.72/285 \\
\hline
NGC~5813 & 1st & 0.71$\pm$0.005 & 0.32$\pm$0.02 & 1308.55/800 & 0.57$^{+0.02}_{-0.04}$ & 0.85$^{+0.02}_{-0.06}$ & 0.38$\pm$0.02 & 1169.86/798\\
\hline
NGC~5846 & 1st & 0.80$\pm$0.02 & 0.20$^{+0.03}_{-0.02}$ & 249.70/140 & 0.78$\pm$0.02 & $<$64.0 & 0.27$^{+0.06}_{-0.04}$ & 169.22/138 \\
 & 2nd & 0.72$\pm$0.02 & 0.30$^{+0.08}_{-0.06}$ & 142.29/138 & 0.56$^{+0.10}_{-0.08}$ & 0.88$^{+0.23}_{-0.11}$ & 0.39$^{+0.15}_{-0.09}$ & 132.06/136\\
\hline
NGC~4325 & 1st & 0.84$^{+0.03}_{-0.02}$ & 0.56$^{+0.28}_{-0.15}$ & 97.75/78 & 0.83$^{+0.02}_{-0.03}$ & 2.15$^{+9.54}_{-0.73}$ & 1.12$^{+2.10}_{-0.54}$ & 90.96/76 \\
\hline
Abell~3581 & 1st & 1.04$\pm$0.03 & 0.17$^{+0.05}_{-0.04}$ & 120.37/134 & 0.76$^{+0.10}_{-0.18}$ & 1.34$^{+0.20}_{-0.13}$ & 0.43$^{+0.29}_{-0.15}$ & 103.77/132\\
 & 2nd & 1.33$^{+0.17}_{-0.05}$ & 0.34$^{+0.14}_{-0.10}$ & 138.18/136 & 1.20$^{+0.12}_{-0.014}$ & 3.06$^{+21.68}_{-0.91}$ & 1.03$^{+1.37}_{-0.60}$ & 121.03/134 \\
 \hline
%Abell~496 & 1st & 1.81$^{+0.08}_{-0.14}$ & 0.56$^{+0.10}_{-0.14}$ & 233.69/210 & 1.30$^{+0.33}_{-0.29}$ & 2.39$^{+0.16}_{-0.21}$ & 0.81$^{+0.22}_{-0.18}$ & 201.23/208 \\
%\hline
NGC~6338 & 1st & 1.25$^{+0.05}_{-0.06}$ & 0.43$^{+0.22}_{-0.14}$ & 66.40/39 & 1.03$^{+0.06}_{-0.15}$ & 2.13$^{+0.67}_{-0.55}$ & 2.62$^{+25.08}_{-1.92}$ & 57.61/37 \\ 
 & 2nd & 1.37$^{+0.31}_{-0.12}$ & 0.63$^{+0.63}_{-0.41}$ & 27.47/39 & $<$0.008 & 1.38$^{+0.30}_{-0.13}$ & $<$0.67 & 27.48/37\\
\hline
IC~1262 & 1st & 1.18$^{+0.06}_{-0.07}$ & 0.23$^{+0.09}_{-0.07}$ & 175.13/179 & 1.02$\pm$0.05 & 3.54$^{+1.59}_{-0.97}$ & $<$1.43 & 145.70/177\\
\hline
%Abell~4059 & 1st & 1.64$\pm$0.04 & 0.53$^{+0.09}_{-0.08}$ & 299.98/275 & 1.00$^{+0.17}_{-0.19}$ & 1.96$^{+0.17}_{-0.19}$ & 0.98$^{+0.35}_{-0.25}$ & 282.77/273\\
%\hline
HCG~62 & 1st & 0.79$\pm$0.01 & 0.65$^{+0.16}_{-0.11}$ & 255.85/219 & 0.78$\pm$0.01 & $<$5.92 & 0.72$^{+0.30}_{-0.13}$ & 252.93/217\\
\hline
Abell~S1101 & 1st & 1.68$^{+0.10}_{-0.08}$ & 0.35$^{+0.12}_{-0.10}$ & 208.83/193 & 0.56$^{+0.19}_{-0.10}$ & 1.75$^{+0.47}_{-0.36}$ & 0.42$^{+0.24}_{-0.13}$ & 205.10/191\\
\hline
\end{tabular}
}
\end{center}
\caption{Single- (1-T) and two-temperature (2-T) fits for the central spectral bins in which an abundance drop is detected or suspected, for each of the 14 sources under study. Column (1) gives the name of the source, column(2) indicates the annulus number (1st is the innermost annulus, 2nd the second innermost and so on), columns (3)--(5) show the results of the 1-T fits, and columns (6)--(9) those of the 2-T fits.}
\label{tab:multitfits}
\end{table*}

\begin{table*}
\begin{center}
\footnotesize{ 
\begin{tabular}{cccccccc}
  \multicolumn{1}{c}{Source name}&\multicolumn{1}{c}{Annulus number}&\multicolumn{3}{c}{2-T fit fixed}&\multicolumn{3}{c}{3-T fit fixed}\\
  \hline
  \multicolumn{1}{c}{}&\multicolumn{1}{c}{}&\multicolumn{1}{c}{$k_{\rm {B}}$T$_{\rm {high}}$}&\multicolumn{1}{c}{$Z_{\rm {Fe}}$} &\multicolumn{1}{c}{$\chi^{2}$/d.o.f.}&\multicolumn{1}{c}{$k_{\rm {B}}$T$_{\rm {high}}$}&\multicolumn{1}{c}{$Z_{\rm {Fe}}$}&\multicolumn{1}{c}{$\chi^{2}$/d.o.f.}\\
  \multicolumn{1}{c}{}&\multicolumn{1}{c}{}&\multicolumn{1}{c}{(keV)}&\multicolumn{1}{c}{(Z$_{\odot}$)}&\multicolumn{1}{c}{}&\multicolumn{1}{c}{(keV)}&\multicolumn{1}{c}{(Z$_{\odot}$)}&\multicolumn{1}{c}{}\\
  \multicolumn{1}{c}{(1)}&\multicolumn{1}{c}{(2)}&\multicolumn{1}{c}{(3)}&\multicolumn{1}{c}{(4)}&\multicolumn{1}{c}{(5)}&\multicolumn{1}{c}{(6)}&\multicolumn{1}{c}{(7)}&\multicolumn{1}{c}{(8)}\\
    \hline
NGC~4636 & 1st & 0.62$^{+0.23}_{-0.04}$ & 0.81$^{+1.28}_{-0.33}$ & 195.74/203 & 0.61$^{+0.32}_{-0.04}$ & $<$1.20 & 194.19/202 \\
 & 2nd & 0.57$\pm$0.57 & 0.55$^{+0.60}_{-0.20}$ & 242.58/203 & 1.13$\pm$0.07 & 0.59$^{+0.78}_{-0.24}$ & 242.37/202 \\
\hline
Abell~1991 & 1st & 2.02$^{+0.06}_{-0.07}$ & 1.00$^{+0.69}_{-0.38}$ & 91.35/94 & 2.02$^{+0.10}_{-0.04}$ & 1.00$^{+0.74}_{-0.33}$ & 91.35/93\\
 & 2nd & 1.83$^{+1.81}_{-0.26}$ & 0.97$^{+1.06}_{-0.52}$ & 94.68/94 & 3.32$^{+0.58}_{-1.75}$ & 0.86$^{+1.77}_{-0.39}$ & 94.16/93 \\
\hline
2A0335+096 & 1st & 1.28$\pm$0.06 & 0.24$^{+0.08}_{-0.06}$ & 470.29/481 & 1.28$\pm$0.06 & 0.24$^{+0.08}_{-0.05}$ & 470.29/480\\ 
 & 2nd & 1.54$\pm$0.11 & 0.38$^{+0.15}_{-0.11}$ & 498.99/482 & 2.62$^{+0.10}_{-0.09}$ & 0.43$^{+0.25}_{-0.10}$ & 489.81/481\\
 & 3rd & 1.62$^{+0.08}_{-0.09}$ & 0.42$^{+0.16}_{-0.12}$ & 505.46/482 & 1.61$^{+0.08}_{-0.07}$ & 0.42$^{+0.15}_{-0.11}$ & 504.53/481 \\
% & 4th & 0.96$^{+0.20}_{-0.23}$ & 0.69$^{+0.21}_{-0.29}$ & 537.66/513 & 1.93$^{+0.20}_{-0.21}$ & 0.69$^{+0.21}_{-0.16}$ & 537.66/512 \\
\hline
Abell~262 & 1st & 1.89$^{+0.04}_{-0.05}$ & 0.79$^{+0.33}_{-0.21}$ & 264.54/222 & 1.94$\pm$0.06 & 0.95$^{+0.47}_{-0.28}$ & 259.77/221 \\
 & 2nd & 1.30$\pm$0.03 & 0.98$^{+0.31}_{-0.21}$ & 278.58/222 & 1.30$\pm$0.03 & 0.98$^{+0.31}_{-0.21}$ & 278.58/221 \\
\hline
NGC~5044 & 1st & 0.90$\pm$0.02 & 0.40$^{+0.07}_{-0.05}$ & 312.65/288 & 0.90$\pm$0.03 & 0.40$^{+0.07}_{-0.05}$ & 312.65/287 \\
\hline
NGC~5813 & 1st & 0.74$\pm$0.01 & 0.36$\pm$0.02 & 1240.42/798 & 0.74$\pm$0.01 & 0.41$^{+0.04}_{-0.03}$ & 1240.42/797 \\
\hline
NGC~5846 & 1st & 0.93$^{+0.05}_{-0.04}$ & 0.28$^{+0.06}_{-0.05}$ & 241.81/137 & 0.93$^{+0.05}_{-0.04}$ & 0.28$\pm$0.05 & 241.81/136 \\
 & 2nd & 0.76$^{+0.15}_{-0.04}$ & 0.35$^{+0.11}_{-0.07}$ & 138.78/137 & 0.76$^{+0.09}_{-0.04}$ & 0.35$^{+0.11}_{-0.08}$ & 138.78/136 \\
\hline
NGC~4325 & 1st & 0.84$^{+0.05}_{-0.02}$ & 0.56$^{+0.28}_{-0.14}$ & 97.76/77 & 0.84$^{+0.05}_{-0.02}$ & 0.56$^{+0.28}_{-0.15}$ & 97.75/76 \\
\hline
Abell~3581 & 1st & 1.30$^{+0.24}_{-0.09}$ & 0.43$^{+0.27}_{-0.15}$ & 104.60/133 & 1.30$^{+0.23}_{-0.09}$ & 0.43$^{+0.23}_{-0.15}$ & 104.60/132\\
 & 2nd & 2.35$^{+0.23}_{-0.25}$ & 0.88$^{+0.77}_{-0.41}$ & 115.86/134 & 2.35$^{+0.26}_{-0.24}$ & 0.87$^{+0.85}_{-0.40}$ & 115.87/133\\ 
\hline 
NGC~6338 & 1st & 2.05$^{+0.13}_{-0.25}$ & $<$2.21 & 57.68/38 & 4.10$^{+0.26}_{-0.49}$ & $<$2.21 & 57.68/37 \\
& 2nd & 1.59$^{+1.85}_{-0.31}$ & $<$1.78 & 26.37/38 & 1.65$^{+1.69}_{-0.29}$ & $<$4.71 & 25.55/37\\
\hline
IC~1262 & 1st & 2.04$^{+0.11}_{-0.14}$ & 0.57$^{+0.39}_{-0.25}$ & 154.61/178 & 4.05$^{+0.20}_{-0.27}$ & 1.46$^{+2.78}_{-0.98}$ & 145.72/177 \\
\hline 
HCG~62 & 1st & 1.56$\pm$0.03 & 0.71$^{+0.25}_{-0.16}$ & 254.98/218 &  1.56$\pm$0.03 & 0.71$^{+0.21}_{-0.13}$ & 254.98/217 \\
\hline 
Abell~S1101 & 1st & 1.73$^{+0.20}_{-0.10}$ & 0.38$^{+0.19}_{-0.11}$ & 206.66/192 & 1.75$^{+0.21}_{-0.05}$ & 0.42$^{+0.25}_{-0.13}$ & 205.23/191\\
\hline
\end{tabular}
}
\end{center}
\caption{Two- (2-T) and three-temperature (3-T) fits with fixed temperature ratios, for the central spectral bins in which an abundance drop is detected or suspected, for each of the 14 sources under study. For the 2-T fits, the second temperature components were fixed to half the value of the higher temperature, $T_{\rm {high}}$. In the 3-T fits, one additional temperature component was fixed to 0.5 times the value of $T_{\rm {high}}$, while the other was fixed at 0.25 times the value of $T_{\rm {high}}$. Column (1) gives the name of the source, column(2) indicates the annulus number (1st is the innermost annulus, 2nd the second innermost and so on), columns (3)--(5) show the results of the 2-T fits with one fixed temperature value, and columns (6)--(8) those of 3-T fits with two fixed temperature values.}
\label{tab:multitfits2}
\end{table*}
\end{document}